\documentclass[draftcls,journal,onecolumn]{IEEEtran}

\usepackage{epsfig,amsfonts,amsmath,graphicx,epsfig}
\usepackage{subfigure,graphics,amssymb,amsxtra,color}
\usepackage{algorithm}
\usepackage{algorithmic}

\usepackage[ansinew]{inputenc}

\newtheorem{lemma}{Lemma}
\newtheorem{theorem}{Theorem}

\newtheorem{proof}{Proof}

\newtheorem{example}{Example}

\newcommand{\HH}{\mathbf{H}}

\newcommand{\x}{\mathbf{x}}
\newcommand{\y}{\mathbf{y}}

\newcommand{\n}{\mathbf{n}}

\begin{document}

\title{Multiple-Antenna Fading Coherent Channels with Arbitrary Inputs: Characterization and Optimization of the Reliable Information Transmission Rate}

\author{Miguel~R.~D.~Rodrigues,~\IEEEmembership{Member,~IEEE}
\thanks{Miguel R. D. Rodrigues was with Instituto de Telecomunica\c{c}\~oes--Porto and Departamento de Ci\^encia de Computadores, Universidade do Porto, Portugal. He is now with the Department of Electronic and Electrical Engineering, University College London, United Kingdom [e-mail: m.rodrigues@ucl.ac.uk]. The work of M. R. D. Rodrigues was supported by Funda\c{c}\~ao para a Ci\^encia e a Tecnologia, Portugal through the research project PTDC/EEA-TEL/100854/2008. This paper was presented in part at the IEEE International Symposium on Information Theory 2011 and 2012.}}

\maketitle


\begin{abstract}
We investigate the constrained capacity of multiple-antenna fading coherent
channels, where the receiver knows the channel state but the transmitter knows only
the channel distribution, driven by arbitrary equiprobable discrete inputs in a regime of high signal-to-noise ratio (${\sf snr}$). In particular, we capitalize on intersections between information theory and estimation theory to conceive expansions to the average minimum-mean squared error (MMSE) and the average mutual information, which leads to an expansion of the constrained capacity, that capture well their behavior in the asymptotic regime of high ${\sf snr}$. We use the expansions to study the constrained capacity of various multiple-antenna fading coherent channels, including Rayleigh fading models, Ricean fading models and antenna-correlated models. The analysis unveils in detail the impact of the number of transmit and receive antennas, transmit and receive antenna correlation, line-of-sight components and the geometry of the signalling scheme on the reliable information transmission rate. We also use the expansions to design key system elements, such as power allocation and precoding schemes, as well as to design space-time signalling schemes for multiple-antenna fading coherent channels. Simulations results demonstrate that the expansions lead to very sharp designs.
\end{abstract}


\newpage

\section{Introduction}
\label{introduction}

In recent years, the surge of interest in multiple-antenna communications systems has been due to the realization that the capacity of the canonical independent and identically distributed (i.i.d.) Rayleigh fading coherent channel, where the receiver knows the exact channel state but the transmitter knows only the channel distribution, scales as~\cite{Foschini96},~\cite{Foschini98},~\cite{Telatar99}:
\begin{align}
C ({\sf snr}) = \min (n_t,n_r) \log ({\sf snr}) + \mathcal{O} (1) \label{scaling_law}
\end{align}
so that, at high signal-to-noise ratio (${\sf snr}$), the potential gain in reliable information transmission rate of a multiple-antenna over a single-antenna system grows (linearly) with the minimum of the number of transmit or receive antennas, $\min (n_t,n_r)$, a quantity naturally known as the multiplexing gain or the degrees of freedom.

Of particular relevance has also been the characterization of the capacity of multi-antenna communications systems that embody prominent channel features that go beyond the canonical model. The crux of the characterizations, which for analytical tractability have been usually pursued in the asymptotic regimes of low- and high-${\sf snr}$, are affine expansions of the capacity in terms of fundamental performance measures. At low-${\sf snr}$, in order to capture the tradeoff between rate, bandwidth and power, it is appropriate to expand the capacity as an affine function of $\frac{E_b}{N_0}\big|_{\text{dB}}$ as follows~\cite{Verdu02},~\cite{Tulino03}:
\begin{align}
C \left(\frac{E_b}{N_0}\right) = \mathcal{S}_0 \cdot \left(\frac{\frac{E_b}{N_0}\big|_{\text{dB}}}{3~\text{dB}} - \frac{{\frac{E_b}{N_0}}_{\text{min}}\big|_{\text{dB}}}{3~\text{dB}}\right) + o \left(\frac{E_b}{N_0} - {\frac{E_b}{N_0}}_{\text{min}}\right) \approx \mathcal{S}_0 \cdot \log_2 \left(\frac{E_b}{N_0}\bigg/{\frac{E_b}{N_0}}_{\text{min}}\right)
\end{align}
where $\frac{E_b}{N_0}\big|_{\text{dB}}$ is the (transmitted) energy per information bit $\frac{E_b}{N_0}$ in dB, ${\frac{E_b}{N_0}}_{\text{min}}\big|_{\text{dB}}$ is the minimum (transmitted) energy per information bit ${\frac{E_b}{N_0}}_{\text{min}}$ required for reliable communication in dB, and $\mathcal{S}_0$ is the capacity slope therein in bit/s/Hz/(3 dB). At high-${\sf snr}$, the capacity is expanded as an affine function of ${\sf snr}|_{\text{dB}}$ as follows~\cite{Shamai01},~\cite{Lozano05}:
\begin{align}
C ({\sf snr}) = \mathcal{S}_{\infty} \left(\frac{{\sf snr}|_{\text{dB}}}{3~\text{dB}} - \mathcal{L}_{\infty}\right) + o(1)
\end{align}
where ${\sf snr}|_{\text{dB}}$ denotes ${\sf snr}$ in dB, $\mathcal{S}_{\infty}$ denotes the high-${\sf snr}$ slope in bits/s/Hz/(3 dB) and $\mathcal{L}_{\infty}$ denotes a zero-order term or a power offset in 3-dB units with respect to a reference channel with the same high ${\sf snr}$ slope but with unfaded and orthogonal dimensions whose expansion intersects the origin at ${\sf snr}|_{\text{dB}} = 0$. Since the low- and high-${\sf snr}$ quantities, $\mathcal{S}_0$ and ${\frac{E_b}{N_0}}_{\text{min}}$ as well as $\mathcal{S}_{\infty}$ and $\mathcal{L}_{\infty}$ are a function of the random channel matrix, recourse to the elegant theory of random matrices~\cite{Tulino04} has disclosed the influence of various factors such as the number of transmit and receive antennas, antenna correlation, antenna polarization, line-of-sight components, spatially colored noise, interference, or even key signal features, on the reliable information transmission rate~\cite{Verdu02},~\cite{Tulino03},~\cite{Lozano05},~\cite{Tulino05},~\cite{Lozano06a},~\cite{Venkatesan03},~\cite{Kim03}, offering a more realistic view of the potential of multiple-antenna communications. In general, the channel capacity, by virtue of the well-known $\log \det (\cdot)$ expression~\cite{Telatar99}, depends ultimately on the distribution of the eigenvalues of certain random matrices linked to the random channel matrix, a distribution which is known in some settings and in some asymptotic regimes~\cite{Tulino04}. Consequently, other characterizations valid for general ${\sf snr}$, rather than only asymptotic ${\sf snr}$ regimes, have also been pursued.

It is well known, assuming that the channel variation over time is stationary and ergodic, that the capacity of the multi-antenna fading coherent channel is achieved by using (complex) Gaussian inputs, i.e., by using codewords whose elements are drawn from a zero-mean circularly symmetric complex Gaussian distribution that satisfy a transmit power constraint. However, it is also very relevant both from the theoretical and perhaps more importantly the practical perspective to study the constrained capacity of multi-antenna fading coherent channels driven by arbitrary (discrete) inputs. This is due to the fact that practical constraints pertaining to the transmission and reception of information often dictate the use of discrete inputs, such as PSK or QAM constellations, \emph{in lieu} of the ideal Gaussian ones. Such a study, in addition to a deeper understanding, could also offer concrete guidelines for signal and system design. 

The characterization of the constrained capacity of systems driven by non-Gaussian inputs poses a myriad of challenges due to the absence of explicit and tractable mutual information expressions. A innovative approach towards the resolution of this class of problems was put forth in~\cite{Lozano06b}, by exploiting connections between key quantities in information theory and estimation theory~\cite{Guo05},~\cite{Palomar06a}. By drawing upon the I-MMSE identity~\cite{Guo05}, Lozano \emph{et al}.~\cite{Lozano06b} have studied optimal power allocation policies for parallel non-interfering Gaussian channels with arbitrary inputs. P\'erez-Cruz \emph{et al}.~\cite{Perez-Cruz10} and Paray\'o and Palomar~\cite{Parayo09a},~\cite{Parayo09b} have in turn studied optimal power allocation and precoding for (interfering) multiple-input multiple-output (MIMO) Gaussian channels with arbitrary inputs (see also~\cite{Lamarca09},~\cite{Xiao08},~\cite{Xiao11},~\cite{Zeng11}, ~\cite{Wang11}). Optimum power allocation for multiuser OFDM systems with arbitrary signal constellations, both in scenarios where the transmitter knows the fading channel state and in scenarios where the transmitter knows only the fading channel distribution, has been addressed using identical techniques in~\cite{Lozano08}. 

This paper pursues the characterization of the reliable information transmission rate of multi-antenna fading coherent channels with arbitrary (discrete) inputs, by capitalizing on the intersections between information theory and estimation theory. The coherent channel model, where the receiver is assumed to know the exact channel state but the transmitter is only assumed to know the channel distribution, is a particularly relevant one because the use of bandwidth limited feedback channels between the receiver and the transmitter only enables the transmission of statistical, rather than instantaneous, channel state information in (fast fading) mobile communications systems. Due to the difficulty in constructing a general analytical characterization valid for all ${\sf snr}$, the analysis focus on the key asymptotic regime of high ${\sf snr}$ in order to shed important insight about the fundamental communication limits. It is important to note, though, that the characterization of the reliable rate in coherent fading channels, where the receiver knows the exact channel state but the transmitter knows only the channel distribution, is considerably more complex than in MIMO Gaussian channels, where the (fixed) channel matrix is known exactly to the receiver and the transmitter (e.g., see~\cite{Perez-Cruz10},~\cite{Parayo09a} and~\cite{Parayo09b}). The most important aspect relates to the fact that the reliable rate is defined by the average of the mutual information with respect to the channel matrix distribution, rather than the mutual information conditioned on the (fixed) channel matrix only, as in~\cite{Perez-Cruz10},~\cite{Parayo09a} and~\cite{Parayo09b}. Consequently, we use the I-MMSE identity as a platform to explore other key analysis techniques, most notably, the machinery of asymptotic analysis and expansions, that lead to the exposure of the behavior of the reliable rate of multi-antenna fading coherent channels driven by arbitrary (discrete) inputs in the asymptotic regime of high ${\sf snr}$. This paves the way to the characterization of the constrained capacity of key multi-antenna fading coherent channel models as well as the design of key system elements.

\vspace{-0.4cm}

\subsection{Contributions}

This paper contains various original contributions, which include:

\begin{itemize}

    \vspace{0.1cm}

\item Analytic characterization of the constrained capacity of multi-antenna fading coherent channels driven by arbitrary discrete inputs in the asymptotic regime of high ${\sf snr}$. The contribution reveals that, whilst at low-${\sf snr}$ the capacity of a multiple-antenna coherent channel driven by Gaussian inputs and the constrained capacity of a multiple-antenna coherent channel driven by (proper complex) non-Gaussian inputs admit a similar characterization~\cite{Verdu02}, at high-${\sf snr}$ the behavior of the reliable rate for systems driven by the capacity-achieving Gaussian inputs and for systems driven by arbitrary discrete inputs is radically different; the contribution also reveals that the asymptotic characterization of the constrained capacity depends on the distribution of certain quadratic forms in Gaussian random variables, rather than the distribution of the singular values or the eigenvalues of certain random matrices. This aspect facilitates considerably the characterization of the constrained capacity of multi-antenna fading coherent channels.

    \vspace{0.1cm}

\item Analysis of the constrained capacity of various multi-antenna fading coherent channel models, including Rayleigh fading models, Ricean fading models and antenna-correlated models. The contribution emphasizes the impact of the channel properties as well as the signal geometry on the constrained capacity of multi-antenna fading coherent channels

    \vspace{0.1cm}

\item Design of system elements for various multi-antenna fading coherent channel models. In particular, the contribution illustrates with some detail the design of power allocation and precoding procedures as well as the construction of space-time schemes for various models.

    \vspace{0.1cm}

\end{itemize}

In addition, the analysis also unveils intimate connections between the asymptotic behavior of key performance measures, namely, the average (non-linear) minimum mean-squared error, the average mutual information and the average error probability, whose interest may transcend the domain of application.

\vspace{-0.4cm}

\subsection{Organization}

This paper is organized as follows: Section \ref{system_model} describes the multiple-antenna fading coherent channel model. Sections \ref{characterization} concentrates on the construction of a high-${\sf snr}$ expansion for the reliable information transmission rate -- the constrained capacity -- of multiple-antenna fading coherent channels driven by arbitrary discrete inputs. The application of the expansions in the characterization of the reliable rate of the multiple-antenna canonical i.i.d. Rayleigh fading coherent channel as well as other multiple-antenna coherent channels that incorporate a variety of features is considered in Sections  \ref{canonical_channel} and \ref{other_channels}, respectively. We study in detail the effect on the reliable rate of Rayleigh fading, Ricean fading, the number of transmit and receive antennas, antenna correlation, as well as the signal properties. In turn, the application of the expansions in the design of communication system elements is considered in Section \ref{designs}. In particular, we study power allocation in a bank of parallel independent fading channels, power allocation and precoding in multiple-antenna fading channels, and space-time signal design. Section \ref{conclusions} summarizes the main conclusions and contributions of the paper.

\vspace{-0.4cm}

\subsection{Notation}

We use the notation: Boldface uppercase letters denote matrices (${\bf X}$), boldface lowercase
letters denote column vectors (${\bf x}$), and italics denote scalars ($x$); the context defines whether the quantities are deterministic or random. The symbols ${\bf I}$, ${\bf 0}$ and ${\sf diag}\left(d_1,d_2,\ldots,d_n\right)$ represent the identity matrix, the null matrix and a diagonal matrix with diagonal elements $d_1,d_2,\ldots,d_n$, respectively. The symbol ${\sf e}_k$ represents a unit vector where the kth entry is equal to one and the other entries are equal to zero. The operators ${\sf det} \left(\cdot\right)$, ${\sf tr} \left(\cdot\right)$, ${\sf rank} \left(\cdot\right)$, ${\sf vec}\left(\cdot\right)$ and $\left\|\cdot\right\|$ represent the determinant, trace, rank, vectorization and Frobenius norm of a matrix, respectively. The operators $\left(\cdot\right)^{\sf T}$ and $\left(\cdot\right)^{\dag}$ represent the transpose and the Hermitian transpose of a matrix, respectively. The operator $\mathbb{E} \left\{\cdot\right\}$ represents the expectation operation. $\mathcal{CN} \left({\bf \mu}, {\bf \Sigma}\right)$ denotes a circularly symmetric complex Gaussian random vector with mean ${\bf \mu}$ and covariance ${\bf \Sigma}$. $\Re \left(\cdot\right)$ and $\Im \left(\cdot\right)$ denote the real part and the imaginary part of a complex number. ${\sf H} \left(\cdot\right)$ represents the entropy of a random variable or a random vector, ${\sf H} \left(\cdot|\cdot\right)$ represents the entropy of a random variable or random vector given another random variable or random vector and ${\sf I}\left(\cdot;\cdot\right)$ represents the mutual information between two random variables or random vectors. We also use the asymptotic notation: $f \left(x\right) = \mathcal{O} \left(g \left(x\right)\right)$ as $x \to x_0$ if $\lim_{x \to x_0} \left|\frac{f (x)}{g (x)}\right| < \infty$ and $f \left(x\right) = o \left(g \left(x\right)\right)$ as $x \to x_0$ if $\lim_{x \to x_0} \left|\frac{f (x)}{g (x)}\right| = 0$.

\vspace{-0.25cm}

\section{System Model}
\label{system_model}

We consider a fading channel with $n_t$ transmit and $n_r$ receive antennas which can be modeled as follows:
\begin{equation}
{\bf y} = \sqrt{{\sf snr}} \cdot {\bf H} {\bf x} + {\bf n} \label{channel_model}
\end{equation}
for a single use of the channel, where $\y \in \mathbb{C}^{n_r}$ represents the vector of complex receive symbols, $\x \in \mathbb{C}^{n_t}$ represents the vector of complex transmit symbols, $\n \sim \mathcal{CN} \left({\bf 0},{\bf I}\right) \in \mathbb{C}^{n_r}$ is a random vector which represents the noise and $\HH \in \mathbb{C}^{n_r \times n_t}$ is a random matrix which represents unit-power random channel gains between the various receive and transmit antennas, so that $\mathbb{E} \left\{ {\sf tr} \left({\bf H} {\bf H}^\dag\right) \right\} = n_t n_r$. We take the input to conform to an equiprobable multi-dimensional constellation with cardinality ${\sf M}$, i.e., ${\bf x} \in \left\{{\bf x}_1,{\bf x}_2,\ldots,{\bf x}_{\sf M}\right\}$ and $\Pr \left({\bf x}_1\right) = \Pr \left({\bf x}_2\right) = \cdots = \Pr \left({\bf x}_{\sf M}\right) = \frac{1}{{\sf M}}$, with 
${\bf \Sigma_x} = \mathbb{E} \left\{{\bf x} {\bf x}^\dag\right\} = \frac{1}{n_t} \cdot {\bf I}$.
We also take the input, the noise and the channel to be independent. Therefore, the signal-to-noise ratio per receive antenna is given by:
\begin{align}
{\sf SNR} &= {\sf snr} \cdot \frac{\mathbb{E} \left\{{\sf tr} \left(\left({\bf H} {\bf x}\right) \left({\bf H} {\bf x}\right)^\dag\right)\right\}}{n_r} = {\sf snr} \cdot \frac{\mathbb{E} \left\{{\sf tr} \left({\bf H} {\bf H}^\dag\right)\right\}}{n_t n_r} = {\sf snr}
\end{align}

We consider a channel matrix that incorporates Rayleigh fading, Ricean fading as well as transmit and receive antenna correlation in a separable correlation model~\cite{Shiu00},~\cite{Chizhik00},~\cite{Driessen99},~\cite{Rashid-Farrokhi01}, given by:
\begin{equation}
{\bf H} = \sqrt{\frac{K}{K+1}} \cdot {\bf H_0} + \sqrt{\frac{1}{K+1}} \cdot {\bf \Theta_R^{\frac{1}{2}}} {\bf H_w} {\bf \Theta_T^{\frac{1}{2}}}
\end{equation}
where ${\bf H_0} \in \mathbb{C}^{n_r \times n_t}$ is a $n_r \times n_t$ deterministic matrix, ${\bf H_w} \in \mathbb{C}^{n_r \times n_t}$ is a $n_r \times n_t$ canonical complex Gaussian random matrix with independent zero-mean and unit-variance circularly symmetric complex Gaussian random entries and ${\bf \Theta_T} \in \mathbb{C}^{n_t \times n_t}$ and ${\bf \Theta_R} \in \mathbb{C}^{n_r \times n_r}$ are $n_t \times n_t$ and $n_r \times n_r$ unit-diagonal Hermitian positive semi-definite  matrices with the correlation coefficients between the $n_t$ transmit and the $n_r$ receive antennas, respectively.
The Ricean K-factor $K$ corresponds to the ratio between the deterministic and random component energies. Since the term ${\bf H_0}$ corresponds to line-of-sight or diffracted components, ${\bf H_0} = {\bf a_R} {\bf a_T^\dag}$ where the vectors ${\bf a_T} \in \mathbb{C}^{n_t}$ and ${\bf a_R} \in \mathbb{C}^{n_r}$ are associated with the transmit and receive array responses to a plane wave so that $\|{\bf a_T}\|^2 = n_t$ and $\|{\bf a_R}\|^2 = n_r$. Note that in channels with Rayleigh fading, where line-of-sight components are absent, $K=0$, whereas in channels with Ricean fading, where line-of-sight components are present, $K \neq 0$; in addition, in channel models without transmit or receive antenna correlation ${\bf \Theta_T} = {\bf I}$ and ${\bf \Theta_R} = {\bf I}$ and in channel models with transmit and receive antenna correlation ${\bf \Theta_T} \neq {\bf I}$ and ${\bf \Theta_R} \neq {\bf I}$.

We assume that the receiver knows the channel matrix realization and that the transmitter knows only the channel matrix distribution. 
We also assume that the sequence of random channel matrices over time is stationary and ergodic. Consequently, the constrained capacity, achieved by coding over multiple fading blocks, is given by:
\begin{align}
\mathbb{E}_{\HH} \left\{{\sf I} \left(\x;\sqrt{{\sf snr}}
\cdot {\bf H} \x + \n \big| {\bf H}\right)\right\}
\end{align}
where ${\sf I} \left(\x;\sqrt{{\sf snr}} \cdot {\bf H} \x + \n \big| {\bf H}\right)$ is the mutual information between input vector ${\bf x}$ and the output vector ${\bf y} = \sqrt{{\sf snr}} \cdot {\bf H} \x + \n$ conditioned on a realization of the channel matrix ${\bf H}$. The goal is to characterize, as well as optimize, the constrained capacity in the asymptotic regime of high ${\sf snr}$.

\vspace{-0.25cm}

\section{Characterization of the Constrained Capacity}
\label{characterization}

We provide a characterization of the constrained capacity of multiple-antenna fading coherent channels with arbitrary
equiprobable discrete inputs in the regime of high ${\sf snr}$. This characterization, which is the crux of the study of the effect of common channel parameters and models on the system performance as well as the design of key system elements in subsequent sections, is based on an asymptotic expansion of the constrained capacity that portrays its behavior as a function of key system parameters in the regime of high ${\sf snr}$.

The definition of the asymptotic behavior is based on the procedure: First, we consider lower and upper bounds to the MMSE associated with the estimation of the noiseless output given the noisy output of the channel model in \eqref{channel_model}, for a fixed channel matrix, given by:
\begin{align}
{\sf mmse} \left({\sf snr};{\bf H}\right) = \mathbb{E} \left\{\left\|{\bf H} {\bf x} - {\bf H} \mathbb{E} \left\{{\bf x}|\sqrt{{\sf snr}} \cdot {\bf H} {\bf x} + {\bf n}\right\}\right\|^2 \big| {\bf H} \right\}
\end{align}
Second, we consider lower and upper bounds to the average value of the MMSE associated with the estimation of the noiseless output given the noisy output of the channel model in \eqref{channel_model}, for a random channel matrix, given by:
\begin{align}
{\sf \overline{mmse}} \left({\sf snr}\right) = \mathbb{E}_{{\bf H}} \left\{{\sf mmse} \left({\sf snr};{\bf H}\right)\right\} = \mathbb{E}_{{\bf H}} \left\{ \mathbb{E} \left\{\left\|{\bf H} {\bf x} - {\bf H} \mathbb{E} \left\{{\bf x}|\sqrt{{\sf snr}} \cdot {\bf H} {\bf x} + {\bf n}\right\}\right\|^2 \big| {\bf H} \right\} \right\}
\end{align}
We then consider upper and lower bounds to the mutual information between the input and the output of the channel model in \eqref{channel_model}, for a fixed channel matrix, given by:
\begin{align}
{\sf I} \left({\sf snr};{\bf H}\right) = {\sf I} \left({\bf x};\sqrt{{\sf snr}} \cdot {\bf H} {\bf x} + {\bf n} | {\bf H} \right)
\end{align}
as well as upper and lower bounds to the average value of the mutual information between the input and the output of the channel model in \eqref{channel_model}, for a random channel matrix, given by:
\begin{align}
{\sf \bar{I}} \left({\sf snr}\right) = \mathbb{E}_{{\bf H}} \left\{{\sf I} \left({\sf snr};{\bf H}\right)\right\}= \mathbb{E}_{{\bf H}} \left\{ {\sf I} \left({\bf x};\sqrt{{\sf snr}} \cdot {\bf H} {\bf x} + {\bf n} | {\bf H} \right) \right\}
\end{align}
by capitalizing on the I-MMSE identity and counterparts~\cite{Guo05},~\cite{Palomar06a}. Finally, we expose the asymptotic behavior as ${\sf snr} \to \infty$ of the average value of the MMSE and the average value of the mutual information, which leads to the asymptotic behavior as ${\sf snr} \to \infty$ of the constrained capacity, by capitalizing on the machinery of asymptotic analysis and asymptotic expansions~\cite{Bleistein86}.


The bounds are expressed in terms of the squared Euclidean distance between the (noiseless) receive vectors ${\bf H} {\bf x}_i$ and ${\bf H} {\bf x}_j$ associated with the transmit vectors $\x_i$ and $\x_j$ given by:
\begin{equation}
{\sf d_{ij}^2} \left({\bf H}\right) = \left\|{\bf H} {\bf x}_i - {\bf H} {\bf x}_j\right\|^2
\end{equation}
as well as their probability density functions $p_{{\sf d_{ij}^2}} \left({\sf d_{ij}^2}\right)$ and the higher-order derivatives $p_{{\sf d_{ij}^2}}^{(n)} \left({\sf d_{ij}^2}\right) = d^n \left(p_{{\sf d_{ij}^2}} \left({\sf d_{ij}^2}\right)\right) \Big/ d \left({\sf d_{ij}^2}\right)^n, n = 1,2,\ldots$.

Let us consider the channel model in \eqref{channel_model} with a fixed channel matrix. It is possible to obtain lower and upper bounds to the MMSE by relying on the use of a genie based estimator and a (sub-optimal) Euclidean distance based estimator, respectively. The bounds, which we express in terms of the squared pairwise Euclidean distances ${\sf d_{ij}^2} \left({\bf H}\right), i \neq j,$ rather than the squared minimum Euclidean distance ${\sf d_{min}^2} \left({\bf H}\right) =\min_{i \neq j} {\sf d_{ij}^2} \left({\bf H}\right)$ are, as opposed to the bounds in~\cite{Perez-Cruz10}, valid for all signal-to-noise ratios. This aspect is particularly relevant because the bounds to the average value of the MMSE follow from the bounds to the MMSE by averaging over the fading statistics.

\vspace{0.25cm}

\begin{lemma}
\label{mmse_deterministic} The MMSE associated with the estimation of the noiseless output given the noisy output of the channel model in \eqref{channel_model}, for a fixed channel matrix, can be bounded as follows:
\begin{equation}
{\sf mmse_{LB}} \left({\sf snr};{\bf H}\right) \leq {\sf
mmse} \left({\sf snr};{\bf H}\right) \leq {\sf mmse_{UB}}
\left({\sf snr};{\bf H}\right)
\end{equation}
where the lower and upper bounds are given by:
\begin{align}
{\sf mmse_{LB}} \left({\sf snr};{\bf H}\right) &= \frac{1}{4 {\sf M} ({\sf M}-1)} \sum_{i=1}^{{\sf M}} \sum_{\substack{j=1 \\ j \neq i}}^{{\sf M}} {\sf d_{ij}^2} \left({\bf H}\right) \cdot \frac{1}{2} \cdot {\sf erfc} \left(\sqrt{\frac{{\sf d_{ij}^2} \left({\bf H}\right) {\sf snr}}{4}}\right) \label{mmse_lb}
\end{align}
\begin{align}
{\sf mmse_{UB}} \left({\sf snr};{\bf H}\right) &= \frac{1}{{\sf M}} \sum_{i=1}^{{\sf M}} \sum_{\substack{j=1 \\ j \neq i}}^{{\sf M}} {\sf d_{ij}^2} \left({\bf H}\right) \cdot \frac{1}{2} \cdot {\sf erfc} \left(\sqrt{\frac{{\sf d_{ij}^2} \left({\bf H}\right) {\sf snr}}{4}}\right) \label{mmse_ub}
\end{align}
\end{lemma}

\vspace{0.25cm}

\begin{proof}
See Appendix A.
\end{proof}

\vspace{0.25cm}

The lower and upper bounds to the mutual information are obtained by using the upper and lower bounds to the MMSE, respectively, in the integral form of the relationship between the mutual information and the MMSE given by~\cite{Guo05}:\footnote{This representation assumes that the matrix ${\bf H}$ is non-singular. If the matrix ${\bf H}$ is singular then ${\sf I} \left({\sf snr};{\bf H}\right) = {\sf H} \left({\bf H} {\bf x}\right) - \int_{{\sf snr}}^{\infty} {\sf mmse} \left(\xi;{\bf H}\right) d \xi$.}
\begin{equation}
{\sf I} \left({\sf snr};{\bf H}\right) = \log {\sf M} - \int_{{\sf snr}}^{\infty} {\sf mmse} \left(\xi;{\bf H}\right) d \xi \label{i-mmse-relationship}
\end{equation}
The bounds, which we also express in terms of the squared pairwise Euclidean distances ${\sf d_{ij}^2} \left({\bf H}\right), i \neq j,$ rather than the squared minimum Euclidean distance ${\sf d_{min}^2} \left({\bf H}\right) = \min_{i \neq j} {\sf d_{ij}^2} \left({\bf H}\right)$ are, as opposed to the bounds in~\cite{Perez-Cruz10}, also valid for all signal-to-noise ratios.

\vspace{0.25cm}

\begin{lemma}
\label{mutual_information_deterministic} The mutual information between the input and the output of the channel model in \eqref{channel_model}, for a fixed channel matrix, can be bounded as follows:
\begin{equation}
{\sf I_{LB}} \left({\sf snr};{\bf H}\right) \leq {\sf I}
\left({\sf snr};{\bf H}\right) \leq {\sf I_{UB}} \left({\sf
snr};{\bf H}\right)
\end{equation}
where the upper and lower bounds are given by:
\begin{align}
{\sf I_{LB}} \left({\sf snr};{\bf H}\right) &= \log {\sf M} - \frac{1}{{\sf M}} \sum_{i=1}^{\sf M} \sum_{\substack{j=1 \\ j \neq i}}^{\sf M} 2 \cdot
e^{-\frac{{\sf d_{ij}^2} \left({\bf H}\right) {\sf snr}}{4}}
\label{i_lb}
\end{align}
\begin{align}
{\sf I_{UB}} \left({\sf snr};{\bf H}\right) &= \log {\sf M} - \frac{1}{{\sf M} ({\sf M}-1)} \sum_{i=1}^{{\sf M}} \sum_{\substack{j=1 \\ j \neq i}}^{{\sf M}} \frac{1}{4} \cdot {\sf erfc} \left(\sqrt{\frac{{\sf d_{ij}^2} \left({\bf H}\right) {\sf snr}}{4}}\right) \label{i_ub}
\end{align}
\end{lemma}

\vspace{0.25cm}

\begin{proof}
See Appendix B.
\end{proof}

\vspace{0.25cm}

Let us now consider the channel model in \eqref{channel_model} with a random channel matrix. It is possible to obtain lower and upper bounds to the average value of the MMSE by averaging over the fading statistics the lower and upper bounds to the MMSE as follows:
\begin{align}
{\sf \overline{mmse}_{LB}} \big({\sf snr}\big) &= \mathbb{E}_{{\bf H}} \left\{{\sf mmse_{LB}} \left({\sf snr};{\bf H}\right)\right\} = \int {\sf mmse_{LB}} \left({\sf snr}; {\bf H}\right) \cdot p_{{\bf H}} ({\bf H}) {\bf d H} \label{integral1}
\end{align}
\begin{align}
{\sf \overline{mmse}_{UB}} \big({\sf snr}\big) &= \mathbb{E}_{{\bf H}} \left\{{\sf mmse_{UB}} \left({\sf snr};{\bf H}\right)\right\} = \int {\sf mmse_{UB}} \left({\sf snr}; {\bf H}\right) \cdot p_{{\bf H}} ({\bf H}) {\bf d H} \label{integral2}
\end{align}
where $p_{{\bf H}} \left(\cdot\right)$ represents the distribution of ${\bf H}$. The determination of insightful closed-form expressions for the integrals in \eqref{integral1} and \eqref{integral2} is not simple. Consequently, rather than attempt to solve the integrals, we will exploit results from asymptotic analysis to determine the asymptotic expansion of the integrals. The asymptotic expansions, which lead to considerable insight, are very useful in the regime of high ${\sf snr}$.

\vspace{0.25cm}

\begin{lemma}
\label{mmse_fading} Assume that $p_{{\sf d_{ij}^2}}^{(n)} \left({\sf d_{ij}^2}\right), n = 0,1,2,\ldots$ are continuous and integrable in $[0,\infty)$. Then, the average value of the MMSE associated with the estimation of the noiseless output given the noisy output of the channel model in \eqref{channel_model}, for a random channel matrix, can be bounded as follows:
\begin{equation}
{\sf \overline{mmse}_{LB}} \left({\sf snr}\right) \leq {\sf \overline{mmse}} \left({\sf snr}\right)
\leq {\sf \overline{mmse}_{UB}} \left({\sf snr}\right)
\end{equation}
where the asymptotic expansion as ${\sf snr} \to \infty$ of the lower and upper bounds are
given by:
\begin{align}
{\sf \overline{mmse}_{LB}} ({\sf snr}) &= \sum_{n=0}^{N} \frac{1}{{\sf snr}^{n+2}} \cdot k_{{\sf LB}_{n+1}} \cdot \left(\sum_{i=1}^{{\sf M}} \sum_{\substack{j=1 \\ j \neq i}}^{{\sf M}} p_{{\sf d_{ij}^2}}^{(n)} (0)\right) + \mathcal{O} \left(\frac{1}{{\sf snr}^{N+3}}\right),~~~N = 0,1,\ldots
\label{mmse_fading_lower_bound_expansion} \\
{\sf \overline{mmse}_{UB}} ({\sf snr}) &= \sum_{n=0}^{N} \frac{1}{{\sf snr}^{n+2}} \cdot k_{{\sf UB}_{n+1}} \cdot \left(\sum_{i=1}^{{\sf M}} \sum_{\substack{j=1 \\ j \neq i}}^{{\sf M}} p_{{\sf d_{ij}^2}}^{(n)} (0)\right) + \mathcal{O} \left(\frac{1}{{\sf snr}^{N+3}}\right),~~~N = 0,1,\ldots
\label{mmse_fading_upper_bound_expansion}
\end{align}
and
\begin{align}
k_{{\sf LB}_n} &=\frac{1}{2 {\sf M} \left({\sf M}-1\right)} \cdot  \frac{n \cdot 4^n}{\sqrt{\pi}} \cdot \frac{\Gamma \left(n+3/2\right)}{\Gamma \left(n+2\right)} \label{constant_mmse_lb}
\end{align}
\begin{align}
k_{{\sf UB}_n} &=\frac{2}{{\sf M}} \cdot  \frac{n \cdot 4^n}{\sqrt{\pi}} \cdot \frac{\Gamma \left(n+3/2\right)}{\Gamma \left(n+2\right)}
\label{constant_mmse_ub}
\end{align}
and $\Gamma \left(\cdot\right)$ is the Gamma function.
\end{lemma}

\vspace{0.25cm}

\begin{proof}
See Appendix C.
\end{proof}

\vspace{0.25cm}

The lower and upper bounds to the average value of the mutual information are obtained by using the upper and lower bounds to the average value of the MMSE, respectively, in the integral form of the relationship between the average mutual information and the average MMSE given by~\cite{Palomar06a}:\footnote{This representation assumes that the matrix ${\bf H}$ is non-singular with probability equal to one. If the matrix ${\bf H}$ is not non-singular with probability equal to one then ${\sf \bar{I}} \big({\sf snr}\big) = \mathbb{E}_{\bf H} \left\{{\sf H} \left({\bf H}{\bf x}|{\bf H}\right)\right\} - \int_{{\sf snr}}^{\infty} {\sf \overline{mmse}} (\xi) d \xi$.}
\begin{equation}
{\sf \bar{I}} \big({\sf snr}\big) = \log {\sf M} - \int_{{\sf snr}}^{\infty} {\sf \overline{mmse}} (\xi) d \xi \label{average-i-average-mmse-relationship}
\end{equation}
Once again, the determination of insightful closed-form expressions for the integrals is not simple. Consequently, rather than attempt to solve the integrals, we will also exploit results from asymptotic analysis to determine the asymptotic expansions of the integrals.

\vspace{0.25cm}

\begin{lemma}
\label{mutual_information_fading} Assume that $p_{{\sf d_{ij}^2}}^{(n)} \left({\sf d_{ij}^2}\right), n = 0,1,2,\ldots$ are continuous and integrable in $[0,\infty)$. Then, the average value of the mutual information between the input and the output of the channel model in \eqref{channel_model}, for a random channel matrix, can be bounded as follows:
\begin{equation}
{\sf \bar{I}_{LB}} \left({\sf snr}\right) \leq {\sf \bar{I}} \left({\sf snr}\right)
\leq {\sf \bar{I}_{UB}} \left({\sf snr}\right)
\end{equation}
where the asymptotic expansion as ${\sf snr} \to \infty$ of the lower and upper bounds are
given by:
\begin{align}
{\sf \bar{I}_{LB}} \left({\sf snr}\right) &= \log {\sf M} - \sum_{n=0}^{N} \frac{1}{{\sf snr}^{n+1}} \cdot k'_{{\sf LB}_{n+1}} \cdot \left(\sum_{i=1}^{{\sf M}} \sum_{\substack{j=1 \\ j \neq i}}^{{\sf M}} p_{{\sf d_{ij}^2}}^{(n)} (0)\right) + \mathcal{O} \left(\frac{1}{{\sf snr}^{N+2}}\right), ~~ N = 0,1,\ldots \label{mutual_information_fading_lower_bound_expansion} \\
{\sf \bar{I}_{UB}} \left({\sf snr}\right) &= \log {\sf M} - \sum_{n=0}^{N} \frac{1}{{\sf snr}^{n+1}} \cdot k'_{{\sf UB}_{n+1}} \cdot \left(\sum_{i=1}^{{\sf M}} \sum_{\substack{j=1 \\ j \neq i}}^{{\sf M}} p_{{\sf d_{ij}^2}}^{(n)} (0)\right) + \mathcal{O} \left(\frac{1}{{\sf snr}^{N+2}}\right), ~~ N = 0,1,\ldots \label{mutual_information_fading_upper_bound_expansion}
\end{align}
and
\begin{align}
k'_{{\sf LB}_n} &= \frac{2}{{\sf M}} \cdot  \frac{4^n}{\sqrt{\pi}} \cdot \frac{\Gamma \left(n+3/2\right)}{\Gamma \left(n+2\right)} \label{constant_mutual_information_lb}
\end{align}
\begin{align}
k'_{{\sf UB}_n} &= \frac{1}{2 {\sf M} \left({\sf M}-1\right)} \cdot  \frac{4^n}{\sqrt{\pi}} \cdot \frac{\Gamma \left(n+3/2\right)}{\Gamma \left(n+2\right)} \label{constant_mutual_information_ub}
\end{align}
and $\Gamma \left(\cdot\right)$ is the Gamma function.
\end{lemma}

\vspace{0.25cm}

\begin{proof}
See Appendix D.
\end{proof}

\vspace{0.25cm}

It is important to remark that, in order to determine the asymptotic expansions of the lower and upper bounds to the average value of the MMSE and the average value of the mutual information, we assume that the functions $p_{{\sf d_{ij}^2}}^{(n)} \left({\sf d_{ij}^2}\right), n = 0, 1,2,\ldots$ are continuous and integrable in $[0,\infty)$. We verify the assumption for the most common fading channel models, including Rayleigh and Ricean fading models, in subsequent sections. \footnote{We will see that the fact that $p_{{\sf d_{ij}^2}}^{(n)} \left({\sf d_{ij}^2}\right), n = 1,2,\ldots$ may be discontinuous at zero is immaterial.}

\vspace{0.25cm}

%

The asymptotic expansions in Lemmas \ref{mmse_fading} and \ref{mutual_information_fading} are the basis of the characterization of the asymptotic behavior of the average value of the MMSE and the average value of the mutual information in the regime of high ${\sf snr}$. Let us define the integer ${\sf d} \geq 1$ as follows:
\begin{align}
{\sf d} = 1 + \min \left\{n \in \mathbb{N}_0: \sum_{i=1}^{{\sf M}} \sum_{\substack{j=1 \\ j \neq i}}^{{\sf M}} p_{{\sf d_{ij}^2}}^{(n)} (0) \neq 0\right\} \label{parameter_d}
\end{align}

We now seek to define the rate at which the average value of the MMSE tends to its limit as ${\sf snr} \to \infty$ as well as the rate at which the average value of the mutual information tends to its limit as ${\sf snr} \to \infty$, i.e.,
\begin{align}
- \lim_{{\sf snr} \to \infty} \frac{\log {\sf \overline{mmse}} \left({\sf snr}\right)}{\log {\sf snr}}
\end{align}
and
\begin{align}
- \lim_{{\sf snr} \to \infty} \frac{\log \left(\log {\sf M} - {\sf \bar{I}} \left({\sf snr}\right)\right)}{\log {\sf snr}}
\end{align}
These rates are trivially bounded as follows:
\begin{align}
- \lim_{{\sf snr} \to \infty} \frac{\log {\sf \overline{mmse}_{UB}} \left({\sf snr}\right)}{\log {\sf snr}} \leq - \lim_{{\sf snr} \to \infty} \frac{\log {\sf \overline{mmse}} \left({\sf snr}\right)}{\log {\sf snr}} \leq - \lim_{{\sf snr} \to \infty} \frac{\log {\sf \overline{mmse}_{LB}} \left({\sf snr}\right)}{\log {\sf snr}}
\end{align}
and
\begin{align}
- \lim_{{\sf snr} \to \infty} \frac{\log \left(\log {\sf M} - {\sf \bar{I}_{LB}} \left({\sf snr}\right)\right)}{\log {\sf snr}} \leq - \lim_{{\sf snr} \to \infty} \frac{\log \left(\log {\sf M} - {\sf \bar{I}} \left({\sf snr}\right)\right)}{\log {\sf snr}} \leq - \lim_{{\sf snr} \to \infty} \frac{\log \left(\log {\sf M} - {\sf \bar{I}_{UB}} \left({\sf snr}\right)\right)}{\log {\sf snr}}
\end{align}
By capitalizing on the asymptotic expansions embodied in Lemmas \ref{mmse_fading} and \ref{mutual_information_fading} it is possible to write as ${\sf snr} \to \infty$:
\begin{align}
\log {\sf \overline{mmse}_{LB}} \left({\sf snr}\right) = \log \left(k_{{\sf LB}_{\sf d}} \cdot \sum_{i=1}^{{\sf M}} \sum_{\substack{j=1 \\ j \neq i}}^{{\sf M}} p_{{\sf d_{ij}^2}}^{({\sf d}-1)} (0)\right) - ({\sf d} + 1) \log {\sf snr} + \mathcal{O} \left(\frac{1}{{\sf snr}}\right)
\end{align}
\begin{align}
\log {\sf \overline{mmse}_{UB}} \left({\sf snr}\right) = \log \left(k_{{\sf UB}_{\sf d}} \cdot \sum_{i=1}^{{\sf M}} \sum_{\substack{j=1 \\ j \neq i}}^{{\sf M}} p_{{\sf d_{ij}^2}}^{({\sf d}-1)} (0)\right) - ({\sf d} + 1) \log {\sf snr} + \mathcal{O} \left(\frac{1}{{\sf snr}}\right)
\end{align}
and
\begin{align}
\log\left(\log {\sf M} - {\sf \bar{I}_{LB}} \left({\sf snr}\right)\right) = \log \left(k'_{{\sf LB}_{\sf d}} \cdot \sum_{i=1}^{{\sf M}} \sum_{\substack{j=1 \\ j \neq i}}^{{\sf M}} p_{{\sf d_{ij}^2}}^{({\sf d}-1)} (0)\right) - {\sf d} \log {\sf snr} + \mathcal{O} \left(\frac{1}{{\sf snr}}\right)
\end{align}
\begin{align}
\log\left(\log {\sf M} - {\sf \bar{I}_{UB}} \left({\sf snr}\right)\right) = \log \left(k'_{{\sf UB}_{\sf d}} \cdot \sum_{i=1}^{{\sf M}} \sum_{\substack{j=1 \\ j \neq i}}^{{\sf M}} p_{{\sf d_{ij}^2}}^{({\sf d}-1)} (0)\right) - {\sf d} \log {\sf snr} + \mathcal{O} \left(\frac{1}{{\sf snr}}\right)
\end{align}
so that
\begin{align}
\label{rate_mmse}
- \lim_{{\sf snr} \to \infty} \frac{\log {\sf \overline{mmse}} \left({\sf snr}\right)}{\log {\sf snr}} = - \lim_{{\sf snr} \to \infty} \frac{\log {\sf \overline{mmse}_{LB}} \left({\sf snr}\right)}{\log {\sf snr}} = - \lim_{{\sf snr} \to \infty} \frac{\log {\sf \overline{mmse}_{UB}} \left({\sf snr}\right)}{\log {\sf snr}} = {\sf d} + 1
\end{align}
and
\begin{align}
\label{rate_mutual_information}
- \lim_{{\sf snr} \to \infty} \frac{\log \left(\log {\sf M} - {\sf \bar{I}} \left({\sf snr}\right)\right)}{\log {\sf snr}}  = - \lim_{{\sf snr} \to \infty} \frac{\log \left(\log {\sf M} - {\sf \bar{I}_{LB}} \left({\sf snr}\right)\right)}{\log {\sf snr}} = - \lim_{{\sf snr} \to \infty} \frac{\log \left(\log {\sf M} - {\sf \bar{I}_{UB}} \left({\sf snr}\right)\right)}{\log {\sf snr}} = {\sf d}
\end{align}

We also seek to define, in addition to the rates at which the average value of the MMSE and the average value of the mutual information tend to their infinite-${\sf snr}$ values, a finer characterization of the high-${\sf snr}$ asymptotic behavior. Towards this end, we define the quantities:
\begin{align}
\overline{\epsilon}_{\sf d} = \limsup_{{\sf snr} \to \infty}~{\sf snr}^{{\sf d} + 1} \cdot {\sf \overline{mmse}} \left({\sf snr}\right) \label{limsup_mmse}
\end{align}
\begin{align}
\underline{\epsilon}_{\sf d} = \liminf_{{\sf snr} \to \infty}~{\sf snr}^{{\sf d} + 1} \cdot {\sf \overline{mmse}} \left({\sf snr}\right) \label{liminf_mmse}
\end{align}
and
\begin{align}
\overline{\epsilon}'_{\sf d} = \limsup_{{\sf snr} \to \infty}~{\sf snr}^{{\sf d}} \cdot \left({\log {\sf M} - \sf \bar{I}} \left({\sf snr}\right)\right) \label{limsup_i}
\end{align}
\begin{align}
\underline{\epsilon}'_{\sf d} = \liminf_{{\sf snr} \to \infty}~{\sf snr}^{{\sf d}} \cdot \left(\log {\sf M} - {\sf \bar{I}} \left({\sf snr}\right)\right) \label{liminf_i}
\end{align}
Note that these quantities, in view of the asymptotic expansions embodied in Lemmas \ref{mmse_fading} and \ref{mutual_information_fading}, can be bounded as follows:
\begin{align}
k_{{\sf LB}_{\sf d}} \cdot \sum_{i=1}^{{\sf M}} \sum_{\substack{j=1 \\ j \neq i}}^{{\sf M}} p_{{\sf d_{ij}^2}}^{({\sf d}-1)} (0) \leq \underline{\epsilon}_{\sf d} \leq \overline{\epsilon}_{\sf d} \leq k_{{\sf UB}_{\sf d}} \cdot \sum_{i=1}^{{\sf M}} \sum_{\substack{j=1 \\ j \neq i}}^{{\sf M}} p_{{\sf d_{ij}^2}}^{({\sf d}-1)} (0) \label{bounds_avg_mmse1}
\end{align}
and
\begin{align}
k'_{{\sf UB}_{\sf d}} \cdot \sum_{i=1}^{{\sf M}} \sum_{\substack{j=1 \\ j \neq i}}^{{\sf M}} p_{{\sf d_{ij}^2}}^{({\sf d}-1)} (0) \leq \underline{\epsilon}'_{\sf d} \leq \overline{\epsilon}'_{\sf d} \leq k'_{{\sf LB}_{\sf d}} \cdot \sum_{i=1}^{{\sf M}} \sum_{\substack{j=1 \\ j \neq i}}^{{\sf M}} p_{{\sf d_{ij}^2}}^{({\sf d}-1)} (0) \label{bounds_avg_mmse2}
\end{align}
Note also that the bounds in \eqref{bounds_avg_mmse1} and \eqref{bounds_avg_mmse2} are finite for key fading models (see Sections \ref{canonical_channel} and \ref{other_channels}).

\vspace{0.25 cm}

The following Theorems, which are based on these considerations, characterize the asymptotic behavior of the average value of the MMSE and the average value of the mutual information in the regime of high ${\sf snr}$.

\vspace{0.25 cm}

\begin{theorem}
\label{mmse_expansion}
Assume that $p_{{\sf d_{ij}^2}}^{(n)} \left({\sf d_{ij}^2}\right), n = 0,1,2,\ldots$ are continuous and integrable in $[0,\infty)$. Then, in the regime of high-${\sf snr}$ the average value of the MMSE associated with the estimation of the noiseless output given the noisy output of the channel model in \eqref{channel_model} can be expanded as follows:
\begin{align}
\label{mmse_expansion_equation}
{\sf \overline{mmse}} \left({\sf snr}\right) = \epsilon_{\sf d} \left({\sf snr}\right) \cdot \frac{1}{{\sf snr}^{{\sf d}+1}} + \mathcal{O} \left(\frac{1}{{\sf snr}^{{\sf d}+2}}\right)
\end{align}
where $\epsilon_{\sf d} \left({\sf snr}\right)$ is a piecewise infinitely differentiable function such that:
\begin{align}
k_{{\sf LB}_{\sf d}} \cdot \sum_{i=1}^{{\sf M}} \sum_{\substack{j=1 \\ j \neq i}}^{{\sf M}} p_{{\sf d_{ij}^2}}^{({\sf d}-1)} (0) \leq \epsilon_{\sf d} \left({\sf snr}\right) \leq k_{{\sf UB}_{\sf d}} \cdot \sum_{i=1}^{{\sf M}} \sum_{\substack{j=1 \\ j \neq i}}^{{\sf M}} p_{{\sf d_{ij}^2}}^{({\sf d}-1)} (0)
\end{align}
in the interval $[{\sf snr_0},\infty)$ for a sufficiently high value of ${\sf snr_0}$ and
\begin{align}
\limsup_{{\sf snr} \to \infty} \epsilon_{\sf d} \left({\sf snr}\right) = \overline{\epsilon}_{\sf d} \qquad \text{and} \qquad \liminf_{{\sf snr} \to \infty} \epsilon_{\sf d} \left({\sf snr}\right) = \underline{\epsilon}_{\sf d}
\end{align}
\end{theorem}

\vspace{0.25cm}

\begin{proof}
See Appendix E.
\end{proof}

\vspace{0.25cm}

\begin{theorem}
\label{mutual_information_expansion}
Assume that $p_{{\sf d_{ij}^2}}^{(n)} \left({\sf d_{ij}^2}\right), n = 0,1,2,\ldots$ are continuous and integrable in $[0,\infty)$. Then, in the regime of high-${\sf snr}$ the average value of the mutual information between the input and the output of the channel model in \eqref{channel_model} can be expanded as follows:
\begin{align}
\label{mutual_information_expansion_equation}
{\sf \bar{I}} \left({\sf snr}\right) = \log {\sf M} - \epsilon'_{\sf d} \left({\sf snr}\right) \cdot \frac{1}{{\sf snr}^{{\sf d}}} + \mathcal{O} \left(\frac{1}{{\sf snr}^{{\sf d}+1}}\right)
\end{align}
where $\epsilon'_{\sf d} \left({\sf snr}\right)$ is a piecewise infinitely differentiable function such that:
\begin{align}
k'_{{\sf UB}_{\sf d}} \cdot \sum_{i=1}^{{\sf M}} \sum_{\substack{j=1 \\ j \neq i}}^{{\sf M}} p_{{\sf d_{ij}^2}}^{({\sf d}-1)} (0) \leq \epsilon'_{\sf d} \left({\sf snr}\right) \leq k'_{{\sf LB}_{\sf d}} \cdot \sum_{i=1}^{{\sf M}} \sum_{\substack{j=1 \\ j \neq i}}^{{\sf M}} p_{{\sf d_{ij}^2}}^{({\sf d}-1)} (0)
\end{align}
in the interval $[{\sf snr_0},\infty)$ for a sufficiently high value of ${\sf snr_0}$ and
\begin{align}
\limsup_{{\sf snr} \to \infty} \epsilon'_{\sf d} \left({\sf snr}\right) = \overline{\epsilon}'_{\sf d} \qquad \text{and} \qquad \liminf_{{\sf snr} \to \infty} \epsilon'_{\sf d} \left({\sf snr}\right) = \underline{\epsilon}'_{\sf d}
\end{align}
\end{theorem}

\vspace{0.25cm}

\begin{proof}
See also Appendix E.
\end{proof}

\vspace{0.25cm}

Note that if
\begin{align}
\label{avg_mmse_dimension1}
\overline{\epsilon}_{\sf d} = \underline{\epsilon}_{\sf d}  = \lim_{{\sf snr} \to \infty}~{\sf snr}^{{\sf d} + 1} \cdot {\sf \overline{mmse}} \left({\sf snr}\right) \triangleq \epsilon_{\sf d}
\end{align}
and
\begin{align}
\label{avg_mmse_dimension2}
\overline{\epsilon}'_{\sf d} = \underline{\epsilon}'_{\sf d} = \lim_{{\sf snr} \to \infty}~{\sf snr}^{{\sf d}} \cdot \left(\log {\sf M} - {\sf \bar{I}} \left({\sf snr}\right)\right) \triangleq \epsilon'_{\sf d}
\end{align}
then the asymptotic expansions exposed in Theorems \ref{mmse_expansion} and \ref{mutual_information_expansion} reduce immediately to:
\begin{align}
\label{mmse_expansion_}
{\sf \overline{mmse}} \left({\sf snr}\right) = \epsilon_{\sf d} \cdot \frac{1}{{\sf snr}^{{\sf d}+1}} + \mathcal{O} \left(\frac{1}{{\sf snr}^{{\sf d}+2}}\right)
\end{align}
and
\begin{align}
\label{mutual_information_expansion_}
{\sf \bar{I}} \left({\sf snr}\right) = \log {\sf M} - \epsilon'_{\sf d} \cdot \frac{1}{{\sf snr}^{{\sf d}}} + \mathcal{O} \left(\frac{1}{{\sf snr}^{{\sf d}+1}}\right)
\end{align}
Therefore, in view of \eqref{average-i-average-mmse-relationship}, \eqref{mmse_expansion_} and \eqref{mutual_information_expansion_} it is possible to establish the relation:
\begin{align}
\epsilon'_{\sf d} = \lim_{{\sf snr} \to \infty}~{\sf snr}^{{\sf d}} \cdot \left(\log {\sf M} - {\sf \bar{I}} \left({\sf snr}\right)\right) = \frac{1}{{\sf d}} \cdot \lim_{{\sf snr} \to \infty}~{\sf snr}^{{\sf d} + 1} \cdot {\sf \overline{mmse}} \left({\sf snr}\right) = \frac{1}{{\sf d}} \cdot \epsilon_{\sf d}
\end{align}
so that, as expected, the scaling constants that define the high-${\sf snr}$ asymptotics of the average value of the MMSE and the average value of the mutual information are also related. Note also that the quantities $\epsilon_{\sf d}$ and $\epsilon'_{\sf d}$ represent a generalization of the MMSE dimension, which, when it exists, defines the high--${\sf snr}$ asymptotics of the MMSE of a random variable observed in zero-mean unit-variance Gaussian noise~\cite{Wu11}. Numerical results suggest that the limits in \eqref{avg_mmse_dimension1} and \eqref{avg_mmse_dimension2} exist for common multiple-antenna fading coherent channel models driven by arbitrary equiprobable discrete inputs, thereby justifying the use of the expansions in \eqref{mmse_expansion_} and \eqref{mutual_information_expansion_} to characterize the asymptotic behavior.


It is also important to note that in general we can only bound $\epsilon_{\sf d}$ and $\epsilon'_{\sf d}$ as in \eqref{bounds_avg_mmse1} and \eqref{bounds_avg_mmse2}, respectively, rather than compute their exact values. 
The upper and lower bounds to $\epsilon_{\sf d}$ and $\epsilon'_{\sf d}$ differ by a factor of $4 \cdot ({\sf M} - 1)$, and so become increasingly loose for multi-dimensional constellations with high cardinality. The lower bound to $\epsilon_{\sf d}$ and the upper bound to $\epsilon'_{\sf d}$, which are due to the genie based estimator, are considerably loose. \footnote{It is possible to construct genie based estimators that lead to tighter bounds. The current genie supplies the receiver with a pair of input vectors for each transmit vector, the true input vector and any of the other input vectors with equal probability. A more appropriate genie supplies the receiver with the true input vector and another suitable input vector for each transmit vector. In particular, in the canonical i.i.d. Rayleigh fading coherent channel, Theorem \ref{characterization_mutual_information_mimo_rayleigh} suggests that the genie ought to minimize as much as possible the sum of the inverse of the squared Euclidean distances between the pairs of supplied input vectors. It is also important to guarantee, in addition, that such a genie construction leads indeed to a lower bound to the MMSE and hence an upper bound to the mutual information. This is met, for example, by constructing a genie that also ensures that the receiver sees all the input vectors with equal probability. We do not pursue this issue further, because the Euclidean distance based estimator upper bound to $\epsilon_{\sf d}$ and lower bound to $\epsilon'_{\sf d}$ are still tighter than the new genie based estimator lower bound to $\epsilon_{\sf d}$ and upper bound to $\epsilon'_{\sf d}$.} In contrast, the upper bound to $\epsilon_{\sf d}$ and the lower bound to $\epsilon'_{\sf d}$, which are due to the Euclidean distance based estimator, are considerably tighter.
In general, the use of the bounds to $\epsilon_{\sf d}$ and $\epsilon'_{\sf d}$ in the expansions in \eqref{mmse_expansion_} and \eqref{mutual_information_expansion_}, respectively, leads to bounds to the true asymptotic expansions. The bounds to the asymptotic expansions and the true asymptotic expansions differ only by an offset in signal-to-noise ratio. The offset in dB between the high-${\sf snr}$ expansion of the average value of the MMSE in \eqref{mmse_expansion_} and the expansions that use the lower and upper bounds to the value of $\epsilon_{\sf d}$ is given respectively by:
\begin{align}
\Delta_{{\sf LB}} \big|_{{\sf dB}} = 10 \cdot \log_{10} \left(\frac{k_{{\sf LB}_{\sf d}} \cdot \sum_{i=1}^{{\sf M}} \sum_{\substack{j=1 \\ j \neq i}}^{{\sf M}} p_{{\sf d_{ij}^2}}^{({\sf d}-1)} (0)}{\epsilon_{\sf d}}\right)^{\frac{1}{{\sf d}+1}}
\end{align}
and
\begin{align}
\Delta_{{\sf UB}} \big|_{{\sf dB}} = 10 \cdot \log_{10} \left(\frac{k_{{\sf UB}_{\sf d}} \cdot \sum_{i=1}^{{\sf M}} \sum_{\substack{j=1 \\ j \neq i}}^{{\sf M}} p_{{\sf d_{ij}^2}}^{({\sf d}-1)} (0)}{\epsilon_{\sf d}}\right)^{\frac{1}{{\sf d}+1}}
\end{align}
In turn, the offset in dB between the high-${\sf snr}$ expansion of the average value of the mutual information in \eqref{mutual_information_expansion_} and the expansions that use the lower and upper bounds to the value of $\epsilon'_{\sf d}$ is given respectively by:
\begin{align}
\Delta'_{{\sf LB}} \big|_{{\sf dB}} = 10 \cdot \log_{10} \left(\frac{k'_{{\sf LB}_{\sf d}} \cdot \sum_{i=1}^{{\sf M}} \sum_{\substack{j=1 \\ j \neq i}}^{{\sf M}} p_{{\sf d_{ij}^2}}^{({\sf d}-1)} (0)}{\epsilon'_{\sf d}}\right)^{\frac{1}{{\sf d}}}
\end{align}
and
\begin{align}
\Delta'_{{\sf UB}} \big|_{{\sf dB}} = 10 \cdot \log_{10} \left(\frac{k'_{{\sf UB}_{\sf d}} \cdot \sum_{i=1}^{{\sf M}} \sum_{\substack{j=1 \\ j \neq i}}^{{\sf M}} p_{{\sf d_{ij}^2}}^{({\sf d}-1)} (0)}{\epsilon'_{\sf d}}\right)^{\frac{1}{{\sf d}}}
\end{align}
Tables \ref{snr_offset1} and \ref{snr_offset2} show the signal-to-noise ratio offset values for particular system configurations. We observe that indeed $\Delta_{{\sf UB}} \big|_{{\sf dB}}$ or $\Delta'_{{\sf LB}} \big|_{{\sf dB}}$ are much lower than $\Delta_{{\sf LB}} \big|_{{\sf dB}}$ or $\Delta'_{{\sf UB}} \big|_{{\sf dB}}$, respectively, which is a manifestation of the fact that the upper bound to $\epsilon_{\sf d}$ and the lower bound to $\epsilon'_{\sf d}$ are considerably tighter that the lower bound to $\epsilon_{\sf d}$ and the upper bound to $\epsilon'_{\sf d}$. We also observe that the signal-to-noise ratio offset values $\Delta_{{\sf UB}} \big|_{{\sf dB}}$ and $\Delta'_{{\sf LB}} \big|_{{\sf dB}}$ decrease with the increase in the number of antennas and, as expected, increase with the increase in the multi-dimensional constellation cardinality. As an example, it is interesting to note that $\Delta_{{\sf UB}} \big|_{{\sf dB}} = 0.8~\sf{dB}$ and $\Delta'_{{\sf LB}} \big|_{{\sf dB}} = 1.2~\sf{dB}$ for a two-transmit two-receive antenna canonical i.i.d. Rayleigh fading coherent channel driven by 16-QAM inputs.

\vspace{0.25cm}

\begin{table}
\caption{Signal-to-noise ratio offset between the true asymptotic expansions and the bounds to the asymptotic expansions for multiple-antenna Rayleigh fading channels driven by 16-QAM inputs}
\begin{center}
  \begin{tabular}{ | c || c | c | c | c | }

    \hline

    ($n_t,n_r$) & $\Delta_{{\sf LB}} \big|_{{\sf dB}}$ & $\Delta_{{\sf UB}} \big|_{{\sf dB}}$ & $\Delta'_{{\sf LB}} \big|_{{\sf dB}}$ & $\Delta'_{{\sf UB}} \big|_{{\sf dB}}$\\

    \hline

    \hline

    (1,1) & $- 6.9~{\sf dB}$ & $2.0~{\sf dB}$ & $3.9~{\sf dB}$ & $- 13.8~{\sf dB}$\\

    \hline

    (1,2) & $-5.0~{\sf dB}$ & $0.9~{\sf dB}$ & $1.4~{\sf dB}$ & $-7.5~{\sf dB}$\\

    \hline

    (1,3) & $-4.0~{\sf dB}$ & $0.5~{\sf dB}$ & $0.7~{\sf dB}$ & $-5.3~{\sf dB}$\\

    \hline

    (2,2) & $-9.2~{\sf dB}$ & $0.8~{\sf dB}$ & $1.2~{\sf dB}$ & $-13.8~{\sf dB}$\\

    \hline

    (3,3) & $-9.8~{\sf dB}$ & $0.8~{\sf dB}$ & $1.0~{\sf dB}$ & $-13.0~{\sf dB}$\\

    \hline

    \end{tabular}
\end{center}
\label{snr_offset1}
\end{table}

\begin{table}
\caption{Signal-to-noise ratio offset between the true asymptotic expansions and the bounds to the asymptotic expansions for single-antenna Rayleigh fading channels driven by various inputs}
\begin{center}
  \begin{tabular}{ | c || c | c | c | c | }

    \hline

     & $\Delta_{{\sf LB}} \big|_{{\sf dB}}$ & $\Delta_{{\sf UB}} \big|_{{\sf dB}}$ & $\Delta'_{{\sf LB}} \big|_{{\sf dB}}$ & $\Delta'_{{\sf UB}} \big|_{{\sf dB}}$\\

    \hline

    \hline

    BPSK & $-2.4~{\sf dB}$ & $0.6~{\sf dB}$ & $1.1~{\sf dB}$ & $-4.9~{\sf dB}$\\

    \hline

    QPSK & $-4.3~{\sf dB}$ & $1.1~{\sf dB}$ & $2.2~{\sf dB}$ & $-8.6~{\sf dB}$\\

    \hline

    16-QAM & $- 6.9~{\sf dB}$ & $2.0~{\sf dB}$ & $3.9~{\sf dB}$ & $- 13.9~{\sf dB}$\\

    \hline

    64-QAM & $-9.4~{\sf dB}$ & $2.6~{\sf dB}$ & $5.3~{\sf dB}$ & $-18.7~{\sf dB}$\\

    \hline

    256-QAM & $-11.7~{\sf dB}$ & $3.3~{\sf dB}$ & $6.6~{\sf dB}$ & $-23.5~{\sf dB}$\\

    \hline

    \end{tabular}
\end{center}
\label{snr_offset2}
\end{table}

\begin{figure}
\centering
\begin{tabular}{c}
\epsfig{file=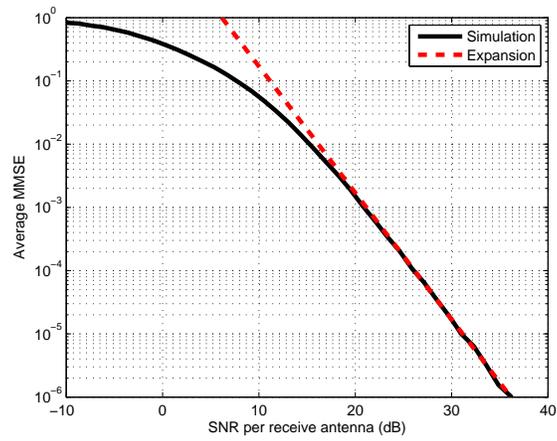,width=0.50\linewidth,clip=} \\
\centerline{(a)} \\
\epsfig{file=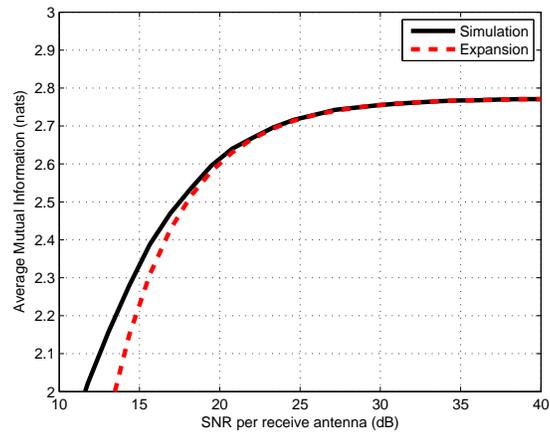,width=0.50\linewidth,clip=} \\
\centerline{(b)} \\
\epsfig{file=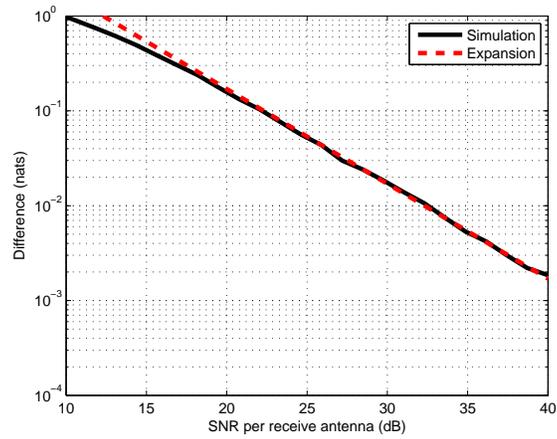,width=0.50\linewidth,clip=} \\
\centerline{(c)} \\
\end{tabular}
\caption{$1 \times 1$ Rayleigh fading coherent channel with a 16-QAM input: a) average MMSE; b) average mutual information; (c) difference between maximum average mutual information and average mutual information.} \label{siso_rayleigh_16qam}
\end{figure}
\begin{figure}
\centering
\begin{tabular}{c}
\epsfig{file=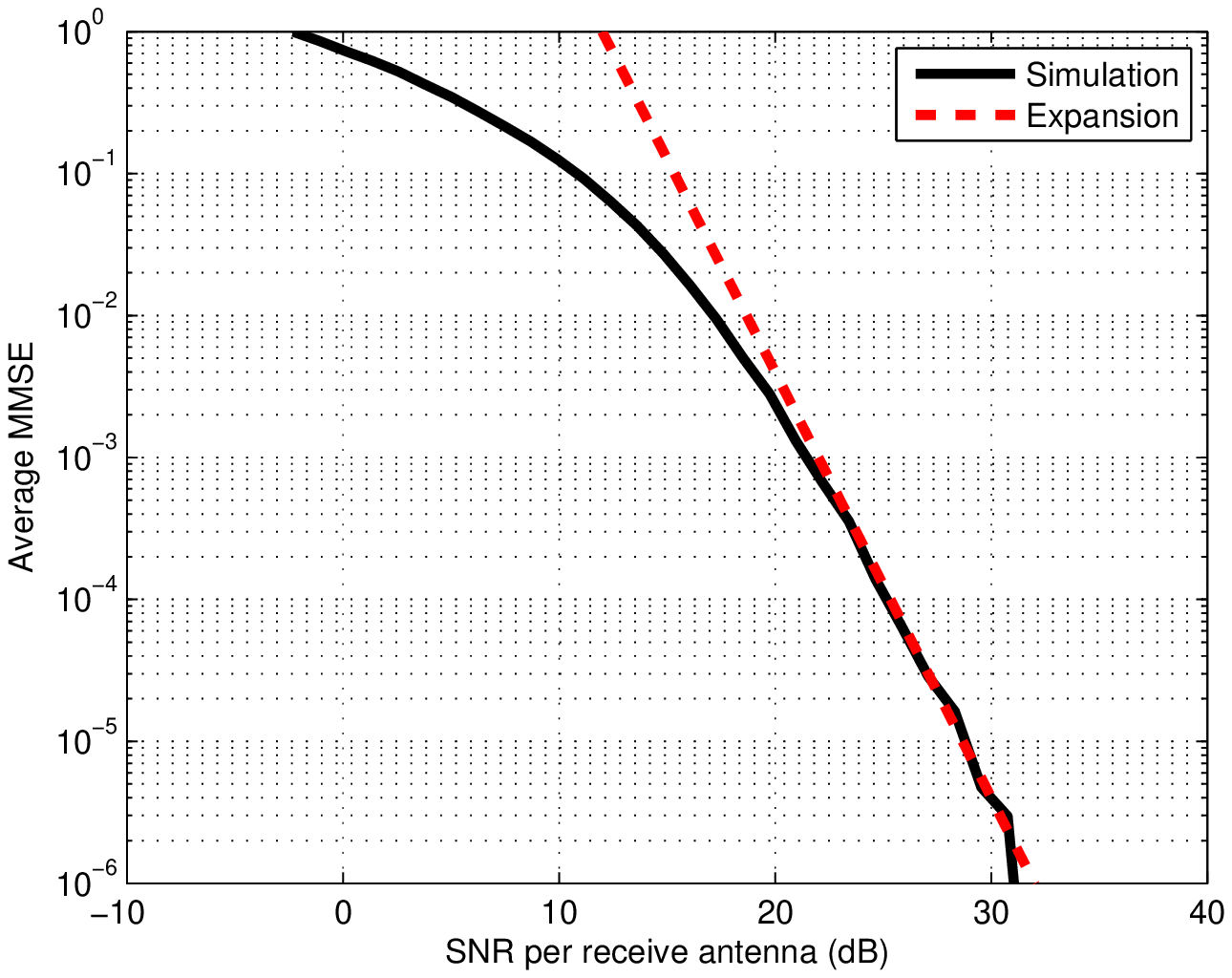,width=0.50\linewidth,clip=} \\
\centerline{(a)} \\
\epsfig{file=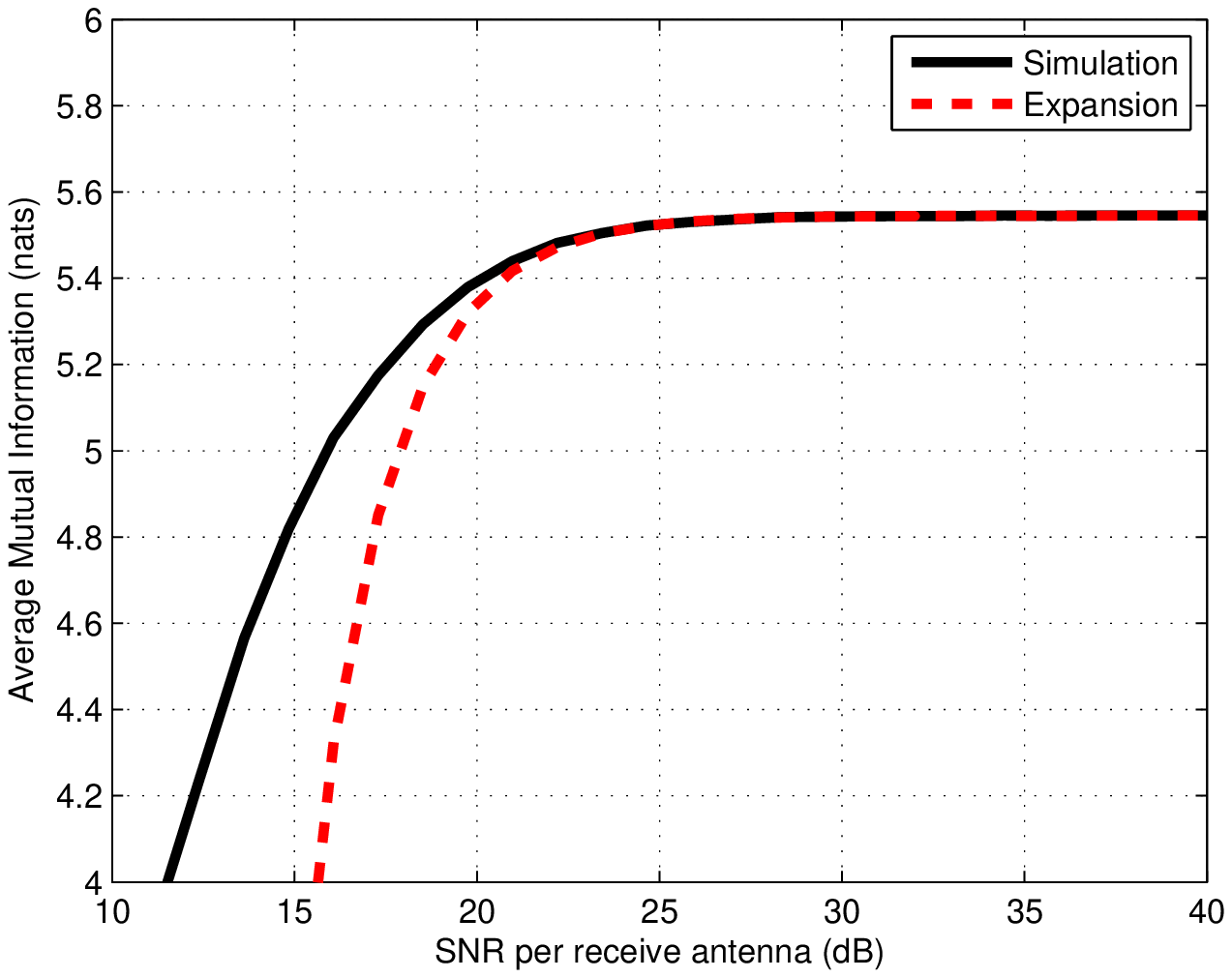,width=0.50\linewidth,clip=} \\
\centerline{(b)} \\
\epsfig{file=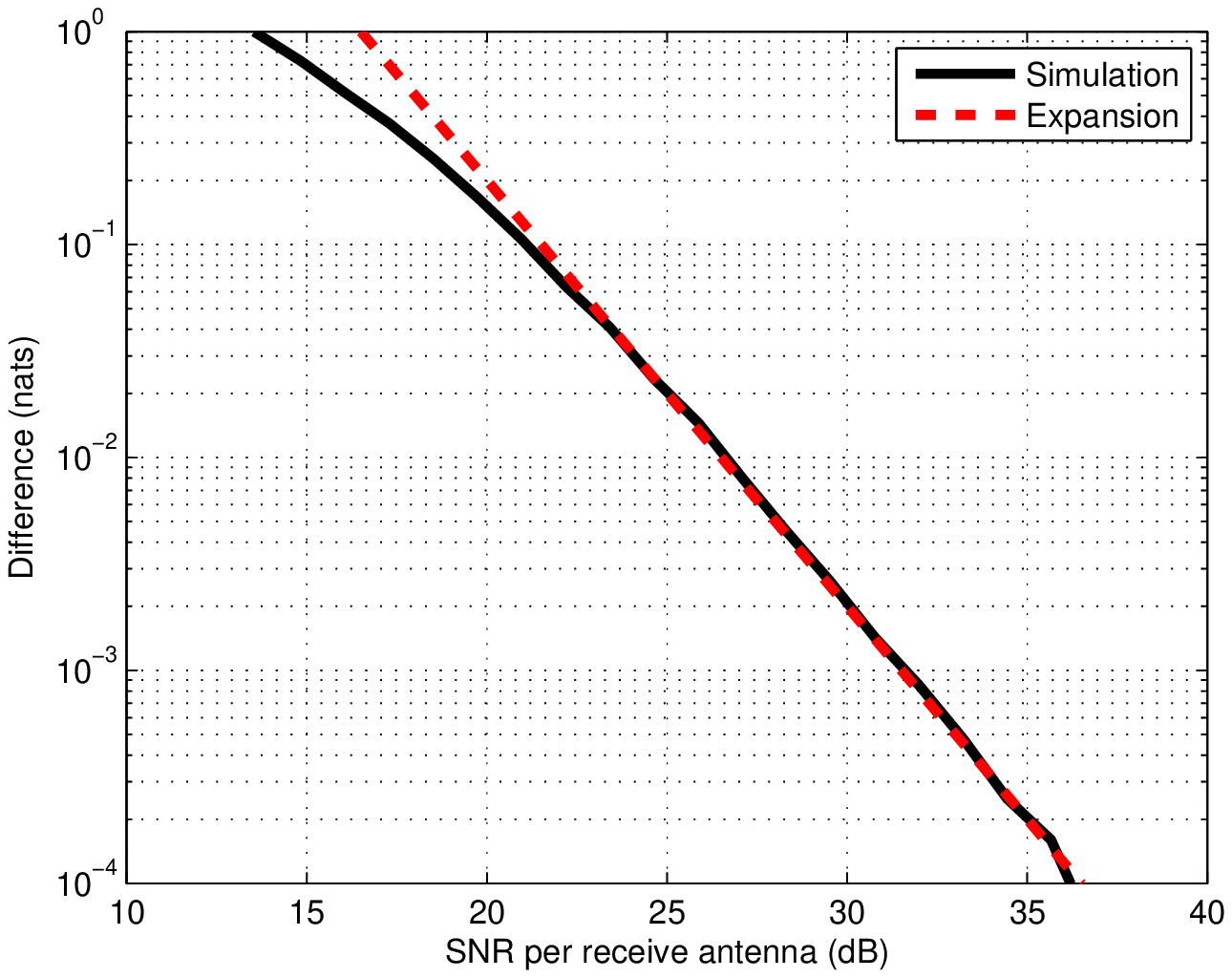,width=0.50\linewidth,clip=} \\
\centerline{(c)} \\
\end{tabular}
\caption{$2 \times 2$ Rayleigh fading coherent channel with 16-QAM inputs: a) average MMSE; b) average mutual information; (c) difference between maximum average mutual information and average mutual information.} \label{mimo22_rayleigh_16qam}
\end{figure}
\begin{figure}
\centering
\begin{tabular}{c}
\epsfig{file=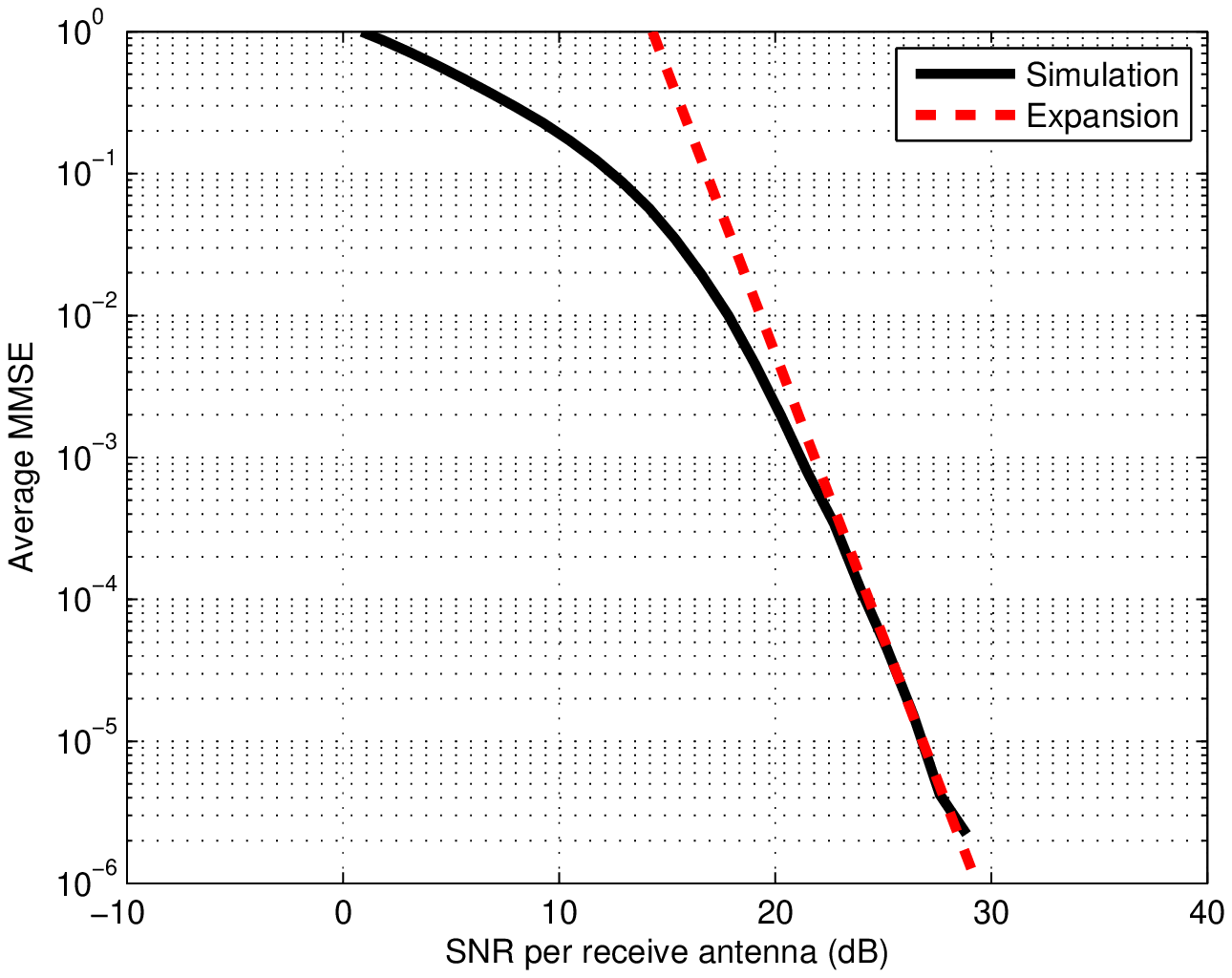,width=0.50\linewidth,clip=} \\
\centerline{(a)} \\
\epsfig{file=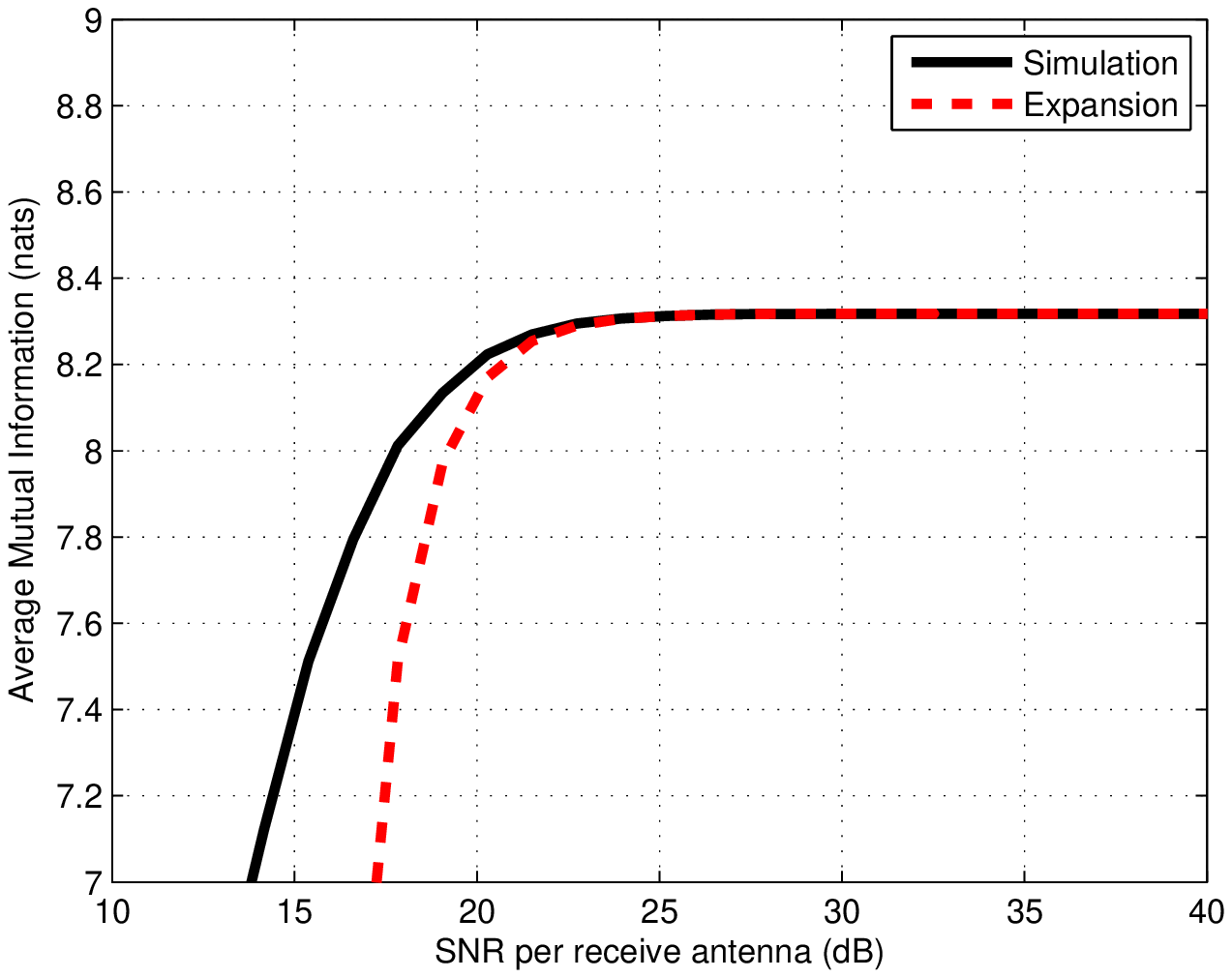,width=0.50\linewidth,clip=} \\
\centerline{(b)} \\
\epsfig{file=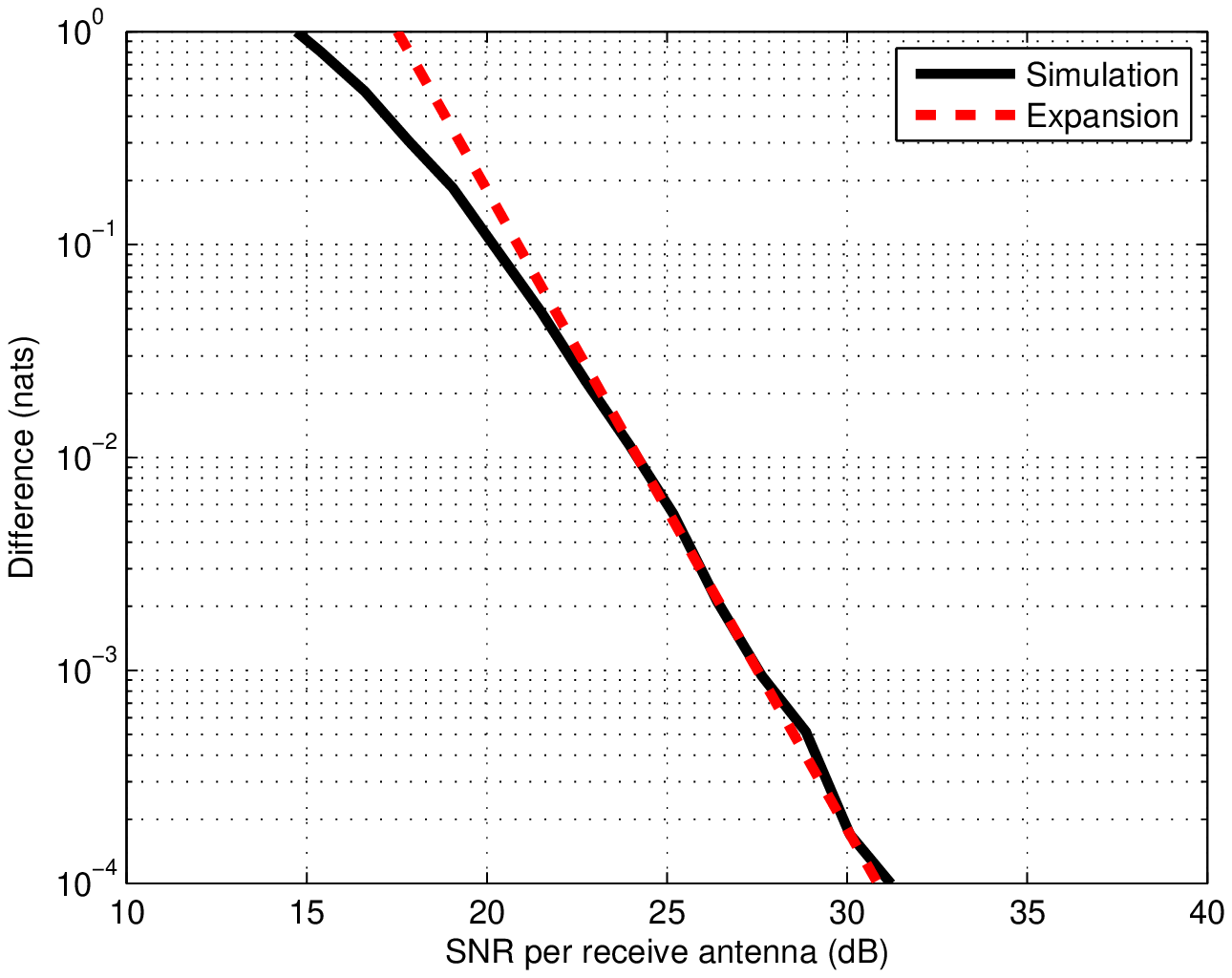,width=0.50\linewidth,clip=} \\
\centerline{(c)} \\
\end{tabular}
\caption{$3 \times 3$ Rayleigh fading coherent channel with 16-QAM inputs: a) average MMSE; b) average mutual information; (c) difference between maximum average mutual information and average mutual information.} \label{mimo33_rayleigh_16qam}
\end{figure}

Figures \ref{siso_rayleigh_16qam}, \ref{mimo22_rayleigh_16qam} and \ref{mimo33_rayleigh_16qam} depict the average value of the MMSE and the average value of the mutual information and the respective asymptotic expansions for key multiple-antenna systems and models. \footnote{We use the exact values of $\epsilon_{\sf d}$ and $\epsilon'_{\sf d}$, obtained through Monte Carlo simulations, in the asymptotic expansions.} We observe a very reasonable match between the exact curve, obtained through Monte Carlo simulations, and the asymptotic expansion in the high-${\sf snr}$ regime. We also observe that the asymptotic expansions characterize perfectly the rates at which the average MMSE and the average mutual information approach their infinite ${\sf snr}$ values.
In general, the use of the high-${\sf snr}$ asymptotic expansions of the average MMSE and the average mutual information results in errors that relate to the fact that: \emph{i}) one can only compute the leading term rather than other possible higher-order terms in the high-${\sf snr}$ asymptotic expansions; and \emph{ii}) one can only compute analytically bounds to $\epsilon_{\sf d}$ and $\epsilon'_{\sf d}$. The first consideration leads to the differences between the asymptotic expansions and the exact curve in Figures \ref{siso_rayleigh_16qam}, \ref{mimo22_rayleigh_16qam} and \ref{mimo33_rayleigh_16qam}. The second consideration leads to the additional signal-to-noise ratio offsets in Tables \ref{snr_offset1} and \ref{snr_offset2}. Nonetheless, it will be shown that the asymptotic expansions still produce very accurate designs in Section \ref{designs}.

\begin{figure}
\centering
\begin{tabular}{c}
\epsfig{file=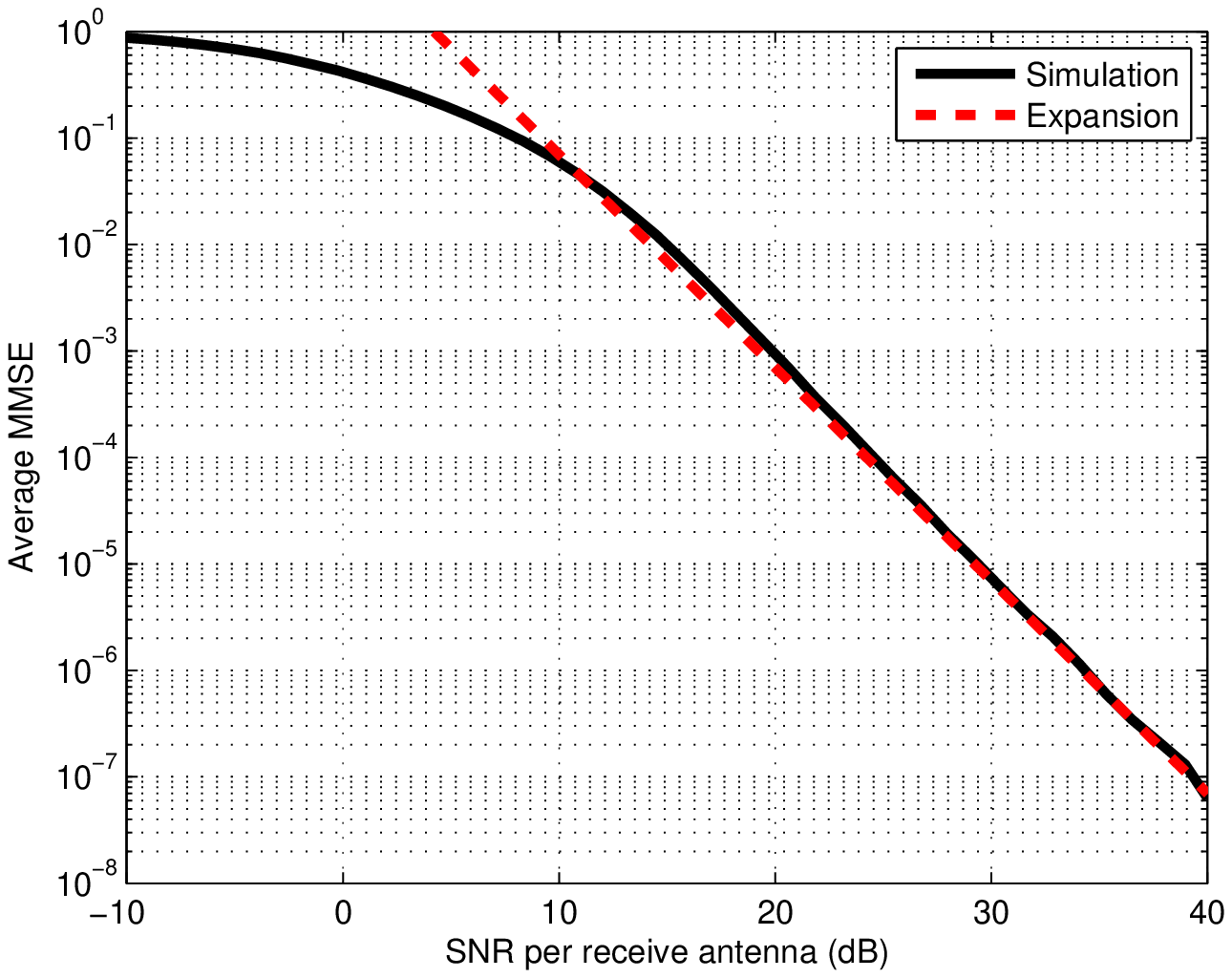,width=0.50\linewidth,clip=} \\
\centerline{(a)} \\
\epsfig{file=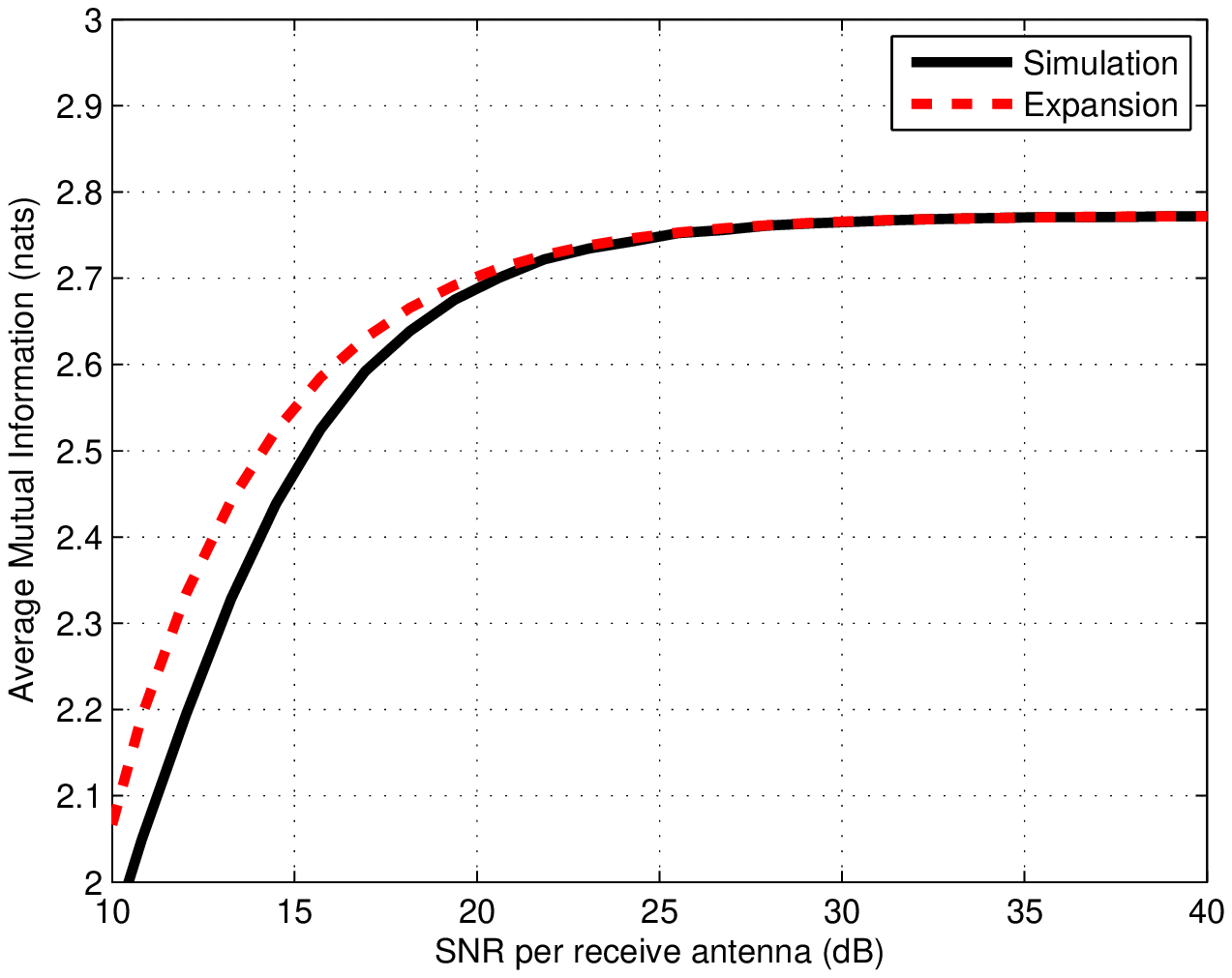,width=0.50\linewidth,clip=} \\
\centerline{(b)} \\
\epsfig{file=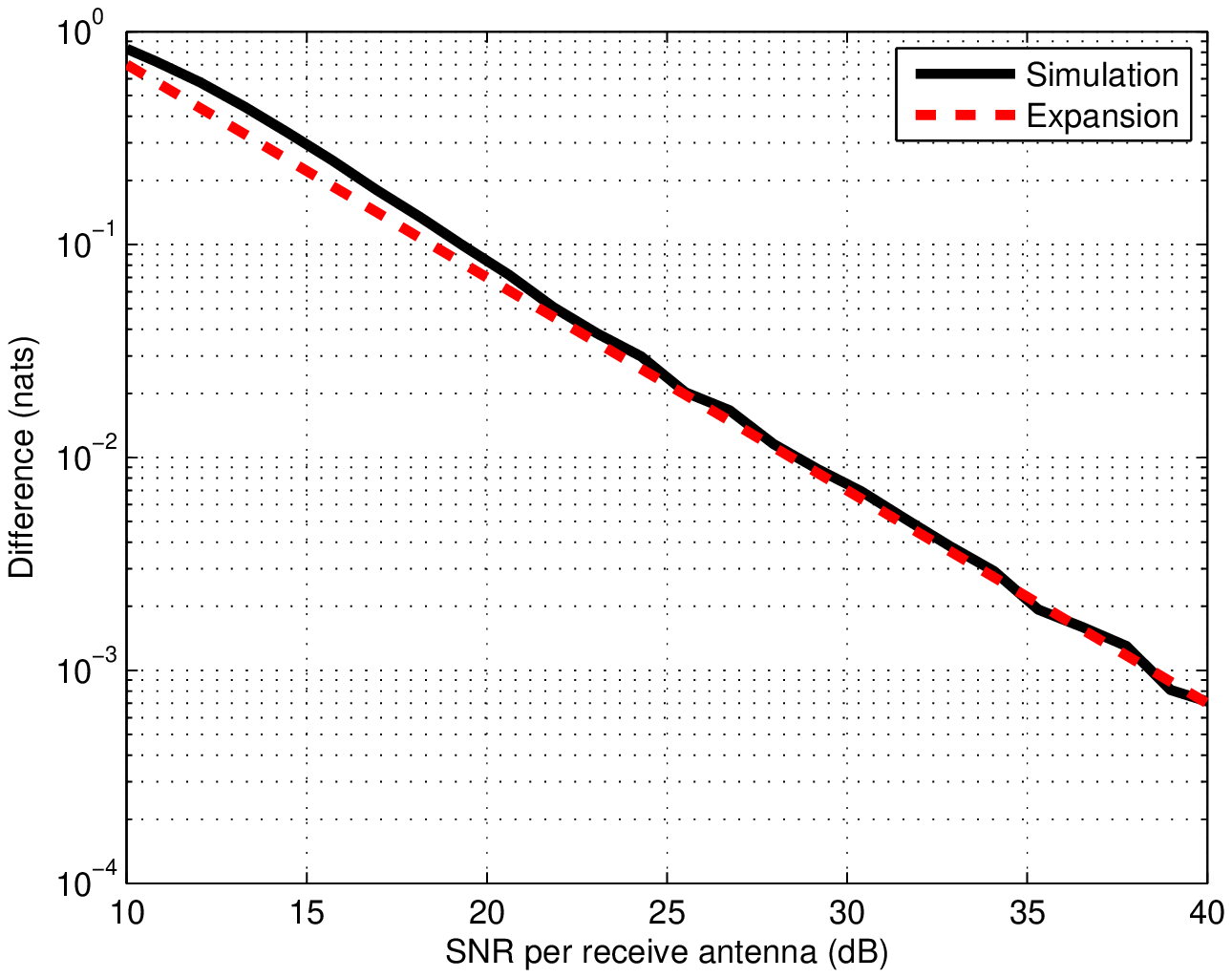,width=0.50\linewidth,clip=} \\
\centerline{(c)} \\
\end{tabular}
\caption{$1 \times 1$ Ricean fading coherent channel with a 16-QAM input ($K = 2$, ${\bf H}_0 = 1$): a) average MMSE; b) average mutual information; (c) difference between maximum average mutual information and average mutual information.} \label{siso_ricean2_16qam}
\end{figure}
\begin{figure}
\centering
\begin{tabular}{c}
\epsfig{file=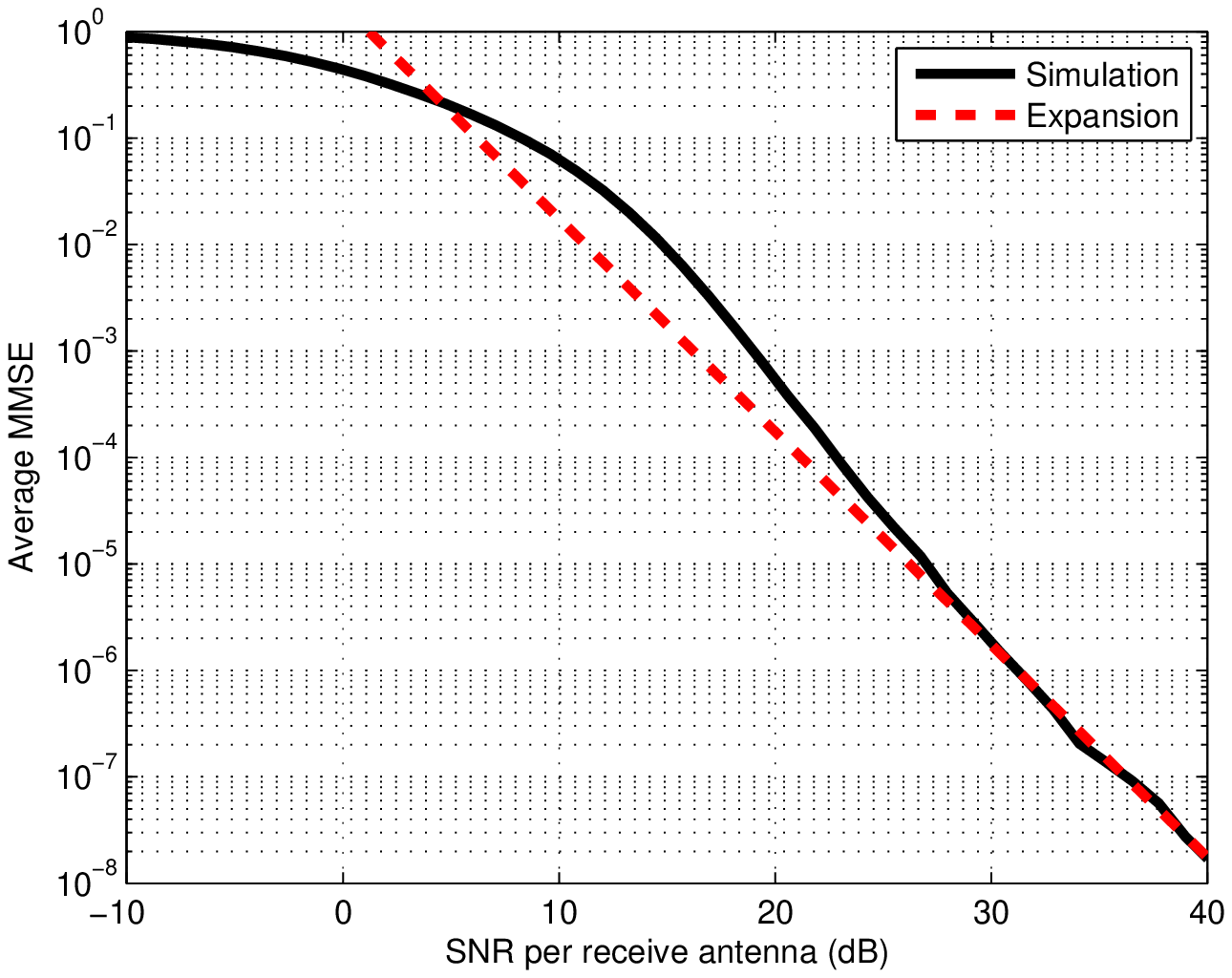,width=0.50\linewidth,clip=} \\
\centerline{(a)} \\
\epsfig{file=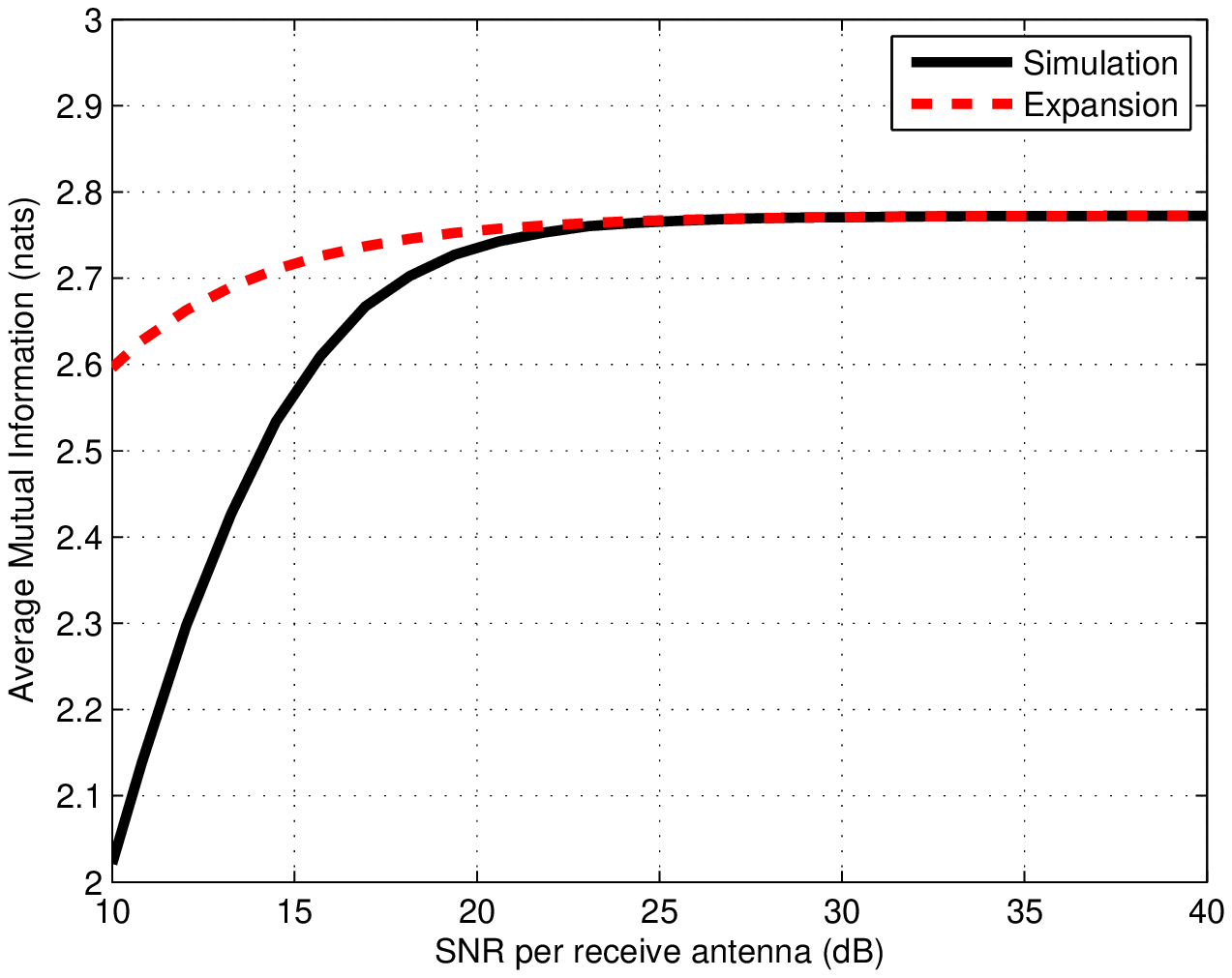,width=0.50\linewidth,clip=} \\
\centerline{(b)} \\
\epsfig{file=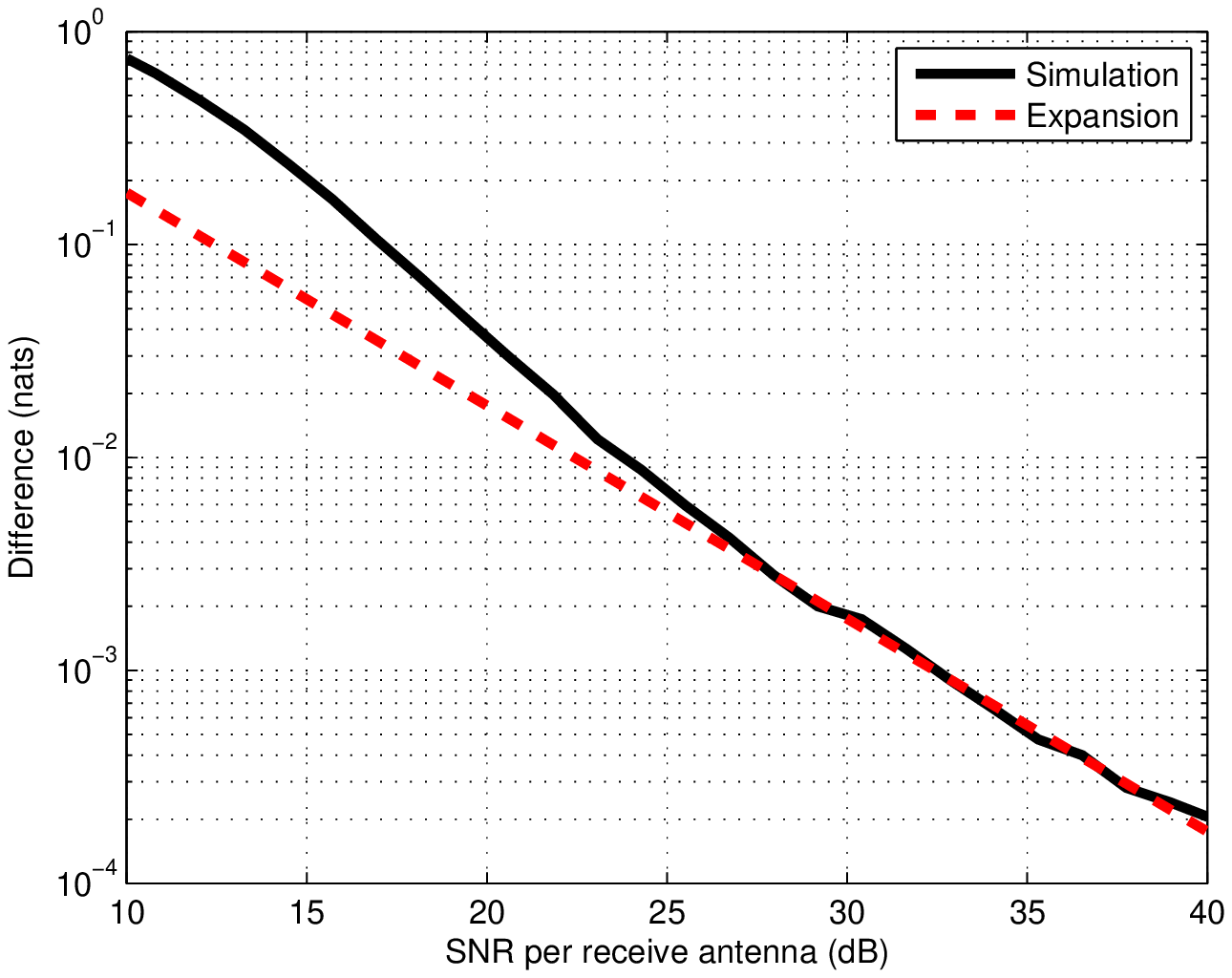,width=0.50\linewidth,clip=} \\
\centerline{(c)} \\
\end{tabular}
\caption{$1 \times 1$ Ricean fading coherent channel with a 16-QAM input ($K = 4$, ${\bf H}_0 = 1$): a) average MMSE; b) average mutual information; (c) difference between maximum average mutual information and average mutual information.} \label{siso_ricean4_16qam}
\end{figure}

The regime of validity of the asymptotic expansions in certain fading models also deserves more attention. The Ricean fading model is an intriguing case study. For example, it is well known that with the increase in the $K$-factor a Ricean fading channel will approach an AWGN-like channel, so that one would expect the average MMSE and the average mutual information to tend to their infinite-${\sf snr}$ values at a much faster rate than ${\sf d} + 1$ and ${\sf d}$, respectively, as put forth in Theorems \ref{mmse_expansion} and \ref{mutual_information_expansion}. Figures \ref{siso_ricean2_16qam} and \ref{siso_ricean4_16qam} overcome this apparent contradiction by showing that the average MMSE and the average mutual information do indeed tend to their infinite-${\sf snr}$ values at rates ${\sf d} + 1$ and ${\sf d}$, respectively. However, the influence of such terms in the expansions only shows up at extremely high signal-to-noise ratios; other possible terms in the expansion -- which mimic more closely the AWGN-like behavior -- play a more prominent role at moderate and high signal-to-noise ratios. Obviously, in such circumstances, typically associated with larger K-factors, the expansions will be of little practical use because they fail to characterize the behavior of the quantities in regimes of interest.



\begin{figure}
\centering
\begin{tabular}{c}
\epsfig{file=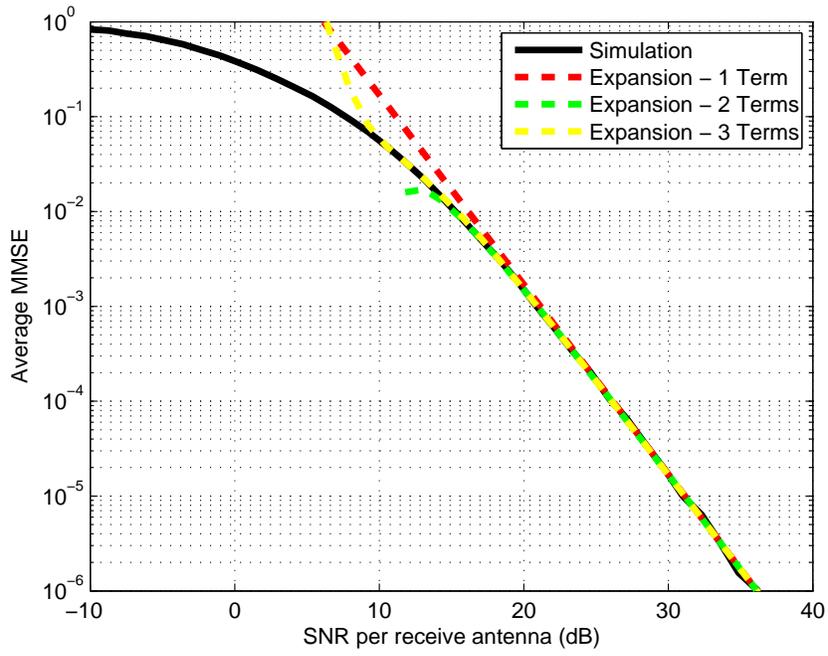,width=0.75\linewidth,clip=} \\
\centerline{(a)} \\
\epsfig{file=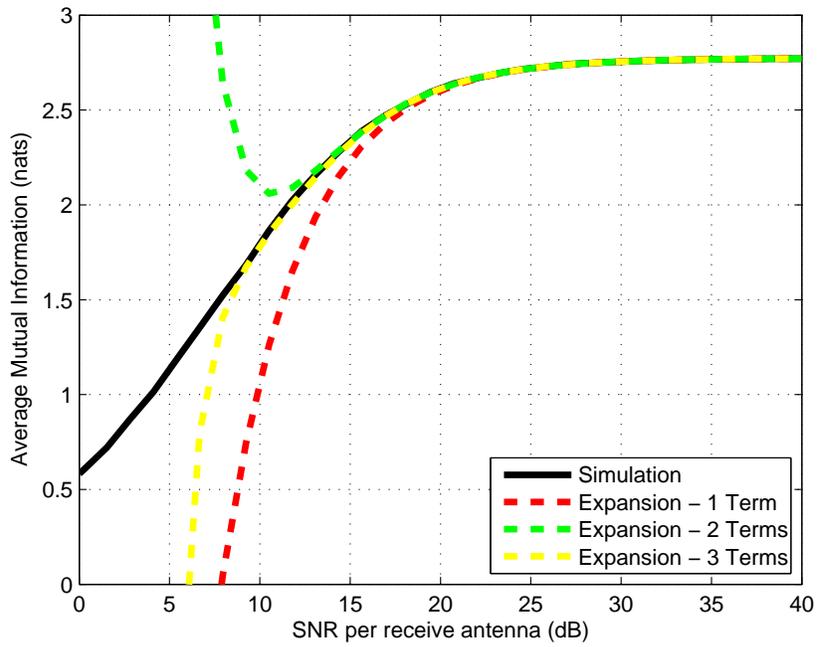,width=0.75\linewidth,clip=} \\
\centerline{(b)} \\
\end{tabular}
\caption{$1 \times 1$ Rayleigh fading coherent channel with a 16-QAM input : a) higher-order expansions of average MMSE; b) higher-order expansions of average mutual information.} \label{siso_rayleigh_16qam_higher_order_expansions}
\end{figure}

\begin{figure}
\centering
\begin{tabular}{c}
\epsfig{file=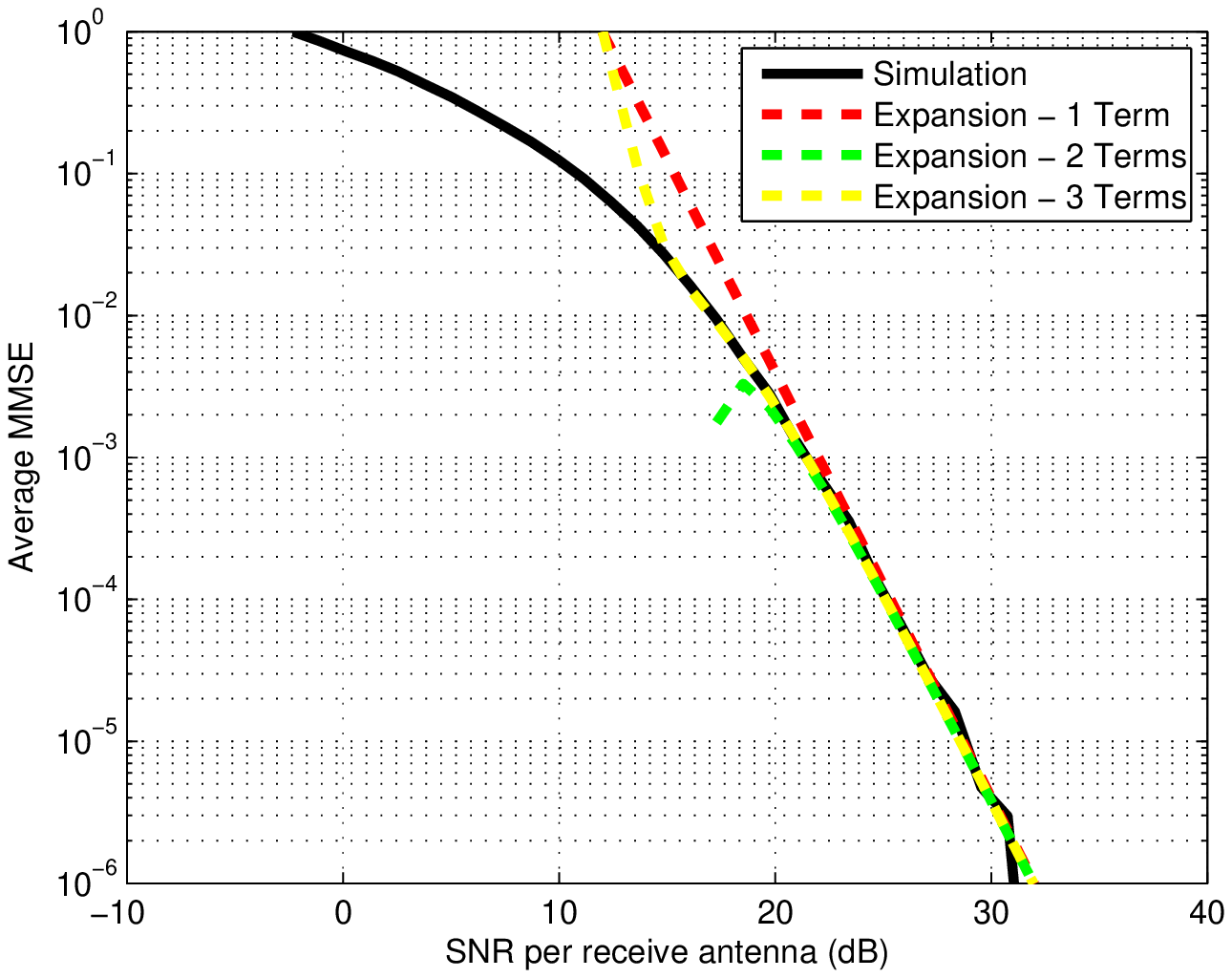,width=0.75\linewidth,clip=} \\
\centerline{(a)} \\
\epsfig{file=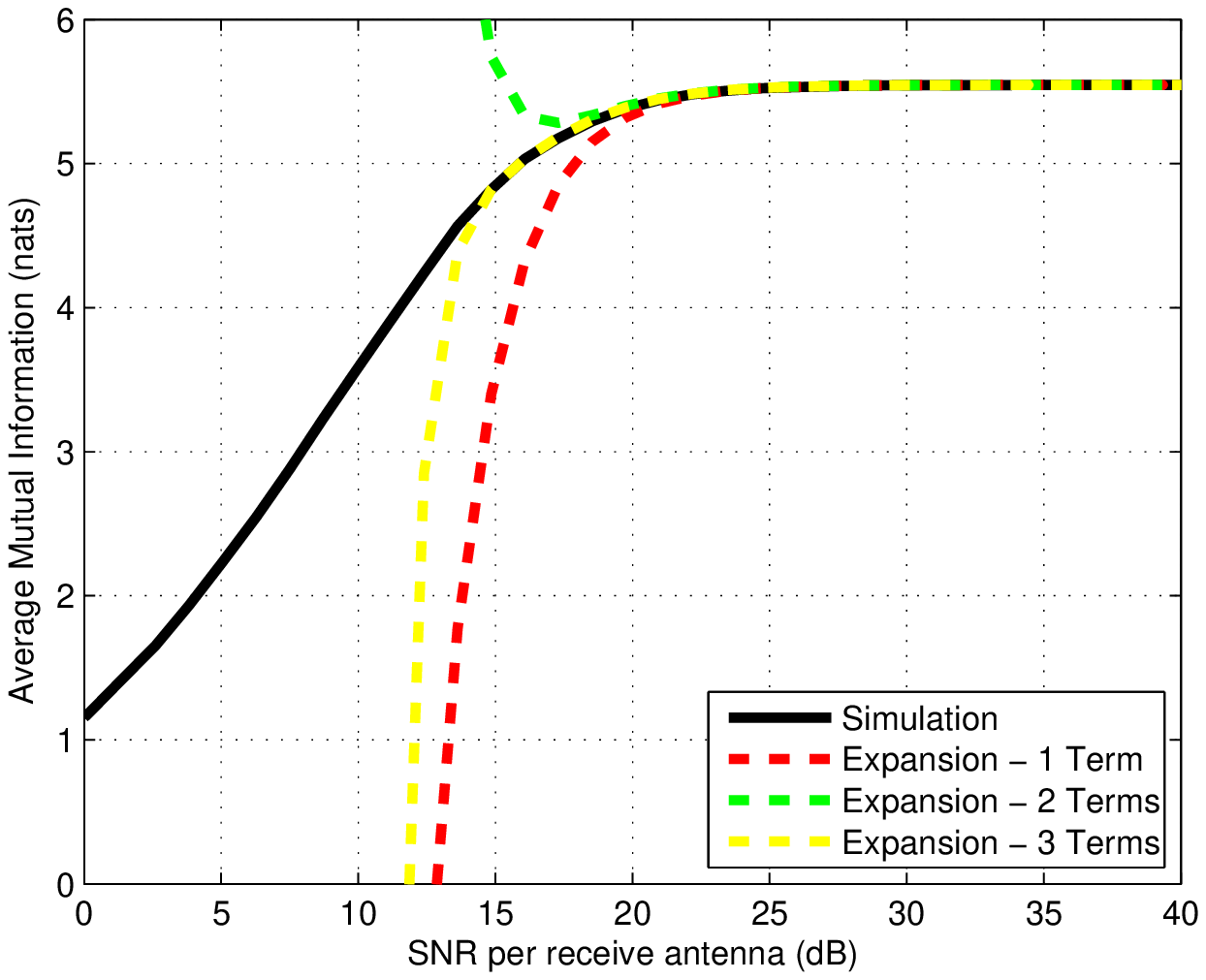,width=0.75\linewidth,clip=} \\
\centerline{(b)} \\
\end{tabular}
\caption{$2 \times 2$ Rayleigh fading coherent channel with 16-QAM inputs : a) higher-order expansions of average MMSE; b) higher-order expansions of average mutual information.} \label{mimo22_rayleigh_16qam_higher_order_expansions}
\end{figure}

Another interesting issue relates to possible generalizations of the expansions in \eqref{mmse_expansion_} and \eqref{mutual_information_expansion_}. In view of the expansions of the upper and lower bounds to the average value of the MMSE as well as the upper and lower bounds to the average value of the mutual information put forth in Lemmas \ref{mmse_fading} and \ref{mutual_information_fading}, and the arguments that lead to Theorems \ref{mmse_expansion} and \ref{mutual_information_expansion}, it is also tempting to conjecture that the average MMSE behaves as:
\begin{align}
{\sf \overline{mmse}} \left({\sf snr}\right) = \sum_{k={\sf d}}^{N} \epsilon_{k} \cdot \frac{1}{{\sf snr}^{k+1}} + o \left(\frac{1}{{\sf snr}^{N+1}}\right) \label{mmse_expansion_conjecture}
\end{align}
for arbitrary $N$, with
\begin{align}
\epsilon_{k} = \lim_{{\sf snr} \to \infty} {\sf snr}^{k+1} \cdot \left({\sf \overline{mmse}} \left({\sf snr}\right) - \sum_{k'={\sf d}}^{k-1} \epsilon_{k'} \cdot \frac{1}{{\sf snr}^{k'+1}}\right)
\end{align}
whereas the average mutual information behaves as:
\begin{align}
{\sf \bar{I}} \left({\sf snr}\right) = \log {\sf M} - \sum_{k={\sf d}}^{N} \epsilon'_{k} \cdot \frac{1}{{\sf snr}^{k}} + o \left(\frac{1}{{\sf snr}^{N}}\right) \label{mutual_information_expansion_conjecture}
\end{align}
for arbitrary $N$ with
\begin{align}
\epsilon'_{k} = \lim_{{\sf snr} \to \infty} {\sf snr}^{k} \cdot \left(\log {\sf M} - {\sf \bar{I}} \left({\sf snr}\right) - \sum_{k'={\sf d}}^{k-1} \epsilon'_{k'} \cdot \frac{1}{{\sf snr}^{k'}}\right)
\end{align}
Expansions \eqref{mmse_expansion_conjecture} and \eqref{mutual_information_expansion_conjecture} would then provide a dissection of the high-${\sf snr}$ behavior of the average value of the MMSE and the average value of the mutual information of arbitrary equiprobable discrete inputs observed through a multiple-antenna fading coherent channel in Gaussian noise in terms of a series of quantities, namely, $\epsilon_{k}$ and $\epsilon'_{k}$ for $k = {\sf d}, {\sf d}+1, {\sf d}+2, \ldots$. Interestingly, though our analytic techniques do not lead to \eqref{mmse_expansion_conjecture} and \eqref{mutual_information_expansion_conjecture}, Figures \ref{siso_rayleigh_16qam_higher_order_expansions} and \ref{mimo22_rayleigh_16qam_higher_order_expansions} show that such expansions provide a more accurate representation of the average value of the MMSE and the average value of the mutual information than the expansions in \eqref{mmse_expansion_} and \eqref{mutual_information_expansion_}.

Finally, it is also of interest to characterize the asymptotic behavior of the probability of error in a multiple-antenna fading coherent channel with arbitrary equiprobable discrete inputs in the regime of high ${\sf snr}$. We consider upper and lower bounds to the (uncoded) probability of error associated with a maximum likelihood detector for the channel model in \eqref{channel_model}, for a fixed channel matrix, given by:
\begin{align}
{\sf P_{e}} \left({\sf snr}; {\bf H}\right) = \Pr \left( \widehat{{\bf x}}_{{\sf ML}} \left(\sqrt{{\sf snr}} \cdot {\bf H} {\bf x} + {\bf n}\right) \neq {\bf x} | {\bf H} \right)
\end{align}
as well as upper and lower bounds to the average value of the (uncoded) probability of error associated with the maximum-likelihood detector for the channel model in \eqref{channel_model}, for a random channel matrix, given by:
\begin{align}
{\sf \bar{P}_{e}} \left({\sf snr}\right) = \mathbb{E}_{{\bf H}} \left\{{\sf P_{e}} \left({\sf snr}; {\bf H}\right)\right\} = \mathbb{E}_{{\bf H}} \left\{\Pr \left( \widehat{{\bf x}}_{{\sf ML}} \left(\sqrt{{\sf snr}} \cdot {\bf H} {\bf x} + {\bf n}\right) \neq {\bf x} | {\bf H} \right)\right\}
\end{align}
where $\widehat{{\bf x}}_{{\sf ML}} \left(\sqrt{{\sf snr}} \cdot {\bf H} {\bf x} + {\bf n}\right)$ corresponds to the maximum-likelihood detector estimate of the input given the output of the channel model in \eqref{channel_model}. We also consider the asymptotic expansions of the upper and lower bounds to the average value of the probability of error. Lemmas \ref{pe_deterministic} and \ref{pe_fading} summarize the results.

\vspace{0.25cm}

\begin{lemma}
\label{pe_deterministic}
The (uncoded) probability of error associated with maximum likelihood detection in the channel model in \eqref{channel_model}, for a fixed channel matrix, can be bounded as follows:
\begin{equation}
{\sf P_{e_{LB}}} \left({\sf snr};{\bf H}\right) \leq {\sf P_{e}}
\left({\sf snr};{\bf H}\right) \leq {\sf P_{e_{UB}}} \left({\sf
snr};{\bf H}\right)
\end{equation}
where the lower and upper bounds are given by:
\begin{align}
{\sf P_{e_{LB}}} \left({\sf snr};{\bf H}\right) &= \frac{1}{{\sf M} ({\sf M}-1)} \sum_{i=1}^{{\sf M}}\sum_{\substack{j=1 \\ j \neq i}}^{{\sf M}} \frac{1}{2} \cdot {\sf erfc} \left(\sqrt{\frac{{\sf d_{ij}^2 \left({\bf H}\right)}{\sf snr}}{4}}\right)  \label{pe_lb}
\end{align}
\begin{align}
{\sf P_{e_{UB}}} \left({\sf snr};{\bf H}\right) &= \frac{1}{{\sf M}} \sum_{i=1}^{{\sf M}}\sum_{\substack{j=1 \\ j \neq i}}^{{\sf M}} \frac{1}{2} \cdot {\sf erfc} \left(\sqrt{\frac{{\sf d_{ij}^2 \left({\bf H}\right)}{\sf snr}}{4}}\right) \label{pe_ub}
\end{align}
\end{lemma}

\vspace{0.25cm}

\begin{proof}
See Appendix F.
\end{proof}

\vspace{0.25cm}

\begin{lemma}
\label{pe_fading} Assume that $p_{{\sf d_{ij}^2}}^{(n)} \left({\sf d_{ij}^2}\right), n = 0,1,2,\ldots$ are continuous and integrable in $[0,\infty)$. Then, the average value of the (uncoded) probability of error associated with maximum likelihood detection in the channel model in \eqref{channel_model}, for a random channel matrix, can be bounded as follows:
\begin{equation}
{\sf \bar{P}_{e_{LB}}} \left({\sf snr}\right) \leq {\sf \bar{P}_{e}} \left({\sf snr}\right)
\leq {\sf \bar{P}_{e_{UB}}} \left({\sf snr}\right)
\end{equation}
where the asymptotic expansion as ${\sf snr} \to \infty$ of the lower and upper bounds are
given by:
\begin{align}
{\sf \bar{P}_{e_{LB}}} \big({\sf snr}\big) & =\sum_{n=0}^{N} \frac{1}{{\sf snr}^{n+1}} \cdot k''_{{\sf LB}_{n+1}} \cdot \left(\sum_{i=1}^{{\sf M}} \sum_{\substack{j=1 \\ j \neq i}}^{{\sf M}} p_{{\sf d_{ij}^2}}^{(n)} (0)\right) + \mathcal{O} \left(\frac{1}{{\sf snr}^{N+2}}\right), \qquad N = 0,1,\ldots \label{pe_fading_lower_bound_expansion} \\
{\sf \bar{P}_{e_{UB}}} \big({\sf snr}\big) & =\sum_{n=0}^{N} \frac{1}{{\sf snr}^{n+1}} \cdot k''_{{\sf UB}_{n+1}} \cdot \left(\sum_{i=1}^{{\sf M}} \sum_{\substack{j=1 \\ j \neq i}}^{{\sf M}} p_{{\sf d_{ij}^2}}^{(n)} (0)\right) + \mathcal{O} \left(\frac{1}{{\sf snr}^{N+2}}\right), \qquad N = 0,1,\ldots \label{pe_fading_upper_bound_expansion}
\end{align}
and
\begin{align}
k''_{{\sf LB}_n} &= \frac{1}{2 {\sf M} ({\sf M}-1)} \cdot \frac{4^n}{\sqrt{\pi}} \cdot \frac{\Gamma \left(n+1/2\right)}{\Gamma \left(n+1\right)}
\end{align}
\begin{align}
k''_{{\sf UB}_n} &= \frac{1}{2 {\sf M}} \cdot \frac{4^n}{\sqrt{\pi}} \cdot \frac{\Gamma \left(n+1/2\right)}{\Gamma \left(n+1\right)}
\end{align}
and $\Gamma \left(\cdot\right)$ is the Gamma function.
\end{lemma}

\vspace{0.25cm}

\begin{proof}
See Appendix G.
\end{proof}

\vspace{0.25cm}

Note that as ${\sf snr} \rightarrow \infty$ the upper and lower bounds to the average value of the probability of error are arbitrarily tight so that:
\begin{align}
{\sf \bar{P}_e} ({\sf snr}) = \mathcal{O} \left(\frac{1}{{\sf snr}^{\sf d}}\right) \neq o \left(\frac{1}{{\sf snr}^{\sf d}}\right) \label{characterization_pe}
\end{align}
We immediately recognize the quantity ${\sf d}$, which also appears in the expansions of the average value of the MMSE and the average value of the mutual information in Theorems \ref{mmse_expansion} and \ref{mutual_information_expansion}, respectively, to be the familiar diversity gain given by~\cite{Proakis08}:
\begin{align}
{\sf d} = - \lim_{{\sf snr} \to \infty} \frac{\log {\sf \bar{P}_e} ({\sf snr})}{\log {\sf snr}}
\end{align}

Interestingly, the high-${\sf snr}$ asymptotic behavior of the average value of the minimum mean-squared error, the average value of the mutual information and also the average value of the probability of error exhibit some common features, with the value of the parameter ${\sf d}$ and the value of the bounds to $\epsilon_{\sf d} \left({\sf snr}\right)$ and $\epsilon'_{\sf d} \left({\sf snr}\right)$ dictated by the value of the probability density function of the squared pairwise Euclidean distances at zero, $p_{{\sf d_{ij}^2}} (0)$, or the value of their higher-order derivatives at zero, $p_{{\sf d_{ij}^2}}^{(n)} (0), n > 0$. This dependency, which has also been very briefly noticed in~\cite{Zheng2003}, is intuitive because one could argue that in fading channels at high-${\sf snr}$ the quantities are affected primarily by the value of the probability of arbitrarily close (noiseless) receive vectors due to unfavorable fading realizations, in the same way that in non-fading channels at high-${\sf snr}$ the minimum mean-squared error, the mutual information, and the probability of error are primarily affected by the value of the Euclidean distance between the (noiseless) receive vectors~\cite{Perez-Cruz10}. As a side remark, we observe that the values of $p_{{\sf d_{ij}^2}} (0)$ or $p_{{\sf d_{ij}^2}}^{(n)} (0), n > 0$, which depend on the channel statistics as well as the system elements, can then be the basis of the characterization or the optimization of the constrained capacity of multiple-antenna fading coherent channels driven by arbitrary equiprobable discrete inputs in the regime of high ${\sf snr}$. Sections \ref{canonical_channel}, \ref{other_channels} and \ref{designs} concentrate on such issues.

\vspace{-0.25cm}

\section{The Canonical i.i.d. Rayleigh Fading Coherent Channel}
\label{canonical_channel}

It is now relevant to characterize the constrained capacity of the canonical i.i.d. Rayleigh fading coherent channel with arbitrary equiprobable discrete inputs in the regime of high-${\sf snr}$. The objective is to understand the effect on the constrained capacity of various system parameters, such as the number of transmit antennas, the number of receive antennas as well as the characteristics of the signalling scheme. 

The following Theorem unveils the high-${\sf snr}$ behavior of the constrained capacity. 
We represent the squared pairwise Euclidean distance between two arbitrary transmit vectors ${\bf x}_i$ and ${\bf x}_j$ by ${\sf \bar{d}_{ij}}^2 = \left\|{\bf x}_i - {\bf x}_j\right\|^2$.

\vspace{0.25cm}

\begin{theorem}
\label{characterization_mutual_information_mimo_rayleigh}
In the regime of high-${\sf snr}$, the constrained capacity of the canonical i.i.d. multiple-transmit--multiple-receive antenna Rayleigh fading coherent channel with arbitrary equiprobable discrete inputs obeys:
\begin{align}
{\sf \bar{I}} ({\sf snr}) = \log {\sf M} - \epsilon'_{n_r} \left({\sf snr}\right) \cdot \frac{1}{{\sf snr}^{n_r}} + \mathcal{O} \left(\frac{1}{{\sf snr}^{n_r+1}}\right)
\end{align}
where
\begin{align}
k'_{UB_{n_r}} \cdot \sum_{i=1}^{{\sf M}} \sum_{\substack{j=1 \\ j \neq i}}^{{\sf M}} \left(\frac{1}{{\sf \bar{d}_{ij}^2}}\right)^{n_r} \leq \epsilon'_{n_r} \left({\sf snr}\right) \leq k'_{LB_{n_r}} \cdot \sum_{i=1}^{{\sf M}} \sum_{\substack{j=1 \\ j \neq i}}^{{\sf M}} \left(\frac{1}{{\sf \bar{d}_{ij}^2}}\right)^{n_r} \label{bounds_mimo_rayleigh}
\end{align}
and $k'_{LB_{n_r}}$ and $k'_{UB_{n_r}}$ are given by \eqref{constant_mutual_information_lb} and \eqref{constant_mutual_information_ub}, respectively.
\end{theorem}

\vspace{0.25cm}

\begin{proof}
The basis of the proof is the calculation of the probability density function of the squared pairwise Euclidean distance between two arbitrary (noiseless) receive vectors given by:
\begin{align}
{\sf d_{ij}^2} = {\sf tr} \left({\bf H_w} {\bf \Delta}_{ij} {\bf H_w^\dag}\right)
\end{align}
where ${\bf \Delta}_{ij} = ({\bf x}_i - {\bf x}_j) ({\bf x}_i - {\bf x}_j)^\dag$. Denote the eigenvalue decomposition of the positive semi-definite matrix ${\bf \Delta}_{ij}$ by ${\bf \Delta}_{ij} = {\bf W}_{ij} {\bf \Lambda}_{ij} {\bf W}_{ij}^\dag$, where ${\bf W}_{ij}$ is a unitary matrix and ${\bf \Lambda}_{ij} = {\sf diag} \big(\lambda_{ij},0,\cdots,0\big)$ is a diagonal matrix with a single non-zero diagonal element $\lambda_{ij} = {\sf tr} \left({\bf \Lambda}_{ij}\right) = {\sf tr} \left({\bf \Delta}_{ij}\right) = {\sf tr} \left(({\bf x}_i - {\bf x}_j) ({\bf x}_i - {\bf x}_j)^\dag\right) = \left\|({\bf x}_i - {\bf x}_j)\right\|^2 = {\sf \bar{d}_{ij}^2}$ (note that there is a single non-zero diagonal element $\lambda_{ij} > 0$ because the matrix ${\bf \Delta}_{ij}$ is unit rank). It is possible to show (see~\cite{Telatar99}) that the distribution of
\begin{align}
{\sf tr} \left({\bf H_w} {\bf \Delta}_{ij} {\bf H_w^\dag}\right) = {\sf tr} \left({\bf H_w} {\bf W}_{ij} {\bf \Lambda}_{ij} {\bf W}_{ij}^\dag {\bf H_w^\dag}\right)
\end{align}
is equal to the distribution of
\begin{align}
{\sf tr} \left({\bf H_w} {\bf \Lambda}_{ij} {\bf H_w^\dag}\right) = \sum_{k=1}^{n_r} \lambda_{ij} |\xi_k|^2
\end{align}
where $\xi_k, k = 1,\ldots,n_r,$ are independent circularly symmetric complex Gaussian random variables with zero-mean and unit-variance. This is a central chi-square (or gamma) distribution with $2n_r$ degrees of freedom with probability density function given by~\cite{Proakis08}:
\begin{equation}
p_{{\sf d_{ij}^2}} \big({\sf d_{ij}^2}\big) = \frac{1}{\lambda_{ij}^{n_r} \left(n_r - 1\right)!} \cdot \big({\sf d_{ij}^2}\big)^{n_r-1} \cdot e^{-\frac{{\sf d_{ij}^2}}{\lambda_{ij}}}, \qquad {\sf d_{ij}^2} \geq 0 \label{pdf_dij_mimo_rayleigh}
\end{equation}

The high-${\sf snr}$ expansion of the constrained capacity follows immediately from Theorem \ref{mutual_information_expansion}, by noting that the function $\frac{1}{\lambda_{ij}^{n_r} \left(n_r - 1\right)!} \cdot \big({\sf d_{ij}^2}\big)^{n_r-1} \cdot e^{-\frac{{\sf d_{ij}^2}}{\lambda_{ij}}}$ and its higher-order derivatives are continuous and integrable on $[0,\infty)$, $p_{{\sf d_{ij}^2}}^{(k)} (0) = 0, k = 0,\ldots,n_r-2,$ and $p_{{\sf d_{ij}^2}}^{(n_r-1)} (0) = \frac{1}{\lambda_{ij}^{n_r}} \neq 0$.
\end{proof}


Theorem \ref{characterization_mutual_information_mimo_rayleigh} defines the impact of the number of transmit and receive antennas as well as the characteristics of the signalling scheme on the constrained capacity of the canonical i.i.d. Rayleigh fading coherent channel with arbitrary equiprobable discrete inputs, in the regime of high ${\sf snr}$. In particular, in view of the bounds in \eqref{bounds_mimo_rayleigh}, this Theorem is consistent with the fact that constellations with poor sphere-packing properties require more signal-to-noise ratio than constellation with good sphere-packing properties in order to achieve a certain target constrained capacity in the regime of high ${\sf snr}$.

\begin{figure}
\centering
\epsfig{file=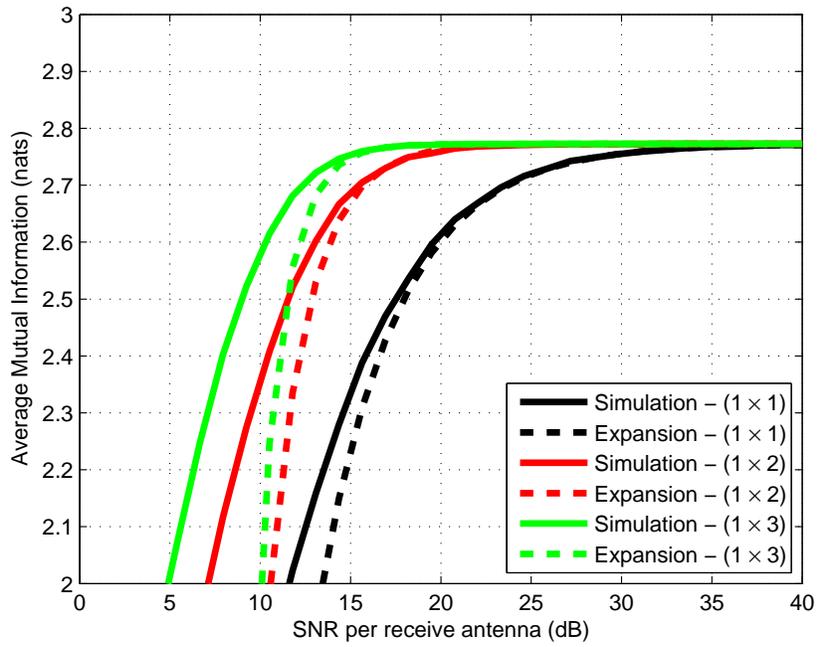,width=0.75\linewidth,clip=} \\
\caption{Comparison of constrained capacity for $1 \times 1$, $1 \times 2$ and $1 \times 3$ Rayleigh fading coherent channels with 16-QAM inputs.} \label{rx_antennas_comparisons}
\end{figure}

\begin{figure}
\centering
\epsfig{file=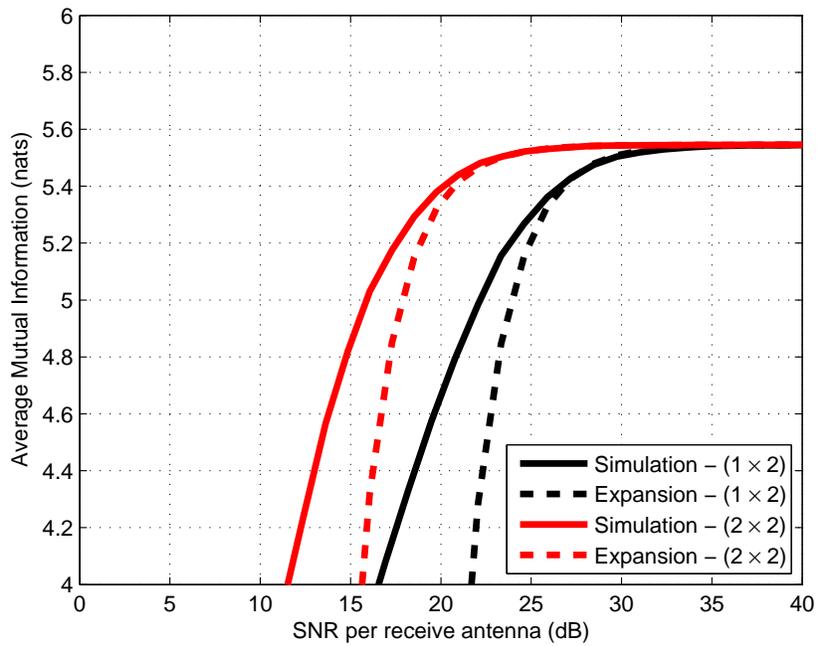,width=0.75\linewidth,clip=} \\
\caption{Comparison of constrained capacity for a $1 \times 2$ Rayleigh fading coherent channel with a 256-QAM input and a $2 \times 2$ Rayleigh fading coherent channel with 16-QAM inputs.} \label{tx_antennas_comparisons}
\end{figure}

The number of receive antennas controls the rate at which the constrained capacity value tends to its infinite-${\sf snr}$ value, so that a multiple-receive antenna system requires less signal-to-noise ratio than a single-receive antenna system in order to achieve a certain target constrained capacity in the regime of high ${\sf snr}$. 
The number of transmit antennas, as opposed to its effect on the capacity of the canonical i.i.d. multiple-antenna Rayleigh fading coherent channel~\cite{Telatar99}, does not affect directly the multiplexing ability of the system because the infinite-${\sf snr}$ constrained capacity value depends solely on the number of constellation points or vectors. Instead, a higher-dimensional (complex) space, in comparison to a lower-dimensional one, enables the construction of more efficiently packed multi-dimensional constellations~~\cite{Conway88}. This, in view of the bounds in \eqref{bounds_mimo_rayleigh}, suggests that a multiple-transmit antenna system requires potentially less signal-to-noise ratio than a single-transmit antenna system in order to achieve a certain target constrained capacity in the regime of high ${\sf snr}$. Figures \ref{rx_antennas_comparisons} and \ref{tx_antennas_comparisons}, which illustrate these aspects, also demonstrate that the asymptotic expansions capture this behavior of the constrained capacity in the regime of high-${\sf snr}$.

Theorem \ref{characterization_mutual_information_mimo_rayleigh} also suggests that appropriate constellations for communication over the canonical i.i.d. multiple-antenna Rayleigh fading coherent channel -- in the sense of optimization of the constrained capacity -- optimize the quantity:
\begin{align}
\sum_{i=1}^{{\sf M}} \sum_{\substack{j=1 \\ j \neq i}}^{{\sf M}} \left(\frac{1}{{\sf \bar{d}_{ij}^2}}\right)^{n_r} \approx {\sf K} \cdot \left(\frac{1}{{\sf \bar{d}_{min}^2}}\right)^{n_r} \label{constellation_design_criteria_mimo_rayleigh}
\end{align}
where ${\sf \bar{d}_{min}^2} = \min_{i \neq j} {\sf \bar{d}_{ij}^2}$ represents the minimum squared Euclidean distance between pairs of constellation points or vectors and ${\sf K}$ represents the number of pairs of points or vectors with pairwise squared Euclidean distance equal to the minimum squared Euclidean distance, a quantity akin to the kissing number. Interestingly, a different compromise appears to exist between ${\sf K}$ and ${\sf \bar{d}_{min}^2}$ in canonical i.i.d Rayleigh fading coherent channels where only the receiver knows the channel state than in fading channels where both the transmitter and the receiver know the channel state. Note that, when the channel matrix is known to the transmitter and the receiver, the high-${\sf snr}$ expansion of the constrained capacity depends exponentially on ${\sf \bar{d}_{min}^2}$~\cite{Perez-Cruz10}.

\vspace{-0.25cm}

\section{Other Fading Coherent Channels}
\label{other_channels}

It is also relevant to characterize the constrained capacity of non-canonical fading coherent channels with arbitrary equiprobable discrete inputs in the regime of high ${\sf snr}$. We study models that arise in many practical scenarios such as: \emph{i}) the antenna-correlated Rayleigh fading coherent channel; and \emph{ii}) the Ricean fading coherent channel. \footnote{We consider for simplicity exclusively Ricean fading channels with no transmit or receive antenna correlation.}


\vspace{-0.4cm}

\subsection{Antenna-Correlated Rayleigh Fading Channels}

We write ${\bf \Delta}_{ij} = ({\bf x}_i - {\bf x}_j) ({\bf x}_i - {\bf x}_j)^\dag$. We denote the eigenvalue decomposition of the positive semi-definite matrix ${\bf \Theta_T^{\frac{1}{2}}} {\bf \Delta}_{ij} {\bf \Theta_T^{\frac{\dag}{2}}}$ by  ${\bf \Theta_T^{\frac{1}{2}}} {\bf \Delta}_{ij} {\bf \Theta_T^{\frac{\dag}{2}}} = {\bf W_{T_{ij}}} {\bf \Lambda_{T_{ij}}} {\bf W_{T_{ij}}^\dag}$, where ${\bf W_{T_{ij}}}$ is a unitary matrix and ${\bf \Lambda_{T_{ij}}} = {\sf diag} \big({\lambda_{T_{ij}}},0,\ldots,0\big)$ is a diagonal matrix with at most a single non-zero diagonal element ${\lambda_{T_{ij}}} \geq 0$ (note that there is at most a single non-zero diagonal element ${\lambda_{T_{ij}}} \geq 0$ because the matrix ${\bf \Theta_T^{\frac{1}{2}}} {\bf \Delta}_{ij} {\bf \Theta_T^{\frac{\dag}{2}}}$ is at most unit rank). We also denote the eigenvalue decomposition of the positive semi-definite matrix ${\bf \Theta_R}$ by ${\bf \Theta_R} = {\bf W_R} {\bf \Lambda_R} {\bf W_R^\dag}$, where ${\bf W_R}$ is a unitary matrix and ${\bf \Lambda_R} = {\sf diag} \big({\lambda_{R_k}}\big)$ is a diagonal matrix with diagonal elements ${\lambda_{R_k}} \geq 0, k = 1,\ldots,n_r$.


The following Theorem, which represents a generalization of Theorem \ref{characterization_mutual_information_mimo_rayleigh}, defines the high-${\sf snr}$ behavior of the constrained capacity of antenna-correlated Rayleigh fading coherent channels. The Theorem concentrates on non-degenerate scenarios where ${\lambda_{T_{ij}}} > 0, \forall~i \neq j$, and ${\lambda_{R_k}} > 0, \forall~k$. \footnote{The Theorem also considers only scenarios where the eigenvalues ${\lambda_{R_k}}, k = 1,\ldots,n_r,$ are either all distinct or all equal (necessarily to one). The Theorem does not consider scenarios where there are groups of identical eigenvalues. This generalization requires considerable algebraic manipulation without adding much relevant insight. Note that $\lambda_{R_1} = \lambda_{R_2} = \cdots = \lambda_{R_{n_r}} = 1 $ corresponds to a scenario where the receive antennas are uncorrelated.}

\vspace{0.25cm}

\begin{theorem}
\label{characterization_mutual_information_mimo_rayleigh_correlation}
In the regime of high-${\sf snr}$, the constrained capacity of the $n_r \times n_t$ antenna-correlated Rayleigh fading coherent channel with arbitrary equiprobable discrete inputs obeys:
\begin{align}
{\sf \bar{I}} ({\sf snr}) = \log {\sf M} - \epsilon'_{n_r} \left({\sf snr}\right) \cdot \frac{1}{{\sf snr}^{n_r}} + \mathcal{O} \left(\frac{1}{{\sf snr}^{n_r+1}}\right)
\end{align}
where
\begin{align}
k'_{UB_{n_r}} \cdot \sum_{i=1}^{{\sf M}} \sum_{\substack{j=1 \\ i \neq j}}^{{\sf M}} \left(\frac{1}{\lambda_{T_{ij}}}\right)^{n_r} \cdot \left(\prod_{k=1}^{n_r} \frac{1}{{\lambda_{R_k}}}\right) \leq \epsilon'_{n_r} \left({\sf snr}\right) \leq k'_{LB_{n_r}} \cdot \sum_{i=1}^{{\sf M}} \sum_{\substack{j=1 \\ i \neq j}}^{{\sf M}} \left(\frac{1}{\lambda_{T_{ij}}}\right)^{n_r} \cdot \left(\prod_{k=1}^{n_r} \frac{1}{{\lambda_{R_k}}}\right) \label{bounds_mimo_rayleigh_correlation}
\end{align}
and $k'_{LB_{n_r}}$ and $k'_{UB_{n_r}}$ are given by \eqref{constant_mutual_information_lb} and \eqref{constant_mutual_information_ub}, respectively.
\end{theorem}

\vspace{0.25cm}

\begin{proof}
The squared pairwise Euclidean distance between two arbitrary (noiseless) receive vectors is given by:
\begin{align}
{\sf d_{ij}^2} = {\sf tr} \Big({\bf \Theta_R^{\frac{1}{2}}} {\bf H_w} {\bf \Theta_T^{\frac{1}{2}}} {\bf \Delta}_{ij} {\bf \Theta_T^{\frac{\dag}{2}}} {\bf H_w^\dag} {\bf \Theta_R^{\frac{\dag}{2}}}\Big)
\end{align}
It is possible to show (see~\cite{Telatar99}) that the distribution of
\begin{align}
{\sf tr} \Big({\bf \Theta_R^{\frac{1}{2}}} {\bf H_w} {\bf \Theta_T^{\frac{1}{2}}} {\bf \Delta}_{ij} {\bf \Theta_T^{\frac{\dag}{2}}} {\bf H_w^\dag} {\bf \Theta_R^{\frac{\dag}{2}}}\Big) =
{\sf tr} \left({\bf W_R} {\bf \Lambda_R^{\frac{1}{2}}} {\bf W_R^\dag} {\bf H_w} {\bf W_{T_{ij}}} {\bf \Lambda_{T_{ij}}} {\bf W_{T_{ij}}^\dag} {\bf H_w^\dag} {\bf W_R} {\bf \Lambda_R^{\frac{1}{2}}} {\bf W_R^\dag}\right)
\end{align}
is equal to the distribution of
\begin{align}
{\sf tr} \left({\bf \Lambda_R^{\frac{1}{2}}} {\bf H_w} {\bf \Lambda_{T_{ij}}} {\bf H_w^\dag} {\bf \Lambda_R^{\frac{1}{2}}}\right) = \sum_{k=1}^{n_r} {\lambda_{T_{ij}}} {\lambda_{R_k}} |\xi_k|^2
\end{align}
where $\xi_k, k = 1,\ldots,n_r,$ are independent circularly symmetric complex Gaussian random variables with zero-mean and unit-variance.

Assume that the eigenvalues of the receive correlation matrix are all equal (necessarily to one), i.e., ${\lambda_{R_1}} = {\lambda_{R_2}} = \cdots = {\lambda_{R_{n_r}}} = 1$. Then, the probability density function of ${\sf d_{ij}^2}$ is given by~\cite{Proakis08}:
\begin{equation}
p_{{\sf d_{ij}^2}} \big({\sf d_{ij}^2}\big) = \frac{1}{{\lambda_{T_{ij}}^{n_r}} \big(n_r-1\big)!} \cdot \big({\sf d_{ij}^2}\big)^{n_r-1} \cdot e^{-\frac{{\sf d_{ij}^2}}{{\lambda_{T_{ij}}}}}, \qquad {\sf d_{ij}^2} \geq 0
\end{equation}
The high-${\sf snr}$ characterization of the constrained capacity follows from Theorem \ref{mutual_information_expansion}, by noting that
\begin{align}
p_{{\sf d_{ij}^2}}^{(k)} (0) = 0, \qquad k = 0,\ldots,n_r-2
\end{align}
and
\begin{align}
p_{{\sf d_{ij}^2}}^{(n_r-1)} (0) = \frac{1}{{\lambda_{T_{ij}}^{n_r}}} \neq 0
\end{align}

Assume now that the eigenvalues of the receive correlation matrix are all distinct, i.e., ${\lambda_{R_1}} \neq {\lambda_{R_2}} \neq \cdots \neq {\lambda_{R_{n_r}}}$. Then, the probability density function of ${\sf d_{ij}^2}$ is given by~\cite{Scheuer88}:
\begin{equation}
p_{{\sf d_{ij}^2}} \big({\sf d_{ij}^2}\big) = \sum_{k=1}^{n_r} \frac{1}{{\lambda_{T_{ij}}} {\lambda_{R_k}} \prod_{\substack{k' = 1 \\ k' \neq k}}^{n_r} \left(1 - \frac{\lambda_{R_{k'}}}{\lambda_{R_{k}}}\right)} \cdot e^{-\frac{{\sf d_{ij}^2}}{{\lambda_{T_{ij}}} {\lambda_{R_k}}}}, \qquad {\sf d_{ij}^2} \geq 0
\end{equation}
The high-${\sf snr}$ characterization of the constrained capacity also follows from Theorem \ref{mutual_information_expansion}, by noting that
\begin{align}
p_{{\sf d_{ij}^2}}^{(k)} (0) = 0, \qquad k = 0,\ldots,n_r-2
\end{align}
and
\begin{align}
p_{{\sf d_{ij}^2}}^{(n_r-1)} (0) = \frac{1}{\lambda_{T_{ij}}^{n_r}} \cdot \prod_{k=1}^{n_r} \frac{1}{{\lambda_{R_k}}} \neq 0
\end{align}

It is also simple to show that the conditions for the application of Theorem \ref{mutual_information_expansion} are satisfied, i.e., the functions:
\begin{align}
\frac{1}{{\lambda_{T_{ij}}^{n_r}} \big(n_r-1\big)!} \cdot \big({\sf d_{ij}^2}\big)^{n_r-1} \cdot e^{-\frac{{\sf d_{ij}^2}}{{\lambda_{T_{ij}}}}}
\end{align}
and
\begin{align}
\sum_{k=1}^{n_r} \frac{1}{{\lambda_{T_{ij}}} {\lambda_{R_k}} \prod_{\substack{k' = 1 \\ k' \neq k}}^{n_r} \left(1 - \frac{\lambda_{R_{k'}}}{\lambda_{R_{k}}}\right)} \cdot e^{-\frac{{\sf d_{ij}^2}}{{\lambda_{T_{ij}}} {\lambda_{R_k}}}}
\end{align}
and their higher-order derivatives are continuous and integrable on $[0,\infty)$.
\end{proof}

\vspace{0.25cm}

It is also instructive to examine the implications of degenerate conditions where ${\lambda_{T_{ij}}} = 0$ for some values of $i$ and $j$ and ${\lambda_{R_k}} = 0$ for some values of $k$. Assume that the receive correlation matrix ${\bf \Theta_R}$ has $n'$ non-zero eigenvalues ${\lambda_{R_k}} > 0$ and $n'' = n_r - n'$ zero eigenvalues ${\lambda_{R_k}} = 0$, so that there are $n''$ perfectly correlated paths. Assume also that some of the matrices ${\bf \Theta_T^{\frac{1}{2}}} {\bf \Delta}_{ij} {\bf \Theta_T^{\frac{\dag}{2}}}$ have eigenvalue ${\lambda_{T_{ij}}} = 0$, so that the vector ${\bf x}_i - {\bf x}_j$ lies in the null space of the transmit correlation matrix ${\bf \Theta_T}$. This is also equivalent to saying that the receiver cannot distinguish some of the transmit vectors. By letting $\log {\sf M'} = {\sf H} \left({\bf \Theta_R^{\frac{1}{2}}} {\bf H_w} {\bf \Theta_T^{\frac{1}{2}}} {\bf x}\big|{\bf H_w}\right) < {\sf H} \left({\bf x}\right) = \log {\sf M}$ , a simple modification of the Theorems reveals that as ${\sf snr} \to \infty$ the constrained capacity behaves as follows:
\begin{align}
{\sf \bar{I}} ({\sf snr}) = \log {\sf M'} - \epsilon'_{n'} \left({\sf snr}\right) \cdot \frac{1}{{\sf snr}^{n'}} + \mathcal{O} \left(\frac{1}{{\sf snr}^{n'+1}}\right)
\end{align}
where
\begin{align}
k'_{UB_{n'}} \cdot \sum_{\lambda_{T_{ij}} > 0} \left(\frac{1}{\lambda_{T_{ij}}}\right)^{n'} \cdot \prod_{\lambda_{R_k} > 0} \frac{1}{{\lambda_{R_k}}} \leq \epsilon'_{n'} \left({\sf snr}\right) \leq k'_{LB_{n'}} \cdot \sum_{\lambda_{T_{ij}} > 0} \left(\frac{1}{\lambda_{T_{ij}}}\right)^{n'} \cdot \prod_{\lambda_{R_k} > 0} \frac{1}{{\lambda_{R_k}}}
\end{align}
and $k'_{LB_{n'}}$ and $k'_{UB_{n'}}$ are given by \eqref{constant_mutual_information_lb} and \eqref{constant_mutual_information_ub}, respectively. This shows that degenerate scenarios affect the constrained capacity infinite-${\sf snr}$ value as well as the rate at which the constrained capacity tends to the infinite-${\sf snr}$ value.

\begin{figure}
\centering
\epsfig{file=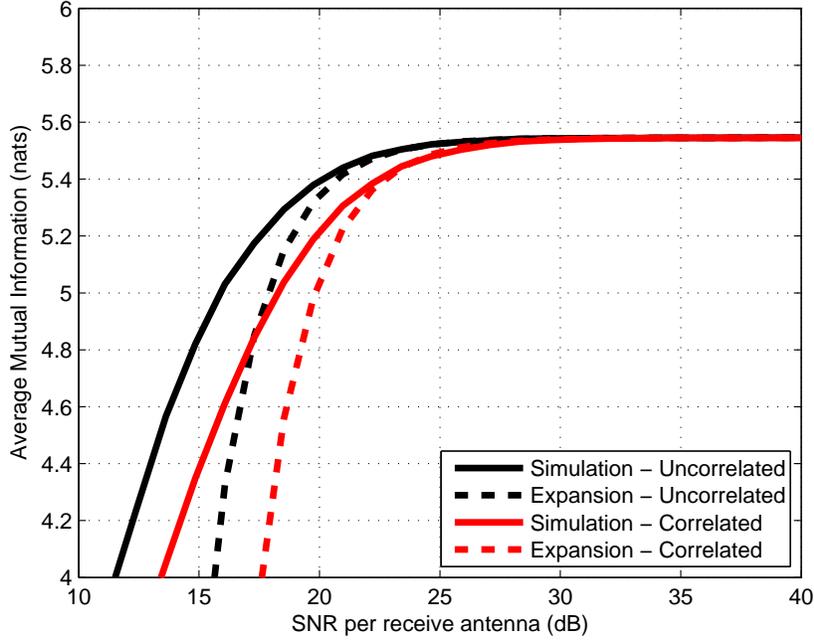,width=0.75\linewidth,clip=} \\
\caption[Text excluding the matrix]{Comparison of constrained capacity for a $2 \times 2$ antenna-uncorrelated Rayleigh fading coherent channel with 16-QAM inputs and a $2 \times 2$ antenna-correlated Rayleigh fading coherent channel with 16-QAM inputs $\left({\bf \Phi_T} = \left[\begin{smallmatrix} 1 & 0.5 \\ 0.5 & 1 \end{smallmatrix}\right] \text{ and } {\bf \Phi_R} = \left[\begin{smallmatrix} 1 & 0.8 \\ 0.8 & 1 \end{smallmatrix}\right]\right)$.} \label{correlation_comparisons}
\end{figure}

In general, and the effect of degenerate conditions apart, the bounds in \eqref{bounds_mimo_rayleigh_correlation} suggest that the constrained capacity of a Rayleigh fading channel with transmit and receive antenna correlation is lower than the constrained capacity of a Rayleigh fading channel without antenna correlation for a certain signal-to-noise ratio in the regime of high ${\sf snr}$. This is due to the fact that the value of the bounds in the correlated scenario in \eqref{bounds_mimo_rayleigh_correlation} is higher than the value of the bounds in the uncorrelated scenario in \eqref{bounds_mimo_rayleigh}. For example, in the presence of receive antenna correlation, it is possible to show that
\begin{align}
\prod_{k=1}^{n_r} \frac{1}{{\lambda_{R_k}}} \geq 1
\end{align}
due to the fact that the function $\prod_{k=1}^{n_r} \frac{1}{{\lambda_{R_k}}}$ is Schur-convex and the vector $\left[\lambda_{R_1}~\lambda_{R_2}~\cdots~\lambda_{R_{n_r}}\right]$ majorizes the vector $\left[1~1~\cdots~1\right]$~\cite{Palomar06b}.
In the presence of transmit antenna correlation, it is possible to show 
that the convex function
\begin{align}
\sum_{i=1}^{{\sf M}} \sum_{\substack{j=1 \\ i \neq j}}^{{\sf M}} \left(\frac{1}{\lambda_{T_{ij}}}\right)^{n_r} &= \sum_{i=1}^{{\sf M}} \sum_{\substack{j=1 \\ i \neq j}}^{{\sf M}} \left(\frac{1}{{\sf tr} \left(({\bf x}_i - {\bf x}_j)^\dag {\bf \Theta_T} ({\bf x}_i - {\bf x}_j)\right)}\right)^{n_r}
\end{align}
is minimized by ${\bf \Theta_T} = {\bf I}$ in the set of unit-diagonal positive semi-definite matrices ${\bf \Theta_T}$. This result assumes that the constellations exhibit certain common symmetries (see also Section \ref{designs}). Then,
\begin{align}
\sum_{i=1}^{{\sf M}} \sum_{\substack{j=1 \\ i \neq j}}^{{\sf M}} \left(\frac{1}{\lambda_{T_{ij}}}\right)^{n_r} &= \sum_{i=1}^{{\sf M}} \sum_{\substack{j=1 \\ i \neq j}}^{{\sf M}} \left(\frac{1}{{\sf tr} \left(({\bf x}_i - {\bf x}_j)^\dag {\bf \Theta_T} ({\bf x}_i - {\bf x}_j)\right)}\right)^{n_r} \nonumber \\
&\geq \sum_{i=1}^{{\sf M}} \sum_{\substack{j=1 \\ i \neq j}}^{{\sf M}} \left(\frac{1}{{\sf tr} \left(({\bf x}_i - {\bf x}_j)^\dag ({\bf x}_i - {\bf x}_j)\right)}\right)^{n_r} = \sum_{i=1}^{{\sf M}} \sum_{\substack{j=1 \\ i \neq j}}^{{\sf M}} \left(\frac{1}{\bar{{\sf d}}_{ij}^2}\right)^{n_r}
\end{align}
This is portrayed in Figure \ref{correlation_comparisons}.

\vspace{-0.4cm}

\subsection{Ricean Fading Channels}


We also write ${\bf \Delta}_{ij} = ({\bf x}_i - {\bf x}_j) ({\bf x}_i - {\bf x}_j)^\dag$. We denote the eigenvalue decomposition of the positive semi-definite matrix ${\bf \Delta}_{ij}$ by ${\bf \Delta}_{ij} = {\bf W}_{ij} {\bf \Lambda}_{ij} {\bf W}_{ij}^\dag$, where ${\bf W}_{ij}$ is a unitary matrix and ${\bf \Lambda}_{ij} = {\sf diag} \big(\lambda_{ij},0,\cdots,0\big)$ is a diagonal matrix with a single non-zero diagonal element $\lambda_{ij} = {\sf tr} \left({\bf \Lambda}_{ij}\right) = {\sf tr} \left({\bf \Delta}_{ij}\right) = {\sf tr} \left(({\bf x}_i - {\bf x}_j) ({\bf x}_i - {\bf x}_j)^\dag\right) = \left\|({\bf x}_i - {\bf x}_j)\right\|^2 = {\sf \bar{d}_{ij}^2}$. 


The following Theorem, which also represents a generalization of Theorem \ref{characterization_mutual_information_mimo_rayleigh}, defines the high-${\sf snr}$ behavior of the constrained capacity of Ricean fading coherent channels.

\vspace{0.25cm}

\begin{theorem}
\label{characterization_mutual_information_mimo_ricean}
In the regime of high-${\sf snr}$, the constrained capacity of the $n_r \times n_t$ Ricean fading coherent channel with arbitrary equiprobable discrete inputs obeys:
\begin{align}
{\sf \bar{I}} ({\sf snr}) = \log {\sf M} - \epsilon'_{n_r} \left({\sf snr}\right) \cdot \frac{1}{{\sf snr}^{n_r}} + \mathcal{O} \left(\frac{1}{{\sf snr}^{n_r+1}}\right)
\end{align}
where
\begin{align}
&k'_{UB_{n_r}} \cdot \sum_{i=1}^{{\sf M}} \sum_{\substack{j=1 \\ j \neq i}}^{{\sf M}} \left(\frac{1}{{\sf \bar{d}_{ij}^2} \cdot \frac{1}{K+1}}\right)^{n_r} \cdot e^{- K \cdot {\sf tr} \left({\bf H_0} {\bf W}_{ij} {\bf e}_1 {\bf e}_1^\dag {\bf W}_{ij}^\dag {\bf H_0^\dag}\right)} \nonumber \\
&\leq \epsilon'_{n_r} \left({\sf snr}\right) \leq \nonumber \\
&k'_{LB_{n_r}} \cdot \sum_{i=1}^{{\sf M}} \sum_{\substack{j=1 \\ j \neq i}}^{{\sf M}} \left(\frac{1}{{\sf \bar{d}_{ij}^2} \cdot \frac{1}{K+1}}\right)^{n_r} \cdot e^{- K \cdot {\sf tr} \left({\bf H_0} {\bf W}_{ij} {\bf e}_1 {\bf e}_1^\dag {\bf W}_{ij}^\dag {\bf H_0^\dag}\right)} \label{bounds_mimo_ricean}
\end{align}
and $k'_{LB_{n_r}}$ and $k'_{UB_{n_r}}$ are given by \eqref{constant_mutual_information_lb} and \eqref{constant_mutual_information_ub}, respectively.
\end{theorem}

\vspace{0.25cm}

\begin{proof}
The squared pairwise Euclidean distance ${\sf d_{ij}^2}$  between two arbitrary (noiseless) receive vectors is given by:
\begin{align}
{\sf d_{ij}^2} = {\sf tr} \left( \left(\sqrt{\frac{K}{K+1}} \cdot {\bf H_0} + \sqrt{\frac{1}{K+1}} \cdot {\bf H_w}\right) {\bf \Delta}_{ij} \left(\sqrt{\frac{K}{K+1}} \cdot {\bf H_0} + \sqrt{\frac{1}{K+1}} \cdot {\bf H_w}\right)^\dag \right)
\end{align}
It is also possible to show (see~\cite{Telatar99}) that the distribution of
\begin{align}
&{\sf tr} \left( \left(\sqrt{\frac{K}{K+1}} \cdot {\bf H_0} + \sqrt{\frac{1}{K+1}} \cdot {\bf H_w}\right) {\bf \Delta}_{ij} \left(\sqrt{\frac{K}{K+1}} \cdot {\bf H_0} + \sqrt{\frac{1}{K+1}} \cdot {\bf H_w}\right)^\dag \right) = \nonumber \\
&={\sf tr} \left( \left(\sqrt{\frac{K}{K+1}} \cdot {\bf H_0} + \sqrt{\frac{1}{K+1}} \cdot {\bf H_w}\right) {\bf W}_{ij} {\bf \Lambda}_{ij} {\bf W}_{ij}^\dag \left(\sqrt{\frac{K}{K+1}} \cdot {\bf H_0} + \sqrt{\frac{1}{K+1}} \cdot {\bf H_w}\right)^\dag \right)
\end{align}
is equal to the distribution of
\begin{align}
&{\sf tr} \left( \left(\sqrt{\frac{K}{K+1}} \cdot {\bf W}_{ij}^\dag {\bf H_0} {\bf W}_{ij} + \sqrt{\frac{1}{K+1}} \cdot {\bf H_w}\right) {\bf \Lambda}_{ij} \left(\sqrt{\frac{K}{K+1}} \cdot {\bf W}_{ij}^\dag {\bf H_0} {\bf W}_{ij} + \sqrt{\frac{1}{K+1}} \cdot {\bf H_w}\right)^\dag \right) = \nonumber \\
&=\sum_{k=1}^{n_r} {\lambda_{ij}} |\xi_k|^2
\end{align}
where $\xi_k, k = 1,\ldots,n_r,$ are independent circularly symmetric complex Gaussian random variables with mean $\sqrt{\frac{K}{K+1}} \cdot {\bf e}_k^\dag {\bf W}_{ij}^\dag {\bf H_0} {\bf W}_{ij} {\bf e}_1$ and variance $\frac{1}{K+1}$, respectively. This is a non-central chi-square (or gamma) distribution with $2n_r$ degrees of freedom with probability density function given by~\cite{Proakis08}:
\begin{equation}
p_{{\sf d_{ij}^2}} \big({\sf d_{ij}^2}\big) = \frac{1}{\lambda_{ij} \cdot \frac{1}{K+1}} \cdot \left(\frac{{\sf d_{ij}^2}}{s^2}\right)^{\frac{n_r-1}{2}}\cdot e^{- \frac{s^2 + {\sf d_{ij}^2}}{\lambda_{ij} \cdot \frac{1}{K+1}}} \cdot I_{n_r-1} \left(\frac{2 \sqrt{s^2 {\sf d_{ij}^2}}}{\lambda_{ij} \cdot \frac{1}{K+1}}\right), \qquad {\sf d_{ij}^2} \geq 0
\end{equation}
where $s^2 = \lambda_{ij} \cdot \frac{K}{K+1}\cdot \left(\sum_{k=1}^{n_r} \left\|{\bf e}_k^\dag {\bf W}_{ij}^\dag {\bf H_0} {\bf W}_{ij} {\bf e}_1\right\|^2 \right) = \lambda_{ij} \cdot \frac{K}{K+1}\cdot {\sf tr} \left({\bf H_0} {\bf W}_{ij} {\bf e}_1 {\bf e}_1^\dag {\bf W}_{ij}^\dag {\bf H_0^\dag}\right)$ and $I_n(\cdot)$ is the $n$th-order modified Bessel function of the first-kind given by:
\begin{equation}
I_n (x) = \sum_{k=0}^{\infty} \frac{1}{k! \Gamma(n + k + 1)} \cdot \left(\frac{x}{2}\right)^{n+2k}, \qquad x \geq 0
\end{equation}

The high-${\sf snr}$ characterization of the constrained capacity also follows from Theorem \ref{mutual_information_expansion}, by noting that
\begin{align}
p_{{\sf d_{ij}^2}}^{(k)} (0) = 0, \qquad k = 0,\ldots,n_r-2
\end{align}
and
\begin{align}
p_{{\sf d_{ij}^2}}^{(n_r-1)} (0) = \frac{1}{\left(\lambda_{ij} \cdot \frac{1}{K+1}\right)^{n_r}} \cdot e^{- \frac{s^2}{\lambda_{ij} \cdot \frac{1}{K+1}}} \neq 0
\end{align}

It is also simple to show that the conditions for the application of Theorem \ref{mutual_information_expansion} are satisfied, i.e., the function:
\begin{align}
\frac{1}{\lambda_{ij} \cdot \frac{1}{K+1}} \cdot \left(\frac{{\sf d_{ij}^2}}{s^2}\right)^{\frac{n_r-1}{2}}\cdot e^{- \frac{s^2 + {\sf d_{ij}^2}}{\lambda_{ij} \cdot \frac{1}{K+1}}} \cdot I_{n_r-1} \left(\frac{2 \sqrt{s^2 {\sf d_{ij}^2}}}{\lambda_{ij} \cdot \frac{1}{K+1}}\right)
\end{align}
and its higher-order derivatives are continuous and integrable on $[0,\infty)$.
\end{proof}

\vspace{0.25cm}

\begin{figure}
\centering
\epsfig{file=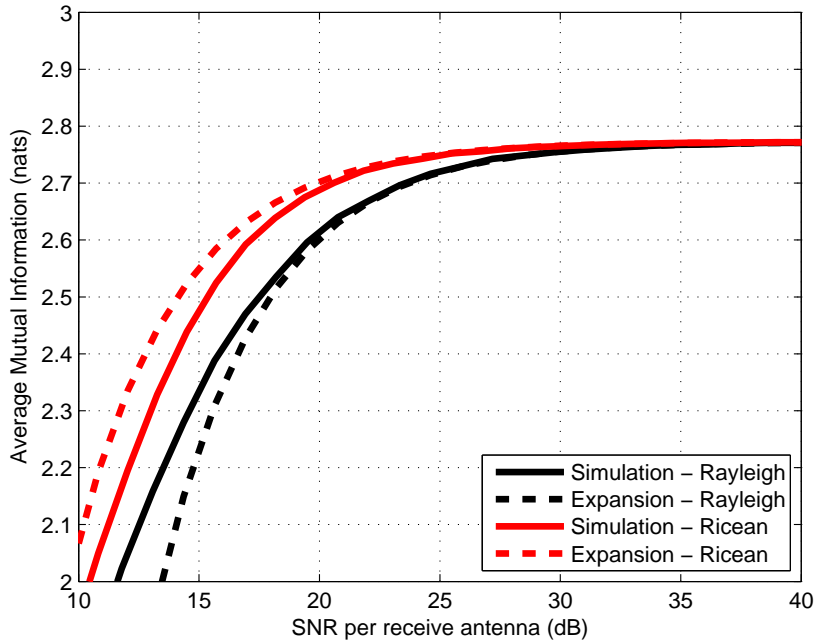,width=0.75\linewidth,clip=} \\
\caption[Text excluding the matrix]{Comparison of constrained capacity for a $1 \times 1$ Rayleigh fading coherent channel with a 16-QAM input and a $1 \times 1$ Ricean fading coherent channel with a 16-QAM input $\left(K = 2, {\bf H}_0 = 1\right)$.} \label{siso_ricean2_comparisons}
\end{figure}

In general, the bounds in \eqref{bounds_mimo_ricean} suggest that the constrained capacity of a Ricean fading channel is normally higher than the constrained capacity of the canonical i.i.d. Rayleigh fading channel for a certain signal-to-noise ratio in the regime of high ${\sf snr}$. 
This is due to the fact that the function
\begin{align}
&\sum_{i=1}^{{\sf M}} \sum_{\substack{j=1 \\ j \neq i}}^{{\sf M}} \left(\frac{1}{{\sf \bar{d}_{ij}^2} \cdot \frac{1}{K+1}}\right)^{n_r} \cdot e^{- K \cdot {\sf tr} \left({\bf H_0} {\bf W}_{ij} {\bf e}_1 {\bf e}_1^\dag {\bf W}_{ij}^\dag {\bf H_0^\dag}\right)} = \nonumber \\
&\sum_{i=1}^{{\sf M}} \sum_{\substack{j=1 \\ j \neq i}}^{{\sf M}} \left(\frac{1}{{\sf \bar{d}_{ij}^2} \cdot \frac{1}{K+1}}\right)^{n_r} \cdot e^{- K \cdot {\sf tr} \left({\bf a_R} {\bf a_T}^\dag {\bf W}_{ij} {\bf e}_1 {\bf e}_1^\dag {\bf W}_{ij}^\dag {\bf a_T} {\bf a_R}^\dag\right)} = \nonumber \\
&\sum_{i=1}^{{\sf M}} \sum_{\substack{j=1 \\ j \neq i}}^{{\sf M}} \left(\frac{1}{{\sf \bar{d}_{ij}^2} \cdot \frac{1}{K+1}}\right)^{n_r} \cdot e^{- n_r \cdot K \cdot {\sf tr} \left({\bf a_T}^\dag {\bf W}_{ij} {\bf e}_1 {\bf e}_1^\dag {\bf W}_{ij}^\dag {\bf a_T}\right)}
\end{align}
is monotonically decreasing in the $K$-factor $K$ provided that ${\sf tr} \left({\bf a_T}^\dag {\bf W}_{ij} {\bf e}_1 {\bf e}_1^\dag {\bf W}_{ij}^\dag {\bf a_T}\right) \geq 1$, $\forall~i \neq j$. This is portrayed in Figure \ref{siso_ricean2_comparisons}.

\begin{figure}
\centering
\epsfig{file=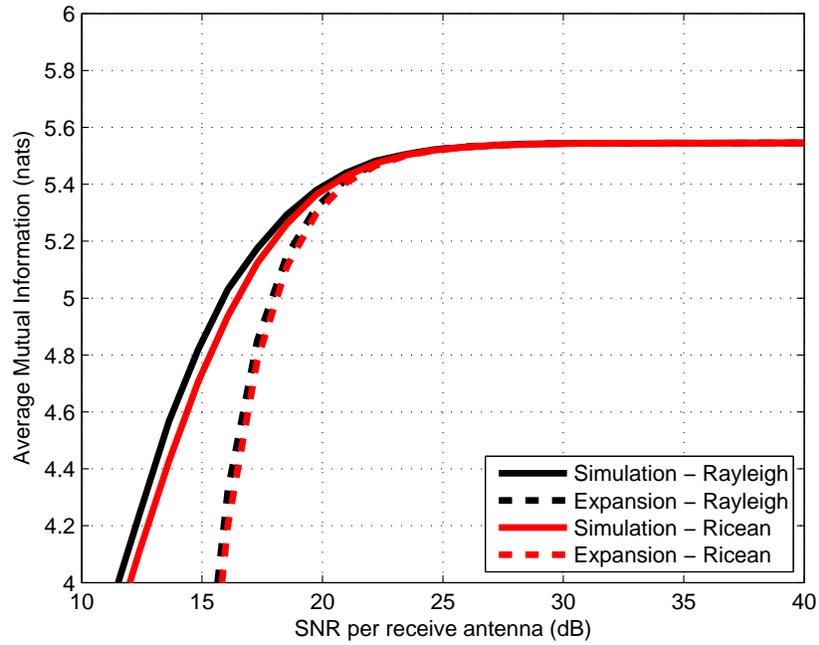,width=0.75\linewidth,clip=} \\
\caption[Text excluding the matrix]{Comparison of constrained capacity for a $2 \times 2$ Rayleigh fading coherent channel with 16-QAM inputs and a $2 \times 2$ Ricean fading coherent channel with 16-QAM inputs $\left(K = 3, {\bf H}_0 = \left[\begin{smallmatrix} 1 & 1 \\ 1 & 1 \end{smallmatrix}\right]\right)$.} \label{ricean3_comparisons}
\end{figure}

However, as shown in Figure \ref{ricean3_comparisons}, and in contrast to the capacity of a multiple-antenna Ricean fading channel~\cite{Kim03}, the constrained capacity of a Ricean fading channel can also be lower than the constrained capacity of a Rayleigh fading channel for a certain signal-to-noise ratio in the regime of high ${\sf snr}$. This aspect is illustrated further by the following toy-examples.


\begin{example}
\label{example_ricean}
Consider a $2 \times 2$ Ricean fading channel with two equiprobable discrete inputs ${\bf x}_1 = [1~0]^{\sf T}$ and ${\bf x}_2 = [0~1]^{\sf T}$, where ${\bf H} = \sqrt{\frac{K}{K+1}} \cdot {\bf H_0} + \sqrt{\frac{1}{K+1}} \cdot {\bf H_w}$ and ${\bf H_0} = {\bf a_R} {\bf a_T}^\dag$ with ${\bf a_R} = [1~1]^{\sf T}$ and ${\bf a_T} = [1~1]^{\sf T}$. Then,
\begin{align}
k'_{UB_{2}} \cdot \frac{1}{2} \cdot \left(K+1\right)^2 \leq \epsilon'_{2} \left({\sf snr}\right) \leq k'_{LB_{2}} \cdot \frac{1}{2} \cdot \left(K+1\right)^2 \label{bound_example_ricean}
\end{align}
with $k'_{LB_{2}}$ and $k'_{UB_{2}}$ given by \eqref{constant_mutual_information_lb} and \eqref{constant_mutual_information_ub}, respectively.
\end{example}


\begin{example}
\label{example_rayleigh}
Consider a $2 \times 2$ Rayleigh fading channel with the two equiprobable discrete inputs ${\bf x}_1 = [1~0]^{\sf T}$ and ${\bf x}_2 = [0~1]^{\sf T}$, where ${\bf H} = {\bf H_w}$. Then,
\begin{align}
k'_{UB_{2}} \cdot \frac{1}{2} \leq \epsilon'_{2} \left({\sf snr}\right) \leq k'_{LB_{2}} \cdot \frac{1}{2} \label{bound_example_rayleigh}
\end{align}
with $k'_{LB_{2}}$ and $k'_{UB_{2}}$ given by \eqref{constant_mutual_information_lb} and \eqref{constant_mutual_information_ub}, respectively.
\end{example}

\begin{figure}
\centering
\epsfig{file=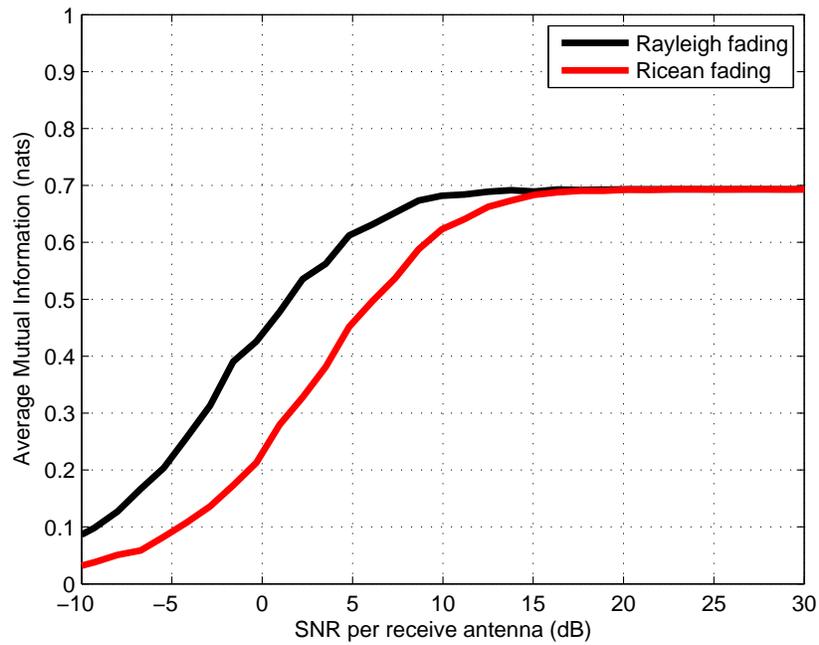,width=0.75\linewidth,clip=} \\
\caption{Comparison of constrained capacity for the fading coherent channels in Examples \ref{example_ricean} and \ref{example_rayleigh} ($K=2$).} \label{ricean_aberration}
\end{figure}

Examples \ref{example_ricean} and \ref{example_rayleigh} suggest that, due to the bounds in \eqref{bound_example_ricean} and \eqref{bound_example_rayleigh}, this Ricean fading channel needs a higher signal-to-noise ratio than the Rayleigh fading channel for $K > 0$, in order to achieve a certain target constrained capacity in the regime of high ${\sf snr}$. Figure \ref{ricean_aberration}, which shows only the simulated constrained capacity, confirms this behavior. This is due to the fact that one cannot rely on the deterministic component but rather, in comparison to a canonical i.i.d. Rayleigh fading channel, on the lower power random component in order to distinguish the transmit vectors. Note that the transmit vector differences ${\bf x}_1 - {\bf x}_2 = [1~-1]^{\sf T}$ and ${\bf x}_2 - {\bf x}_1 = [-1~1]^{\sf T}$ are orthogonal to the transmit array response ${\bf a_T} = [1~1]^{\sf T}$.

\vspace{-0.4cm}

\section{Designs}
\label{designs}


The focus now is on the design of schemes for communication over multiple-antenna fading coherent channels driven by equiprobable discrete inputs, in the regime of high ${\sf snr}$. In particular, we conceive designs for: \emph{i}) optimal power allocation over a bank of parallel independent fading coherent channels; \emph{ii}) optimal precoding for multiple-antenna fading coherent channels; and \emph{iii}) space-time coding for the multiple-antenna canonical i.i.d. Rayleigh fading coherent channel. The design principle is based on the optimization of the lower bound to the asymptotic expansion of the constrained capacity, rather than the exact asymptotic expansion of the constrained capacity. It is interesting to note though that the procedure leads to very sharp designs. It is also interesting to note that some of the design criteria coincide with standard design criteria in the literature, most notably the space-time coding criteria, which have been derived based on other principles.

\vspace{-0.4cm}

\subsection{Power Allocation in a Bank of Parallel Independent Fading Channels}

We consider a bank of $K$ parallel independent fading channels which can be modeled as follows:
\begin{equation}
y_k = \sqrt{{\sf snr}} \cdot h_k ~ \sqrt{p_k} ~ x_k + n_k, \qquad k = 1,\ldots,K
\end{equation}
for a single channel use, where $y_k \in \mathbb{C}$ is the $k$th sub-channel complex receive symbol, $x_k \in \mathbb{C}$ is the $k$th sub-channel complex transmit symbol, $h_k \in \mathbb{C}$ is the $k$th sub-channel random complex fading coefficient and $n_k \in \mathbb{C}$ is a circularly symmetric complex Gaussian noise random variable with zero-mean and unit-variance. The variable $p_k \in \mathbb{R}_{0}^{+}$ corresponds to the power injected into the $k$th sub-channel. The scaling factor ${\sf snr}$ relates to the signal-to-noise ratio. We assume that the transmit symbols $x_k, k = 1,\ldots,K,$ conform to equiprobable unit-power discrete constellations with cardinality ${\sf M_k}, k=1,\ldots,K$. We also assume that all the random variables are independent.

The constrained capacity of the bank of parallel independent fading channels is given by:
\begin{align}
{\sf \bar{I}} \left({\sf snr}\right) = \sum_{k=1}^{K} {\sf \bar{I}}_k \left({\sf snr}\right) = \sum_{k=1}^{K} \mathbb{E}_{h_k} \left\{{\sf I}(x_k;\sqrt{{\sf snr}} \cdot h_k ~ \sqrt{p_k} ~ x_k + n_k|h_k)\right\} \label{constrained_capacity_bank_independent_channels}
\end{align}
The objective is to determine the power allocation procedure that maximizes a lower bound to the constrained capacity in the asymptotic regime of high ${\sf snr}$, subject to the total power constraint:
\begin{align}
\sum_{k=1}^{K} p_k \leq {\sf P}
\end{align}
with $p_k \geq 0, k=1,\ldots,K$.

Next, we consider a bank of parallel independent channels subject to either Rayleigh fading or Ricean fading. We denote the squared pairwise Euclidean distance between two distinct transmit symbols in a particular sub-channel $k$, $x_{k_i}$ and $x_{k_j}$, by:
\begin{align}
{\sf \bar{d}_{k_{ij}}^2} = \left\|x_{k_i} - x_{k_j}\right\|^2
\end{align}
We also denote the squared pairwise Euclidean distance between two distinct noiseless receive symbols in a particular sub-channel $k$, $h_k \sqrt{p_k} x_{k_i}$ and $h_k \sqrt{p_k} x_{k_j}$, by:
\begin{align}
{\sf d_{k_{ij}}^2} = \left\|h_k \sqrt{p_k} \left(x_{k_i} - x_{k_j}\right)\right\|^2 = p_k \left\|h_k\right\|^2 \left\|x_{k_i} - x_{k_j}\right\|^2 = p_k \left|h_k\right|^2 {\sf \bar{d}_{k_{ij}}^2}
\end{align}

\subsubsection{Rayleigh fading case} In the Rayleigh fading case, the complex fading coefficients $h_k \sim \mathcal{CN} \big(0,\sigma_{h_k}^2\big), k=1,\ldots,K$. 
Therefore, the probability density function of the squared pairwise Euclidean distance between two distinct noiseless receive symbols in a particular sub-channel $k$ is given by:
\begin{align}
\label{pdf_dist_siso_rayleigh}
p_{{\sf d_{k_{ij}}^2}} \left({\sf d_{k_{ij}}^2}\right) = \frac{1}{{\sf \bar{d}_{k_{ij}}^2} \sigma_{h_k}^2 p_k} \cdot e^{- \frac{{\sf d_{k_{ij}}^2}}{{\sf \bar{d}_{k_{ij}}^2} \sigma_{h_k}^2 p_k}}, \qquad {\sf d_{k_{ij}}^2} \geq 0
\end{align}

The asymptotic characterization of the constrained capacity, which can be found by using Theorem \ref{mutual_information_expansion} together with the properties of \eqref{pdf_dist_siso_rayleigh}, is given by:
\begin{align}
{\sf \bar{I}} \left({\sf snr}\right) = \log {\sf M} - \epsilon'_1 \left({\sf snr}\right) \cdot \frac{1}{{\sf snr}} + \mathcal{O} \left(\frac{1}{{\sf snr}^2}\right) \label{constrained_capacity_bank_rayleigh1}
\end{align}
where ${\sf M} = \prod_{k=1}^{K} {\sf M_k}$ and
\begin{align}
\sum_{k=1}^{K} k'_{{\sf UB}_{1_k}} \sum_{i=1}^{{\sf M_k}} \sum_{\substack{j=1 \\ j \neq i}}^{{\sf M_k}} \frac{1}{{\sf \bar{d}_{k_{ij}}^2} \sigma_{h_k}^2 p_k} \leq \epsilon'_1 \left({\sf snr}\right) \leq \sum_{k=1}^{K} k'_{{\sf LB}_{1_k}} \sum_{i=1}^{{\sf M_k}} \sum_{\substack{j=1 \\ j \neq i}}^{{\sf M_k}} \frac{1}{{\sf \bar{d}_{k_{ij}}^2} \sigma_{h_k}^2 p_k} \label{constrained_capacity_bank_rayleigh2}
\end{align}
and
\begin{align}
k'_{{\sf UB}_{1_k}} = \frac{1}{2 {\sf M_k} \left({\sf M_k}-1\right)} \cdot \frac{1}{\sqrt{\pi}} \cdot 4^n \cdot \frac{\Gamma \left(5/2\right)}{\Gamma \left(3\right)} \label{constant_ub_bank}
\end{align}
\begin{align}
k'_{{\sf LB}_{1_k}} = \frac{2}{{\sf M_k}} \cdot \frac{1}{\sqrt{\pi}} \cdot 4^n \cdot \frac{\Gamma \left(5/2\right)}{\Gamma \left(3\right)} \label{constant_lb_bank}
\end{align}

As ${\sf snr} \to \infty$, it follows from \eqref{constrained_capacity_bank_rayleigh1} and the upper bound to $\epsilon'_1 \left({\sf snr}\right)$ in \eqref{constrained_capacity_bank_rayleigh2} that the power allocation procedure that maximizes the lower bound to the constrained capacity is given by:
\begin{equation}
\label{optimal_power_bank_rayleigh}
p_k^* = \alpha \cdot \frac{1}{\sqrt{\frac{1}{{\sf M_k}} \sum_{i=1}^{{\sf M_k}} \sum_{\substack{j=1 \\ j \neq i}}^{{\sf M_k}} {\sf \bar{d}_{k_{ij}}^2}  \sigma_{h_k}^2}} + o \left(1\right)
\end{equation}
with
\begin{equation}
\label{optimal_power_bank_rayleigh1}
\alpha^{-1} =\frac{1}{{\sf P}} \cdot \sum_{k=1}^{K} \frac{1}{\sqrt{\frac{1}{{\sf M_k}} \sum_{i=1}^{{\sf M_k}} \sum_{\substack{j=1 \\ j \neq i}}^{{\sf M_k}} {\sf \bar{d}_{k_{ij}}^2}  \sigma_{h_k}^2}}
\end{equation}

Note that, given equal sub-channel constellations, the higher the average sub-channel strength (i.e., the higher $\sigma_{h_k}^2$), then the lower the allocated power. Note also that the power allocation policy embodied in \eqref{optimal_power_bank_rayleigh} and \eqref{optimal_power_bank_rayleigh1} in fact corresponds to the power allocation policy put forth in~\cite{Lozano08}.
%
%
%
%
%
%

\begin{figure}
\centering
\epsfig{file=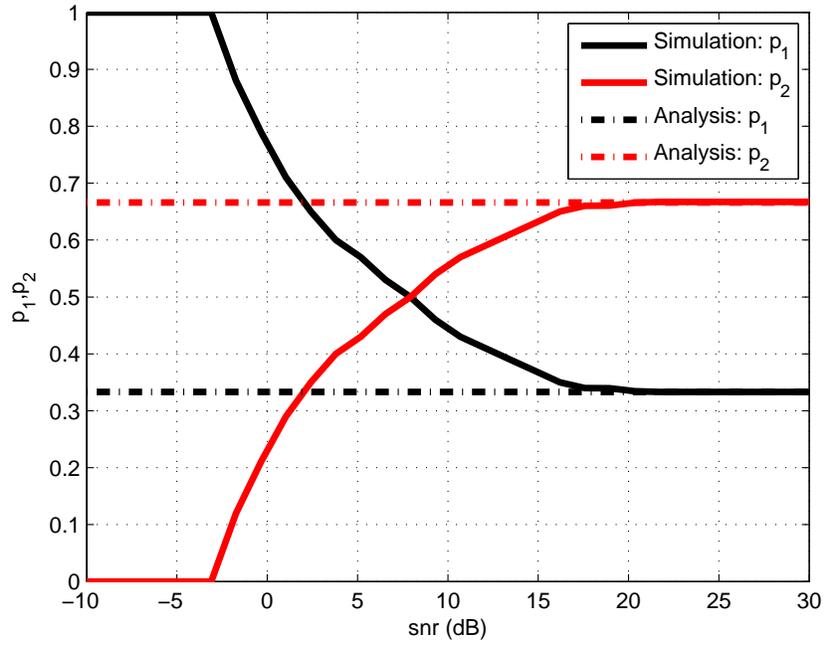,width=0.75\linewidth,clip=} \\
\caption{Optimal power allocation in a bank of two parallel independent Rayleigh fading coherent channels driven by 16-QAM inputs ($\sigma_{h_1}^2=4$ and $\sigma_{h_2}^2=1$).} \label{opt_power_alloc_parallel_rayleigh}
\end{figure}

\begin{figure}
\centering
\epsfig{file=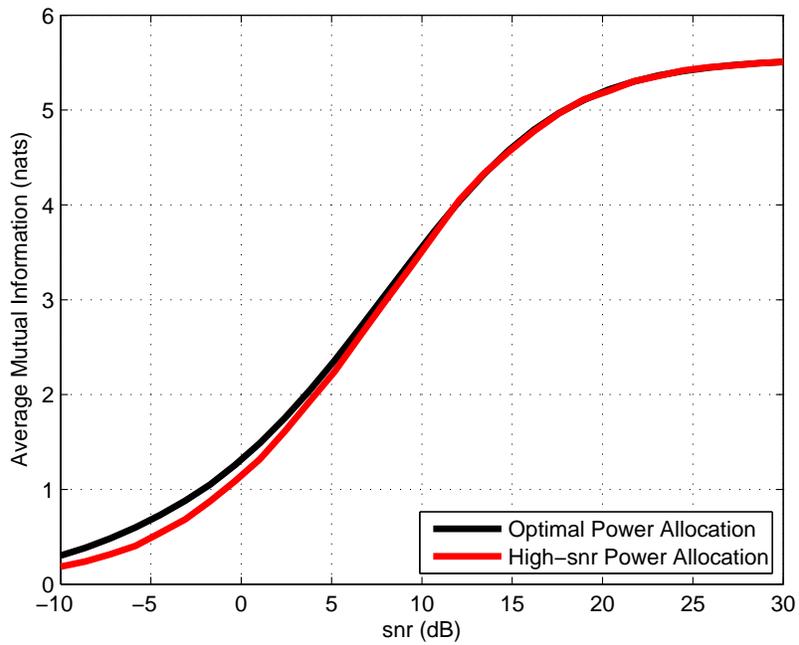,width=0.75\linewidth,clip=} \\
\caption{Average mutual information in a bank of two parallel independent Rayleigh fading coherent channels driven by 16-QAM inputs ($\sigma_{h_1}^2=4$ and $\sigma_{h_2}^2=1$).} \label{avg_mutual_information_parallel_rayleigh}
\end{figure}

Figures \ref{opt_power_alloc_parallel_rayleigh} and \ref{avg_mutual_information_parallel_rayleigh} compare analysis to simulation for a bank of two parallel independent Rayleigh fading coherent channels driven by 16-QAM inputs. We observe that the optimal power allocation policy, obtained via direct optimization of the Monte-Carlo simulated constrained capacity, tends with the increase of the signal-to-noise ratio to the high-${\sf snr}$ power allocation policy embodied in \eqref{optimal_power_bank_rayleigh} and \eqref{optimal_power_bank_rayleigh1}. We also observe that the constrained capacity associated with the optimal power allocation policy tends with the increase of the signal-to-noise ratio to the constrained capacity associated with the high-${\sf snr}$ power allocation policy. It is important to note though that the designs, which are shown to be very good in the regime of high ${\sf snr}$ in Figures \ref{opt_power_alloc_parallel_rayleigh} and \ref{avg_mutual_information_parallel_rayleigh}, are not sharp for parallel independent channels driven by distinct inputs. This is due to the fact that one can only bound rather than compute the exact value of the quantity akin to the MMSE dimension, which differs for different inputs.

\begin{figure}
\centering
\epsfig{file=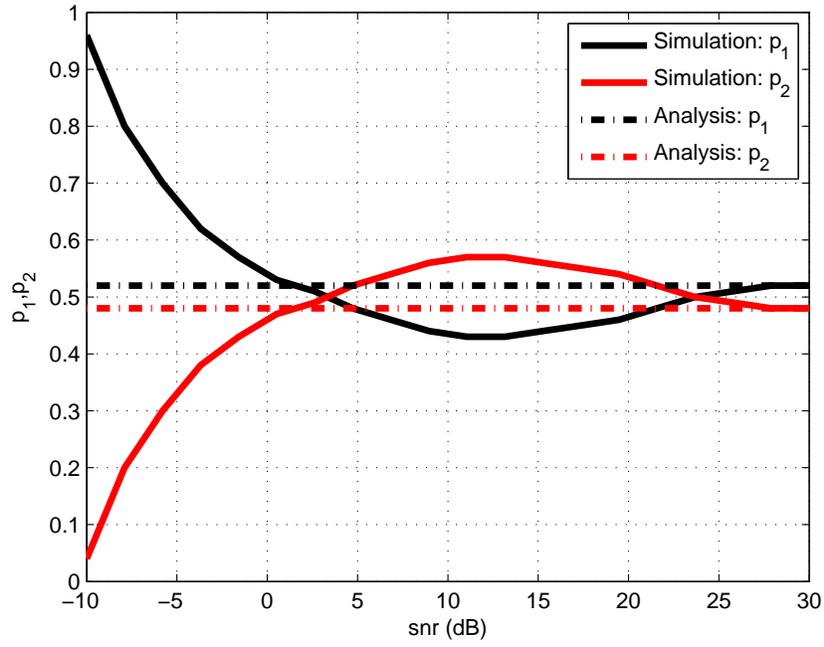,width=0.75\linewidth,clip=} \\
\caption{Optimal power allocation in a bank of two parallel independent Ricean fading coherent channels driven by 16-QAM inputs ($\mu_{h_1}=1+j$, $\mu_{h_2}=1+j$, $\sigma_{h_1}^2=4$ and $\sigma_{h_2}^2=1$).} \label{opt_power_alloc_parallel_ricean}
\end{figure}

\begin{figure}
\centering
\epsfig{file=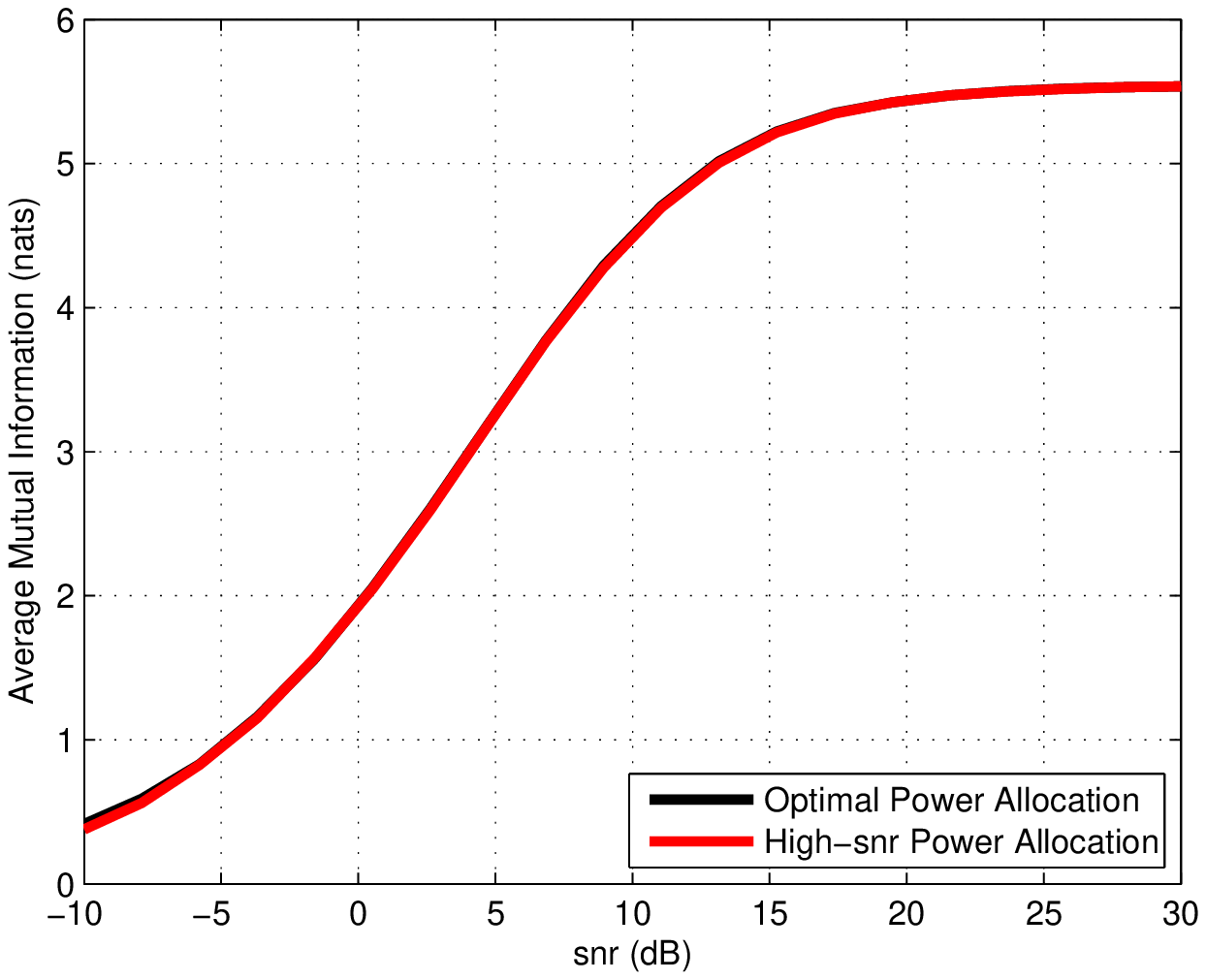,width=0.75\linewidth,clip=} \\
\caption{Average mutual information in a bank of two parallel independent Ricean fading coherent channels driven by 16-QAM inputs ($\mu_{h_1}=1+j$, $\mu_{h_2}=1+j$, $\sigma_{h_1}^2=4$ and $\sigma_{h_2}^2=1$).} \label{avg_mutual_information_parallel_ricean}
\end{figure}


\subsubsection{Ricean fading case} In the Ricean fading case, the complex fading coefficients $h_k \sim \mathcal{CN} \big(\mu_{h_k},\sigma_{h_k}^2\big), k=1,\ldots,K$. 
Therefore, the probability density function of the squared pairwise Euclidean distance between two distinct noiseless receive symbols in a particular sub-channel $k$ is given by:
\begin{equation}
\label{pdf_dist_siso_ricean}
p_{{\sf d_{k_{ij}}^2}} \big({\sf d_{k_{ij}}^2}\big) = \frac{1}{{\sf \bar{d}_{k_{ij}}^2} \sigma_{h_k}^2 p_k} \cdot e^{- \frac{s_k^2 + {\sf d_{k_{ij}}^2}}{{\sf \bar{d}_{k_{ij}}^2} \sigma_{h_k}^2 p_k}} \cdot I_0 \left(\frac{2 \sqrt{s_k^2 {\sf d_{k_{ij}}^2}}}{{\sf \bar{d}_{k_{ij}}^2} \sigma_{h_k}^2 p_k}\right), \qquad {\sf d_{k_{ij}}^2} \geq 0
\end{equation}
where $s_k^2 = {\sf \bar{d}_{k_{ij}}}^2 \cdot p_k \cdot \left(\Re\{\mu_{h_k}\}^2 + \Im\{\mu_{h_k}\}^2\right)$ and
\begin{equation}
I_0 (x) = \sum_{k=0}^{\infty} \frac{1}{k! \Gamma(k + 1)} \cdot \left(\frac{x}{2}\right)^{2k}, \qquad x \geq 0
\end{equation}
The asymptotic characterization of the constrained capacity, which can also be found by using Theorem \ref{mutual_information_expansion} together with the properties of \eqref{pdf_dist_siso_ricean}, is given by:
\begin{align}
{\sf \bar{I}} \left({\sf snr}\right) = \log {\sf M} - \epsilon'_1 \left({\sf snr}\right) \cdot \frac{1}{{\sf snr}} + \mathcal{O} \left(\frac{1}{{\sf snr}^2}\right) \label{constrained_capacity_bank_ricean1}
\end{align}
where ${\sf M} = \prod_{k=1}^{K} {\sf M_k}$ and
\begin{align}
&\sum_{k=1}^{K} k'_{{\sf UB}_{1_k}} \sum_{i=1}^{{\sf M_k}} \sum_{\substack{j=1 \\ j \neq i}}^{{\sf M_k}} \frac{1}{{\sf \bar{d}_{k_{ij}}^2} \sigma_{h_k}^2 p_k} \cdot e^{- \frac{\Re\left\{\mu_{h_k}\right\}^2 + \Im\left\{\mu_{h_k}\right\}^2}{\sigma_{h_k}^2}} \nonumber \\
&\leq \epsilon'_1 \left({\sf snr}\right) \leq \nonumber \\
&\sum_{k=1}^{K} k'_{{\sf LB}_{1_k}} \sum_{i=1}^{{\sf M_k}} \sum_{\substack{j=1 \\ j \neq i}}^{{\sf M_k}} \frac{1}{{\sf \bar{d}_{k_{ij}}^2} \sigma_{h_k}^2 p_k} \cdot e^{- \frac{\Re\left\{\mu_{h_k}\right\}^2 + \Im\left\{\mu_{h_k}\right\}^2}{\sigma_{h_k}^2}} \label{constrained_capacity_bank_ricean2}
\end{align}
and
\begin{align}
k'_{{\sf UB}_{1_k}} = \frac{1}{2 {\sf M_k} \left({\sf M_k}-1\right)} \cdot \frac{1}{\sqrt{\pi}} \cdot 4^n \cdot \frac{\Gamma \left(5/2\right)}{\Gamma \left(3\right)}
\end{align}
\begin{align}
k'_{{\sf LB}_{1_k}} = \frac{2}{{\sf M_k}} \cdot \frac{1}{\sqrt{\pi}} \cdot 4^n \cdot \frac{\Gamma \left(5/2\right)}{\Gamma \left(3\right)}
\end{align}

As ${\sf snr} \to \infty$, it follows from \eqref{constrained_capacity_bank_ricean1} and the upper bound to $\epsilon'_1 \left({\sf snr}\right)$ in \eqref{constrained_capacity_bank_ricean2} that the power allocation procedure that maximizes the lower bound to the constrained capacity is given by:
\begin{equation}
\label{optimal_power_bank_ricean}
p_k^* = \alpha \cdot \frac{e^{- \frac{\Re\left\{\mu_{h_k}\right\}^2 + \Im\left\{\mu_{h_k}\right\}^2}{2\sigma_{h_k}^2}}}{\sqrt{\frac{1}{{\sf M_k}} \sum_{i=1}^{{\sf M_k}} \sum_{\substack{j=1 \\ j \neq i}}^{{\sf M_k}} {\sf \bar{d}_{k_{ij}}^2} \sigma_{h_k}^2}} + o \left(1\right)
\end{equation}
with
\begin{equation}
\label{optimal_power_bank_ricean1}
\alpha^{-1} =\frac{1}{{\sf P}} \cdot \sum_{k=1}^{K} \frac{e^{- \frac{\Re\left\{\mu_{h_k}\right\}^2 + \Im\left\{\mu_{h_k}\right\}^2}{2\sigma_{h_k}^2}}}{\sqrt{\frac{1}{{\sf M_k}} \sum_{i=1}^{{\sf M_k}} \sum_{\substack{j=1 \\ j \neq i}}^{{\sf M_k}} {\sf \bar{d}_{k_{ij}}^2} \sigma_{h_k}^2}}
\end{equation}
Note now that the Ricean K-factor $\left(\Re\left\{\mu_{h_k}\right\}^2 + \Im\left\{\mu_{h_k}\right\}^2\right)\big/\sigma_{h_k}^2$ has a direct impact on the power allocation procedure (compare \eqref{optimal_power_bank_rayleigh} and \eqref{optimal_power_bank_rayleigh1} to \eqref{optimal_power_bank_ricean} and \eqref{optimal_power_bank_ricean1}).

Figures \ref{opt_power_alloc_parallel_ricean} and \ref{avg_mutual_information_parallel_ricean}, which compare analysis to simulation for a bank of two parallel independent Ricean fading coherent channels driven by 16-QAM inputs, also demonstrate that the optimal designs tend with the increase of the signal-to-noise ratio to the high-${\sf snr}$ designs. \footnote{Note that a more substantial difference between the constrained capacity associated with the optimal power allocation and the constrained capacity associated with the high-${\sf snr}$ power allocation is expected for a bank of parallel independent fading coherent channels with more than two sub-channels. The simulations, however, would be very time consuming.} It is also important to note that the designs, which are shown to be very good in the regime of high ${\sf snr}$ in Figures \ref{opt_power_alloc_parallel_ricean} and \ref{avg_mutual_information_parallel_ricean}, are also not sharp for parallel independent channels driven by distinct inputs, as discussed previously.

\vspace{-0.4cm}

\subsection{Precoding in a Multiple-Antenna Fading Channel}

We consider a linearly-precoded multiple-antenna fading channel which can be modeled as follows:
\begin{equation}
{\bf y} = \sqrt{{\sf snr}} \cdot {\bf H} {\bf P} {\bf x} + {\bf n}
\end{equation}
for a single channel use, where $\y \in \mathbb{C}^{n_r}$ is the vector of complex receive symbols, $\x \in \mathbb{C}^{n_t}$ is the vector of complex transmit symbols, ${\bf H} \in \mathbb{C}^{n_r \times n_t}$ is the random channel fading matrix (with $\mathbb{E} \left\{ {\sf tr} \left({\bf H} {\bf H}^\dag\right) \right\} = n_t n_r$), and $\n \in \mathbb{C}^{n_r}$ is a vector of independent circularly symmetric complex Gaussian noise random variables with zero-mean and unit-variance. The matrix ${\bf P} \in \mathbb{C}^{n_t \times n_t}$ represents a linear precoder.\footnote{We consider without any loss of generality the matrix ${\bf P}$ to be square.} The scaling factor ${\sf snr}$ relates to the signal-to-noise ratio. We assume that the transmit vector conforms to an equiprobable multi-dimensional constellation with cardinality ${\sf M}$, i.e., ${\bf x} \in \left\{{\bf x}_1,{\bf x}_2,\ldots,{\bf x}_{\sf M}\right\}$ and $\Pr \left({\bf x}_1\right) = \Pr \left({\bf x}_2\right) = \cdots = \Pr \left({\bf x}_{\sf M}\right) = \frac{1}{{\sf M}}$, with ${\bf \Sigma_x} = \mathbb{E} \left\{{\bf x} {\bf x}^\dag\right\} = \frac{1}{n_t} \cdot {\bf I}$. We also assume that all the random variables are independent.

The objective is to determine the precoder that maximizes a lower bound to the constrained capacity in the asymptotic regime of high ${\sf snr}$, subject to the total power constraint:
\begin{align}
{\sf tr} \left({\bf P}{\bf P}^\dag\right) \leq {\sf P}
\end{align}

We illustrate the design procedure for the canonical i.i.d. Rayleigh fading coherent channel and the antenna-correlated Rayleigh fading coherent channel. In particular, the availability of closed-form expressions for the bounds to the asymptotic expansions of the constrained capacity, which embody the effect of the precoder, leads to simple design methods based on numerical procedures -- and occasionally analytic results -- rather than time-consuming Monte Carlo simulation procedures.




\subsubsection{The Canonical i.i.d. Rayleigh Fading Channel} In this channel model, where ${\bf H} = {\bf H_w}$,  the squared pairwise Euclidean distance between two arbitrary noiseless receive vectors is given by:
\begin{align}
{\sf d_{ij}^2} = {\sf tr} \left({\bf H_w} {\bf P} \left({\bf x}_i-{\bf x}_j\right)\left({\bf x}_i-{\bf x}_j\right)^\dag {\bf P}^\dag {\bf H_w^\dag}\right) \label{squared_distance_precoding_canonical}
\end{align}
and its probability density function is given by:
\begin{align}
p_{{\sf d_{ij}^2}} \big({\sf d_{ij}^2}\big) = \frac{1}{\left(n_r - 1\right)!} \cdot \frac{1}{\left({\sf tr} \left(\left({\bf x}_i - {\bf x}_j\right)^\dag {\bf P}^\dag {\bf P} \left({\bf x}_i - {\bf x}_j\right)\right)\right)^{n_r}} \cdot \big({\sf d_{ij}^2}\big)^{n_r-1} \cdot \nonumber \\
\cdot e^{-\frac{{\sf d_{ij}^2}}{{\sf tr} \left(\left({\bf x}_i - {\bf x}_j\right)^\dag {\bf P}^\dag {\bf P} \left({\bf x}_i - {\bf x}_j\right)\right)}}, \qquad {\sf d_{ij}^2} \geq 0 \label{pdf_precoding_canonical}
\end{align}
This, together with Theorem \ref{mutual_information_expansion}, leads to the high-${\sf snr}$ characterization of the constrained capacity, which is a function of the precoder, given by:
\begin{align}
{\sf \bar{I}} ({\sf snr}) = \log {\sf M} - \epsilon'_{n_r} \left({\sf snr}\right) \cdot \frac{1}{{\sf snr}^{n_r}} + \mathcal{O} \left(\frac{1}{{\sf snr}^{n_r+1}}\right)
\end{align}
where
\begin{align}
& k'_{UB_{n_r}} \cdot \sum_{i=1}^{{\sf M}} \sum_{\substack{j=1 \\ j \neq i}}^{{\sf M}} \left(\frac{1}{{\sf tr} \left(\left({\bf x}_i - {\bf x}_j\right)^\dag {\bf P}^\dag {\bf P} \left({\bf x}_i - {\bf x}_j\right)\right)}\right)^{n_r} \nonumber \\
& \leq \epsilon'_{n_r} \left({\sf snr}\right) \leq \nonumber \\
& k'_{LB_{n_r}} \cdot \sum_{i=1}^{{\sf M}} \sum_{\substack{j=1 \\ j \neq i}}^{{\sf M}} \left(\frac{1}{{\sf tr} \left(\left({\bf x}_i - {\bf x}_j\right)^\dag {\bf P}^\dag {\bf P} \left({\bf x}_i - {\bf x}_j\right)\right)}\right)^{n_r}
\end{align}
and
\begin{align}
k'_{{\sf UB}_{n_r}} = \frac{1}{2 {\sf M} \left({\sf M}-1\right)} \cdot \frac{1}{\sqrt{\pi}} \cdot 4^n \cdot \frac{\Gamma \left(n_r+3/2\right)}{\Gamma \left(n_r+2\right)} \label{constant_ub_bank}
\end{align}
\begin{align}
k'_{{\sf LB}_{n_r}} = \frac{2}{{\sf M}} \cdot \frac{1}{\sqrt{\pi}} \cdot 4^n \cdot \frac{\Gamma \left(n_r+3/2\right)}{\Gamma \left(n_r+2\right)} \label{constant_lb_bank}
\end{align}

Therefore, we pose the optimization problem, which is equivalent to the maximization of the lower bound (or the upper bound) to the constrained capacity of the canonical i.i.d. Rayleigh fading coherent channel in the asymptotic regime of high ${\sf snr}$ subject to a total power constraint, given by:
\begin{align}
\label{optimization_problem_precoder1}
\min_{{\bf P}} \sum_{i=1}^{{\sf M}} \sum_{\substack{j=1 \\ j \neq i}}^{{\sf M}} \left(\frac{1}{{\sf tr} \left(\left({\bf x}_i - {\bf x}_j\right)^\dag {\bf P}^\dag {\bf P} \left({\bf x}_i - {\bf x}_j\right)\right)}\right)^{n_r}
\end{align}
with:
\begin{align}
\label{optimization_problem_precoder2}
{\sf tr} \left({\bf P}{\bf P}^\dag\right) \leq {\sf P}
\end{align}

This optimization problem leads immediately to a simple precoder design procedure, that bypasses the need for time-consuming Monte Carlo simulations, based on numerical or analytical techniques. We illustrate the analysis by assuming that the individual precoder inputs conform to the same equiprobable symmetric constellation, e.g., some PSK or some QAM constellation. \footnote{Note that this analysis only seems to be applicable to scenarios where the individual precoder inputs conform to the same equiprobable symmetric constellation, rather than scenarios where the individual precoder inputs conform to distinct constellations.}

By introducing the change of variables ${\bf Z} = {\bf P}^\dag {\bf P}$, we pose an optimization problem equivalent to the optimization problem in \eqref{optimization_problem_precoder1} and \eqref{optimization_problem_precoder2} as follows:
\begin{align}
\min_{{\bf Z}}~~~\sum_{\substack{i,j=1 \\ i \neq j}}^{{\sf M}} \left(\frac{1}{{\sf tr} \left(\left({\bf x}_i - {\bf x}_j\right)^\dag {\bf Z} \left({\bf x}_i - {\bf x}_j\right)\right)}\right)^{n_r}
\end{align}
subject to:
\begin{align}
{\sf tr} \left({\bf Z}\right) \leq {\sf P}
\end{align}
\begin{align}
{\bf Z} \succeq {\bf 0}
\end{align}

Note that this represents a convex optimization problem because: \emph{i}) the objective function is convex in ${\bf Z} \succeq {\bf 0}$ (${\bf Z} \neq {\bf 0}$); \emph{ii}) the constraint set is convex. Define the Lagrangian of the optimization problem as follows:
\begin{align}
\mathcal{L} \left({\bf Z},{\bf \Psi},\lambda\right) = \sum_{\substack{i,j=1 \\ i \neq j}}^{{\sf M}} \left(\frac{1}{{\sf tr} \left(\left({\bf x}_i - {\bf x}_j\right) {\bf Z} \left({\bf x}_i - {\bf x}_j\right)^\dag\right)}\right)^{n_r} - \lambda \cdot \left({\sf P} - {\sf tr} \left({\bf Z}\right)\right) - {\sf tr} \left({\bf \Psi} {\bf Z}\right)
\end{align}
The Karush-Kuhn-Tucker conditions, which are both necessary and sufficient, state that the optimal solution ${\bf Z}^*$ is such that:
\begin{align}
&\nabla_{\bf Z} \mathcal{L} \left({\bf Z},{\bf \Psi},\lambda\right) |_{{\bf Z} = {\bf Z}^*} = \nonumber \\
&= - n_r \cdot \sum_{\substack{i,j=1 \\ i \neq j}}^{{\sf M}} \left(\frac{1}{{\sf tr} \left(\left({\bf x}_i - {\bf x}_j\right) {\bf Z}^* \left({\bf x}_i - {\bf x}_j\right)^\dag\right)}\right)^{n_r+1} \!\!\!\!\! \cdot \left({\bf x}_i - {\bf x}_j\right) \left({\bf x}_i - {\bf x}_j\right)^\dag + \lambda \cdot {\bf I} - {\bf \Psi} = 0
\end{align}
\begin{align}
{\sf tr} \left({\bf \Psi} {\bf Z}^*\right) = 0, \qquad {\bf \Psi} \succeq {\bf 0}, \qquad {\bf Z}^* \succeq {\bf 0}
\end{align}
\begin{align}
\lambda \cdot \left({\sf P} - {\sf tr}\left({\bf Z}^*\right)\right) = 0, \qquad \lambda \geq 0
\end{align}

It is possible to show that ${\bf Z}^* = \frac{{\sf P}}{n_t} \cdot {\bf I}$ without any loss of generality, because ${\bf Z}^* = \frac{{\sf P}}{n_t} \cdot {\bf I}$ together with
\begin{align}
\lambda = n_r \cdot \sum_{\substack{i,j=1 \\ j \neq i}}^{{\sf M}} \left(\frac{n_t/{\sf P}}{{\sf tr} \left(\left({\bf x}_i - {\bf x}_j\right) \left({\bf x}_i - {\bf x}_j\right)^\dag\right)}\right)^{n_r+1} \!\!\!\!\! \cdot \left|{\bf x}_i (k) - {\bf x}_j (k)\right|^2
\end{align}
which does not depend on $k$ due to the symmetry conditions, and ${\bf \Psi} = {\bf 0}$ satisfy the Karush-Kuhn-Tucker conditions. Note that ${\bf x} (k)$ represents the kth element of the vector ${\bf x}$. It is only necessary to prove that:
\begin{align}
n_r \cdot \sum_{\substack{i,j=1 \\ j \neq i}}^{{\sf M}} \left(\frac{n_t/{\sf P}}{{\sf tr} \left(\left({\bf x}_i - {\bf x}_j\right) \left({\bf x}_i - {\bf x}_j\right)^\dag\right)}\right)^{n_r+1} \cdot \left({\bf x}_i - {\bf x}_j\right) \left({\bf x}_i - {\bf x}_j\right)^\dag = \lambda \cdot {\bf I} \label{precoder_power_allocation_proof}
\end{align}

Note that the element in the $m$-th row and $n$-th column of the matrix on the left hand side of \eqref{precoder_power_allocation_proof} is given by:
\begin{align}
\sum_{\substack{i,j=1 \\ j \neq i}}^{{\sf M}} \left(\frac{n_t/{\sf P}}{{\sf tr} \left(\left({\bf x}_i - {\bf x}_j\right) \left({\bf x}_i - {\bf x}_j\right)^\dag\right)}\right)^{n_r+1} \cdot \left({\bf x}_i (m) - {\bf x}_j (m)\right) \left({\bf x}_i (n) - {\bf x}_j (n)\right)^\dag
\end{align}

In view of the symmetry conditions, for fixed $m$ and $n$ it is possible to divide the set of pairs of indices $\left\{(i,j): i,j = 1,\ldots,{\sf M}, i \neq j\right\}$, which contains ${\sf M} \cdot \left({\sf M} - 1\right)$ indices pairs, into $\frac{1}{2} \cdot {\sf M} \cdot \left({\sf M}-1\right)$ sets of pairs of indices $\left\{(i_1,j_1),(i_2,j_2)\right\}$,  which contain only two indices pairs, with the property that ${\bf x}_{i_1} (m) = - {\bf x}_{i_2} (m)$, ${\bf x}_{j_1} (m) = - {\bf x}_{j_2} (m)$, ${\bf x}_{i_1} (k) = {\bf x}_{i_2} (k)$, $\forall~k \neq m$, and ${\bf x}_{j_1} (k) = {\bf x}_{j_2} (k)$, $\forall~k \neq m$. 
Then, it follows immediately that:
\begin{align}
\left(\frac{n_t/{\sf P}}{{\sf tr} \left(\left({\bf x}_{i_1} - {\bf x}_{j_1}\right) \left({\bf x}_{i_1} - {\bf x}_{j_1}\right)^\dag\right)}\right)^{n_r+1} \cdot \left({\bf x}_{i_1} (m) - {\bf x}_{j_1} (m)\right) \left({\bf x}_{i_1} (n) - {\bf x}_{j_1} (n)\right)^\dag &+ \nonumber \\
\left(\frac{n_t/{\sf P}}{{\sf tr} \left(\left({\bf x}_{i_2} - {\bf x}_{j_2}\right) \left({\bf x}_{i_2} - {\bf x}_{j_2}\right)^\dag\right)}\right)^{n_r+1} \cdot \left({\bf x}_{i_2} (m) - {\bf x}_{j_2} (m)\right) \left({\bf x}_{i_2} (n) - {\bf x}_{j_2} (n)\right)^\dag &= 0
\end{align}
and that
\begin{align}
\sum_{\substack{i,j=1 \\ j \neq i}}^{{\sf M}} \left(\frac{n_t/{\sf P}}{{\sf tr} \left(\left({\bf x}_i - {\bf x}_j\right) \left({\bf x}_i - {\bf x}_j\right)^\dag\right)}\right)^{n_r+1} \cdot \left({\bf x}_i (m) - {\bf x}_j (m)\right) \left({\bf x}_i (n) - {\bf x}_j (n)\right)^\dag = 0
\end{align}

Note also that the $n$-the diagonal element of the matrix on the left hand side of \eqref{precoder_power_allocation_proof} is given by:
\begin{align}
\sum_{i=1}^{{\sf M}} \sum_{\substack{j=1 \\ j \neq i}}^{{\sf M}} \left(\frac{n_t/{\sf P}}{{\sf tr} \left(\left({\bf x}_i - {\bf x}_j\right) \left({\bf x}_i - {\bf x}_j\right)^\dag\right)}\right)^{n_r+1} \cdot \left({\bf x}_i (n) - {\bf x}_j (n)\right) \left({\bf x}_i (n) - {\bf x}_j (n)\right)^\dag
\end{align}
In view of the symmetry conditions, it also follows immediately that this quantity is independent of $n$.

The fact that ${\bf Z}^* = \frac{{\sf P}}{n_t} \cdot {\bf I}$ without any loss of generality leads directly to the form of the precoder that maximizes the lower bound to the constrained capacity as follows:
\begin{align}
{\bf P}^* = \frac{{\sf P}}{n_t} \cdot {\bf Q} {\bf I}
\end{align}
where ${\bf Q}$ is any unitary matrix. This result is very intuitive due to the symmetry of the system model.

\begin{figure}
\centering
\epsfig{file=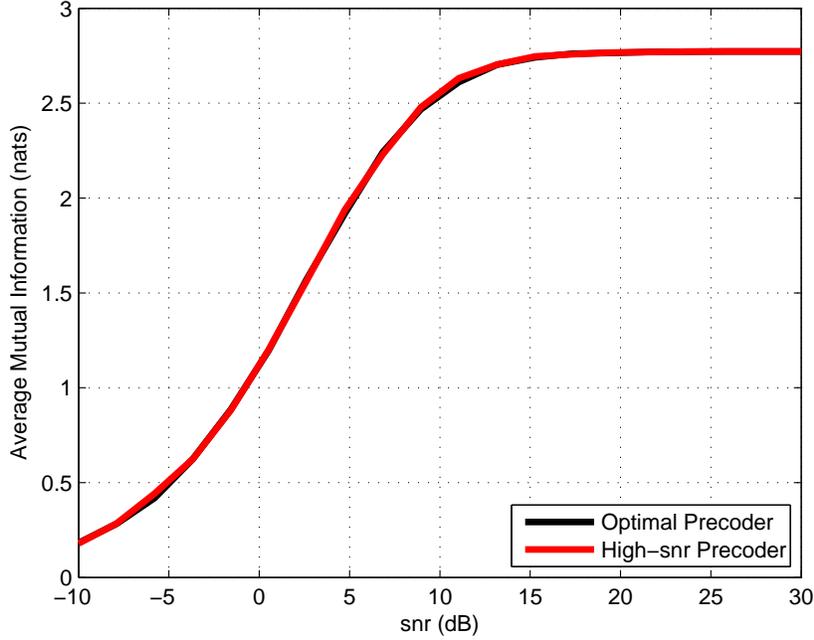,width=0.75\linewidth,clip=} \\
\caption{Average mutual information for a $2 \times 2$ canonical i.i.d. Rayleigh fading coherent channel driven by QPSK inputs.} \label{avg_mutual_information_precoder_canonical_rayleigh}
\end{figure}

Figure \ref{avg_mutual_information_precoder_canonical_rayleigh} shows that the constrained capacity achieved by the optimal precoder, obtained via direct optimization of the Monte-Carlo simulated constrained capacity, appears to be equal to the constrained capacity achieved by the high-${\sf snr}$ isotropic precoder across the signal-to-noise ratio range. This suggests that the approach is sharp in the sense that it captures well the effect of the precoder on the constrained capacity of the canonical i.i.d. Rayleigh fading coherent channel.

\subsubsection{The Antenna-Correlated Rayleigh Fading Channel} In this channel model, where ${\bf H} = {\bf \Theta_R^{\frac{1}{2}}} {\bf H_w} {\bf \Theta_T^{\frac{1}{2}}}$,  the squared pairwise Euclidean distance between two arbitrary noiseless receive vectors is given by:
\begin{align}
{\sf d_{ij}^2} = {\sf tr} \left({\bf \Theta_R^{\frac{1}{2}}} {\bf H_w} {\bf \Theta_T^{\frac{1}{2}}} {\bf P} \left({\bf x}_i-{\bf x}_j\right)\left({\bf x}_i-{\bf x}_j\right)^\dag {\bf P}^\dag {\bf \Theta_T^{\frac{\dag}{2}}} {\bf H_w^\dag} {\bf \Theta_R^{\frac{\dag}{2}}}\right) \label{squared_distance_precoding_correlated}
\end{align}
and its probability density function is given by:
\begin{align}
p_{{\sf d_{ij}^2}} \big({\sf d_{ij}^2}\big) = \frac{1}{\left(n_r - 1\right)!} \cdot \frac{1}{{\left({\sf tr} \left(\left({\bf x}_i - {\bf x}_j\right)^\dag {\bf P}^\dag {\bf \Theta_T} {\bf P} \left({\bf x}_i - {\bf x}_j\right)\right)\right)^{n_r}}} \cdot \big({\sf d_{ij}^2}\big)^{n_r-1} \cdot \nonumber \\
\cdot e^{-\frac{{\sf d_{ij}^2}}{{\sf tr} \left(\left({\bf x}_i - {\bf x}_j\right)^\dag {\bf P}^\dag {\bf \Theta_T} {\bf P} \left({\bf x}_i - {\bf x}_j\right)\right)}}, \qquad {\sf d_{ij}^2} \geq 0 \label{pdf_precoding_correlated1}
\end{align}
when the eigenvalues of the receive correlation matrix, ${\lambda_{R_1}},{\lambda_{R_2}},\ldots,{\lambda_{R_{n_r}}}$, are all equal (necessarily to one) or
\begin{align}
p_{{\sf d_{ij}^2}} \big({\sf d_{ij}^2}\big) = \sum_{k=1}^{n_r} \frac{1}{{\lambda_{R_k}} \prod_{\substack{k' = 1 \\ k' \neq k}}^{n_r} \left(1 - \frac{\lambda_{R_{k'}}}{\lambda_{R_{k}}}\right) \cdot {\sf tr} \left(\left({\bf x}_i - {\bf x}_j\right)^\dag {\bf P}^\dag {\bf \Theta_T} {\bf P} \left({\bf x}_i - {\bf x}_j\right)\right)} \cdot \nonumber \\
\cdot e^{-\frac{{\sf d_{ij}^2}}{{\lambda_{R_k}} \cdot {\sf tr} \left(\left({\bf x}_i - {\bf x}_j\right)^\dag {\bf P}^\dag {\bf \Theta_T} {\bf P} \left({\bf x}_i - {\bf x}_j\right)\right)}}, \qquad {\sf d_{ij}^2} \geq 0 \label{pdf_precoding_correlated2}
\end{align}
when the eigenvalues of the receive correlation matrix, ${\lambda_{R_1}},{\lambda_{R_2}},\ldots,{\lambda_{R_{n_r}}}$, are all distinct\footnote{We once again concentrate on non-degenerate scenarios where ${\lambda_{T_{ij}}} > 0, \forall~i \neq j$, and ${\lambda_{R_k}} > 0, \forall~k$, and on scenarios where the eigenvalues ${\lambda_{R_k}}, k = 1,\ldots,n_r,$ are either all distinct or all equal (necessarily to one).}.

This, together with Theorem \ref{mutual_information_expansion}, also leads to the high-${\sf snr}$ characterization of the constrained capacity, which is a function of the precoder, given by:
\begin{align}
{\sf \bar{I}} ({\sf snr}) = \log {\sf M} - \epsilon'_{n_r} \left({\sf snr}\right) \cdot \frac{1}{{\sf snr}^{n_r}} + \mathcal{O} \left(\frac{1}{{\sf snr}^{n_r+1}}\right)
\end{align}
where
\begin{align}
& k'_{UB_{n_r}} \cdot \sum_{i=1}^{{\sf M}} \sum_{\substack{j=1 \\ i \neq j}}^{{\sf M}} \left(\frac{1}{{\sf tr} \left(\left({\bf x}_i - {\bf x}_j\right)^\dag {\bf P}^\dag {\bf \Theta_T} {\bf P} \left({\bf x}_i - {\bf x}_j\right)\right)}\right)^{n_r} \cdot \frac{1}{{\sf det} \left({\bf \Theta_R}\right)} \nonumber \\
& \leq \epsilon'_{n_r} \left({\sf snr}\right) \leq \nonumber \\
& k'_{LB_{n_r}} \cdot \sum_{i=1}^{{\sf M}} \sum_{\substack{j=1 \\ i \neq j}}^{{\sf M}} \left(\frac{1}{{\sf tr} \left(\left({\bf x}_i - {\bf x}_j\right)^\dag {\bf P}^\dag {\bf \Theta_T} {\bf P} \left({\bf x}_i - {\bf x}_j\right)\right)}\right)^{n_r} \cdot \frac{1}{{\sf det} \left({\bf \Theta_R}\right)}
\end{align}
and
\begin{align}
k'_{{\sf UB}_{n_r}} = \frac{1}{2 {\sf M} \left({\sf M}-1\right)} \cdot \frac{1}{\sqrt{\pi}} \cdot 4^n \cdot \frac{\Gamma \left(n_r+3/2\right)}{\Gamma \left(n_r+2\right)} \label{constant_ub_bank}
\end{align}
\begin{align}
k'_{{\sf LB}_{n_r}} = \frac{2}{{\sf M}} \cdot \frac{1}{\sqrt{\pi}} \cdot 4^n \cdot \frac{\Gamma \left(n_r+3/2\right)}{\Gamma \left(n_r+2\right)} \label{constant_lb_bank}
\end{align}

Therefore, we pose the optimization problem, which is equivalent to the maximization of the lower bound (or the upper bound) to the constrained capacity of the antenna-correlated Rayleigh fading coherent channel in the asymptotic regime of high ${\sf snr}$ subject to a total power constraint, given by:
\begin{align}
\label{precoder_optimization_correlated1}
\min_{{\bf P}} \sum_{i=1}^{{\sf M}} \sum_{\substack{j=1 \\ i \neq j}}^{{\sf M}} \left(\frac{1}{{\sf tr} \left(\left({\bf x}_i - {\bf x}_j\right)^\dag {\bf P}^\dag {\bf \Theta_T} {\bf P} \left({\bf x}_i - {\bf x}_j\right)\right)}\right)^{n_r} \cdot \frac{1}{{\sf det} \left({\bf \Theta_R}\right)}
\end{align}
with:
\begin{align}
\label{precoder_optimization_correlated2}
{\sf tr} \left({\bf P}{\bf P}^\dag\right) \leq {\sf P}
\end{align}

This optimization problem, akin to the previous one, also leads immediately to a simple precoder design procedure based on numerical techniques. Most notably, it is possible to prove from \eqref{precoder_optimization_correlated1} and \eqref{precoder_optimization_correlated2}, using the techniques in \cite{Palomar06b}, that the matrix containing the left singular vectors of the precoder matrix corresponds to the matrix containing the right singular vectors of the transmit-antennas correlation matrix, i.e., the optimal precoder diagonalizes the transmit-antenna correlation matrix. It is also possible to prove from \eqref{precoder_optimization_correlated1} and \eqref{precoder_optimization_correlated2} that, upon setting the matrix containing the left singular vectors of the precoder matrix to be equal to the matrix containing the right singular vectors of the transmit-antenna correlation matrix, the optimization problem becomes concave in the precoder squared singular values. These facts have also be been recently established by different means in \cite{Zeng12}.

\begin{figure}
\centering
\epsfig{file=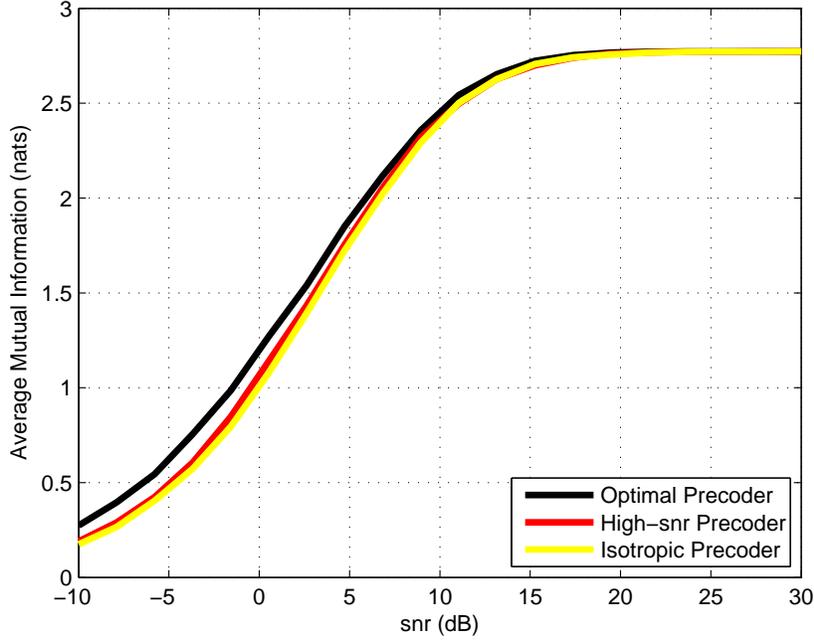,width=0.75\linewidth,clip=} \\
\caption[Text excluding the matrix]{Average mutual information for a $2 \times 2$ antenna-correlated Rayleigh fading coherent channel driven by QPSK inputs $\left({\bf \Phi_T} = \left[\begin{smallmatrix} 1 & 0.5 \\ 0.5 & 1 \end{smallmatrix}\right] \text{ and } {\bf \Phi_R} = \left[\begin{smallmatrix} 1 & 0.8 \\ 0.8 & 1 \end{smallmatrix}\right]\right)$.}
\label{avg_mutual_information_precoder_correlated_rayleigh}
\end{figure}

Figure \ref{avg_mutual_information_precoder_correlated_rayleigh} shows that the constrained capacity achieved by the optimal precoder design, obtained via direct optimization of the Monte-Carlo simulated constrained capacity, tends with the increase of the signal-to-noise ratio to the constrained capacity achieved by the high-${\sf snr}$ precoder design, obtained via the optimization problem in \eqref{precoder_optimization_correlated1} and \eqref{precoder_optimization_correlated2}. This suggests once again that the design approach is sharp in the sense that it also captures well the effect of the precoder on the constrained capacity of the antenna-correlated Rayleigh fading coherent channel.

\vspace{-0.4cm}

\subsection{Space-Time Coding in the Multiple-Antenna Canonical i.i.d. Rayleigh Fading Channel}

We now consider a more general multiple-antenna fading channel, which encompasses communication over $t$ symbol intervals, given by:
\begin{equation}
{\bf Y}=\sqrt{{\sf snr}} \cdot {\bf H_w} {\bf X} + {\bf N} \label{new_channel_model}
\end{equation}
where ${\bf Y} \in \mathbb{C}^{n_r \times t}$ denotes the receive codeword matrix, ${\bf X} \in \mathbb{C}^{n_t \times t}$ denotes the transmit codeword matrix, ${\bf H_w} \in \mathbb{C}^{n_r \times n_t}$ is the canonical i.i.d. complex Gaussian random channel fading matrix, and ${\bf N} \in \mathbb{C}^{n_r \times t}$ is the noise matrix consisting of independent circularly symmetric complex Gaussian random variables with zero-mean and unit-variance. We assume that the channel matrix is constant for the duration of a codeword, changing from codeword to codeword in a stationary and ergodic manner. We also assume that the receiver knows the exact channel matrix realization but the transmitter knows only the channel matrix distribution. The space-time codeword matrices conform to an equiprobable multi-dimensional constellation with cardinality ${\sf M}$, i.e., ${\bf X} \in \left\{{\bf X}_1,{\bf X}_2,\ldots,{\bf X}_{\sf M}\right\}$ and $\Pr \left({\bf X}_1\right) = \Pr \left({\bf X}_2\right) = \cdots = \Pr \left({\bf X}_{\sf M}\right) = \frac{1}{{\sf M}}$.

By capitalizing on Theorem \ref{mutual_information_expansion}, the high-${\sf snr}$ expansion of the constrained capacity in nats per channel use, where a channel use encompasses $t$ symbol intervals, obeys:
\begin{align}
{\sf \bar{I}} ({\sf snr}) = \log  {\sf M} - \epsilon'_{\sf d} \left({\sf snr}\right) \cdot \frac{1}{{\sf snr}^{\sf d}} + \mathcal{O} \left(\frac{1}{{\sf snr}^{{\sf d}+1}}\right) \label{constrained_capacity_space_time1}
\end{align}
where
\begin{align}
{\sf d} = 1 + \min \left\{n \in \mathbb{N}_0: \sum_{i=1}^{{\sf M}} \sum_{\substack{j=1 \\ j \neq i}}^{{\sf M}} p_{{\sf d_{ij}^2}}^{(n)} (0) \neq 0\right\} \label{constrained_capacity_space_time2}
\end{align}
and
\begin{align}
k'_{{\sf UB}_{\sf d}} \cdot \sum_{i=1}^{{\sf M}} \sum_{\substack{j=1 \\ j \neq i}}^{{\sf M}} p_{{\sf d_{ij}^2}}^{({\sf d}-1)} (0) \leq \epsilon'_{\sf d} \left({\sf snr}\right) \leq k'_{{\sf LB}_{\sf d}} \cdot \sum_{i=1}^{{\sf M}} \sum_{\substack{j=1 \\ j \neq i}}^{{\sf M}} p_{{\sf d_{ij}^2}}^{({\sf d}-1)} (0) \label{constrained_capacity_space_time3}
\end{align}
\begin{align}
k'_{{\sf LB}_{\sf d}} &= \frac{2}{{\sf M}} \cdot  \frac{4^{\sf d}}{\sqrt{\pi}} \cdot \frac{\Gamma \left({\sf d}+3/2\right)}{\Gamma \left({\sf d}+2\right)} \label{constrained_capacity_space_time4}
\end{align}
\begin{align}
k'_{{\sf UB}_{\sf d}} &= \frac{1}{2 {\sf M} \left({\sf M}-1\right)} \cdot  \frac{4^{\sf d}}{\sqrt{\pi}} \cdot \frac{\Gamma \left({\sf d}+3/2\right)}{\Gamma \left({\sf d}+2\right)} \label{constrained_capacity_space_time5}
\end{align}
and $p_{{\sf d_{ij}^2}} (\cdot)$ and $p_{{\sf d_{ij}^2}}^{(n)} (\cdot)$ denote the probability density function and the higher-order derivatives of the probability density function of the squared Euclidean distance between pairs of arbitrary noiseless receive codeword matrices.

It is evident that the exact value of ${\sf d}$ and the bounds to the value of $\epsilon'_{\sf d} \left({\sf snr}\right)$ are connected to the properties of the set of space-time codewords. The objective is to unveil the structure of the set of space-time codewords that lead to: \emph{i}) the maximization of the exact value of ${\sf d}$; and \emph{ii}) the minimization of the upper bound (as well as the lower bound) to the value of $\epsilon'_{\sf d} \left({\sf snr}\right)$. This leads to the maximization of the lower bound to the constrained capacity in the asymptotic regime of high-${\sf snr}$ regime.

The following Theorem summarizes the space-time code design criteria for the canonical i.i.d. multiple-antenna Rayleigh fading coherent channel, with two transmit antennas and one or two receive antennas. We denote ${\bf \Delta}_{ij} = ({\bf X}_i - {\bf X}_j) ({\bf X}_i - {\bf X}_j)^\dag$. We also denote the eigenvalue decomposition of the matrix ${\bf \Delta}_{ij}$ by ${\bf \Delta}_{ij} = {\bf W} {\bf \Lambda}_{ij} {\bf W}^\dag$, where ${\bf W}$ is a unitary matrix, ${\bf \Lambda}_{ij} = {\sf diag} \big(\lambda_{ij}\left(1\right),\ldots,\lambda_{ij}\left(r_{ij}\right),0,\ldots,0\big)$ is a diagonal matrix, $\lambda_{ij}\left(1\right) \geq \cdots \geq \lambda_{ij}\left(r_{ij}\right) > 0$ are the non-zero eigenvalues of ${\bf \Delta}_{ij}$, and $r_{ij} = {\sf rank} \left({\bf \Delta}_{ij}\right) \leq \min \left(n_t,t\right)$ is the rank of ${\bf \Delta}_{ij}$.  We let $R_{ij} \leq r_{ij}$ represent the number of distinct eigenvalues in $\lambda_{ij}\left(1\right), \ldots, \lambda_{ij}\left(r_{ij}\right)$. We also let $\lambda'_{ij}\left(1\right),\ldots,\lambda'_{ij}\left(R_{ij}\right)$ and $r'_{ij} \left(1\right),\ldots, r'_{ij} \left(R_{ij}\right)$ represent the distinct eigenvalues and their multiplicities, respectively, where $r'_{ij} \left(1\right) + \cdots + r'_{ij} \left(R_{ij}\right) = r_{ij}$.

\vspace{0.25cm}

\begin{theorem}
\label{space_time_code_criteria}
Consider the canonical i.i.d. multiple-antenna Rayleigh fading coherent channel model in \eqref{new_channel_model} with $n_t = 2$ and $n_r = 1$ or $n_r = 2$, where ${\bf X} \in \left\{{\bf X}_1,{\bf X}_2,\ldots,{\bf X}_{\sf M}\right\}$ and $\Pr \left({\bf X} = {\bf X}_1\right)=\Pr \left({\bf X} = {\bf X}_2\right)=\cdots=\Pr \left({\bf X} = {\bf X}_{\sf M}\right) = \frac{1}{{\sf M}}$. Then,
\begin{itemize}

\item The value of ${\sf d}$ is maximized by the set of space-time codewords that maximize:
    \begin{align}
    r_{\sf min} = \min_{i \neq j} r_{ij}
    \end{align}

\item The value of the bounds to ${\epsilon'_{\sf d}} \left({\sf snr}\right)$ are maximized by the set of space-time codewords that maximize:
    \begin{align}
    \sum_{\{i,j\} \in \Omega}~\prod_{r=1}^{r_{ij}} \left(\frac{1}{{\lambda}_{ij} (r)}\right)^{n_r}
    \end{align}

\end{itemize}

where $\Omega = \left\{\{i,j\} \in \left\{1,\ldots,{\sf M}\right\} \times \left\{1,\ldots,{\sf M}\right\}: r_{ij} = r_{\sf min}, i \neq j\right\}$.
\end{theorem}

\vspace{0.25cm}

\begin{proof}
Let us consider the squared Euclidean distance between a pair of arbitrary (noiseless) receive codeword matrices associated with the pair of transmit codeword matrices ${\bf X}_i$ and ${\bf X}_j$ given by:
\begin{align}
{\sf d_{ij}^2} = {\sf tr} \Big({\bf H_w} {\bf \Delta}_{ij} {\bf H_w}^\dag\Big)
\end{align}
It is straightforward to show that the distribution of
\begin{align}
{\sf tr} \Big({\bf H_w} {\bf \Delta}_{ij} {\bf H_w^\dag}\Big) = {\sf tr} \Big({\bf H_w} {\bf W} {\bf \Lambda}_{ij} {\bf W}^\dag {\bf H_w^\dag}\Big)
\end{align}
is equal to the distribution of
\begin{align}
{\sf tr} \Big({\bf H_w} {\bf \Lambda}_{ij} {\bf H_w^\dag}\Big) = \sum_{m=1}^{n_r} \sum_{n=1}^{r_{ij}} \lambda_{ij} \big(n\big) |\xi_{mn}|^2
\end{align}
where $\xi_{mn}$ are independent circularly symmetric complex Gaussian random variables with zero-mean and unit-variance. Its probability density function is given by~\cite{Hammarwall08},~\cite{Bjornson09}:
\begin{align}
p_{\sf d_{ij}^2} \left({\sf d_{ij}^2}\right) = \frac{1}{\prod_{r=1}^{R_{ij}} {\lambda'}_{ij}^{n_r r'_{ij} (r)} (r)} \cdot \sum_{r = 1}^{R_{ij}} \sum_{r' = 1}^{n_r r'_{ij} (r)} \frac{\Psi_{r,r'} \left(n_r r'_{ij}(1),\ldots,n_r r'_{ij} \left(R_{ij}\right)\right)}{\left(n_r r'_{ij} (r) - r'\right)!} \cdot \nonumber \\
\cdot \left(- {\sf d_{ij}^2}\right)^{n_r r'_{ij} (r) - r'} \cdot e^{-\frac{{\sf d_{ij}^2}}{\lambda'_{ij} (r)}}, \qquad {\sf d_{ij}^2} \geq 0 \label{pdf_space_time1}
\end{align}
where
\begin{align}
\Psi_{r,r'} \left(n_r r'_{ij}(1),\ldots,n_r r'_{ij} \left(R_{ij}\right)\right) = (-1)^{n_r r'_{ij} (r) - 1} \sum_{k_1,\ldots,k_{R_{ij}} \in \Omega_{r,r'}} \sum_{\substack{k = 1 \\ k \neq r}}^{R_{ij}} \left(\substack{k_k + n_r r'_{ij} (k) -1 \\ k_k}\right) \cdot \nonumber \\
\cdot \left(\frac{1}{\lambda'_{ij} (k)} - \frac{1}{\lambda'_{ij} (r)}\right)^{-\left(n_r r'_{ij} (k) - k_k\right)} \label{pdf_space_time2}
\end{align}
with
\begin{align}
\Omega_{r,r'} = \left\{k_1,\ldots,k_{R_{ij}} \in \mathbb{Z}^{R_{ij}}: k_1 + \cdots + k_{R_{ij}} = r' - 1, k_r = 0, k_k \geq 0~\forall~k\right\} \label{pdf_space_time3}
\end{align}

Let us consider the scenario where $n_t = 2$ and $n_r = 1$. It is possible to show from \eqref{pdf_space_time1}, \eqref{pdf_space_time2} and \eqref{pdf_space_time3} that if $\lambda_{ij} (1) > 0$ and $\lambda_{ij} (2) > 0$, which implies that $r_{ij} = 2$, then
\begin{align}
p_{{\sf d_{ij}^2}} (0) = 0
\end{align}
and
\begin{align}
p_{{\sf d_{ij}^2}}^{(1)} (0) = \frac{1}{\lambda_{ij} (1) \lambda_{ij} (2)} \neq 0
\end{align}
whereas if $\lambda_{ij} (1) > 0$ and $\lambda_{ij} (2) = 0$, which implies that $r_{ij} = 1$, then
\begin{align}
p_{{\sf d_{ij}^2}} (0) = \frac{1}{\lambda_{ij} (1)} \neq 0
\end{align}


Let us now consider the scenario where $n_t = 2$ and $n_r = 2$. It is also possible to show from \eqref{pdf_space_time1}, \eqref{pdf_space_time2} and \eqref{pdf_space_time3} that if $\lambda_{ij} (1) > 0$ and $\lambda_{ij} (2) > 0$ then
\begin{align}
p_{{\sf d_{ij}^2}}^{(n)} (0) = 0, \qquad n = 0, 1, 2
\end{align}
and
\begin{align}
p_{{\sf d_{ij}^2}}^{(3)} (0) = \frac{1}{\lambda_{ij}^2 (1) \lambda_{ij}^2 (2)} \neq 0
\end{align}
where as if $\lambda_{ij} (1) > 0$ and $\lambda_{ij} (2) = 0$ then
\begin{align}
p_{{\sf d_{ij}^2}} (0) = 0
\end{align}
and
\begin{align}
p_{{\sf d_{ij}^2}}^{(1)} (0) = \frac{1}{\lambda_{ij}^2 (1)} \neq 0
\end{align}

Therefore, the Theorem follows immediately from the high-${\sf snr}$ characterization in \eqref{constrained_capacity_space_time1}, \eqref{constrained_capacity_space_time2}, \eqref{constrained_capacity_space_time3}, \eqref{constrained_capacity_space_time4} and \eqref{constrained_capacity_space_time5}.

\end{proof}

\vspace{0.25cm}

In principle, it is also possible to generalize the result for canonical i.i.d. multiple-antenna Rayleigh fading coherent channels with an arbitrary number of transmit and receive antennas. However, this requires considerable algebraic manipulation due to the form of the probability density function in \eqref{pdf_space_time1}, \eqref{pdf_space_time2} and \eqref{pdf_space_time3}.

Note that if $n_t \geq t$, then the maximum possible value for $r_{ij}$ is $t$; in contrast, if $n_t \leq t$, then the maximum possible value of $r_{ij}$ is $n_t$. Note also that if $r_{ij} = n_t,~\forall~i \neq j$, then the maximization of:
\begin{align}
\sum_{\{i,j\} \in \Omega}~\prod_{r=1}^{r_{ij}} \left(\frac{1}{{\lambda}_{ij} (r)}\right)^{n_r}
\end{align}
corresponds to the maximization of:
\begin{align}
\sum_{i = 1}^{{\sf M}} \sum_{\substack{j = 1 \\ j \neq i}}^{{\sf M}} \left(\frac{1}{\det \left({\bf \Delta}_{ij}\right)}\right)^{n_r}
\end{align}

We conclude that the space-time code design criteria embodied in Theorem \ref{space_time_code_criteria} are akin to the conventional rank and determinant design criteria in~\cite{Tarokh98}, which have been derived from the pairwise error probability view point. Since the average error probability, the average MMSE and the average mutual information exhibit identical high-${\sf snr}$ behavior, the design criteria put forth in Theorem \ref{space_time_code_criteria}, in addition to maximizing the lower bound to the average mutual information, also minimize the upper bounds to the average MMSE and the average error probability.

\vspace{-0.25cm}

\section{Conclusions}
\label{conclusions}

By drawing on the I-MMSE identity and counterparts, together with key results in asymptotic analysis and expansions, it has been possible to put forth a high-${\sf snr}$ characterization of the constrained capacity of multiple-antenna fading coherent channels driven by arbitrary equiprobable discrete inputs. This characterization has enabled the study of the effect of various system models, parameters, and elements on the constrained capacity of multiple-antenna fading coherent channels. Key contributions include:

\begin{itemize}

\item In the antenna-uncorrelated Rayleigh fading coherent channel, we have analyzed the impact on the high-${snr}$ behavior of the constrained capacity of the number of transmit antennas, the number of receive antennas and the characteristics of the multi-dimensional constellation. It has been observed that in the regime of high-${\sf snr}$ the constrained capacity increases with the increase in the number of transmit antennas and the number of receive antennas, as well as with the use of multi-dimensional constellations with better sphere-packing properties. In particular, it has been observed that in the regime of high-${\sf snr}$ the number of receive antennas control the rate at which the constrained capacity value tends to its infinite-${\sf snr}$ value whereas the number of transmit antennas control the constrained capacity power offset, by enabling the construction of multi-dimensional constellations with better sphere-packing properties. It has also been observed that the geometry of the multi-dimensional constellation, most notably, its sphere-packing properties, also control the constrained capacity power offset. It has also been emphasized that in the regime of high-${\sf snr}$ the role of the transmit and receive antennas on the behavior of the constrained capacity of canonical i.i.d. Rayleigh fading channels is very different from their role on the behavior of the capacity of the canonical i.i.d. Rayleigh fading coherent channel, which is achieved by Gaussian inputs.

    \vspace{0.1cm}

\item In the antenna-correlated Rayleigh fading channel, we have analyzed the impact on the high-${snr}$ behavior of the constrained capacity of the number of transmit antennas, the number of receive antennas, transmit and receive antenna correlation and the characteristics of the multi-dimensional constellation. It has been observed that in the regime of high-${\sf snr}$ the constrained capacity also increases with the increase in the number of transmit antennas and the number of receive antennas, as well as with the use of multi-dimensional constellations with better sphere-packing properties in the coordinate system induced by the transmit correlation matrix. It has been found that the presence of transmit and receive antenna correlation has a negative impact on the constrained capacity in the regime of high-${\sf snr}$. It has also been found that degenerate conditions, i.e., perfectly correlated paths, have an impact on the constrained capacity infinite-${\sf snr}$ value as well as the rate at which the constrained capacity value tends to its infinite-${\sf snr}$ value.

    \vspace{0.1cm}

\item In Ricean fading channels line-of-sight components also have profound implications on the behavior of the constrained capacity in the regime of high ${\sf snr}$. Of particular interest, it has been observed that, whereas in a single-transmit single-receive antenna Ricean fading coherent channel the constrained capacity increases with the increase in the K-factor in the regime of high-${\sf snr}$, in a multiple-transmit multiple-receive antenna Ricean fading channel the constrained capacity may decrease with the increase of the K-factor in the regime of high-${\sf snr}$. This result, which does not occur for Gaussian inputs, is due to the nature of the interaction of the multi-dimensional constellation with the channel model.

   \vspace{0.1cm}

\item The high-${\sf snr}$ characterization of the constrained capacity has also enabled the design of elements for multiple-antenna fading coherent channel models. We have considered power allocation in a bank of parallel independent fading coherent channels, precoding for multiple-antenna Rayleigh fading coherent channels and space-time coding for multiple-antenna Rayleigh fading coherent channels, showing that the expansions lead to very sharp designs.

\end{itemize}

\vspace{0.1cm}

Of particular relevance, the construction of the high-${\sf snr}$ constrained capacity characterization has also disclosed intimate connections between the high-${\sf snr}$ asymptotic behavior of the average minimum mean-squared error, the average mutual information and the average error probability in multiple-antenna fading coherent channels driven by arbitrary equiprobable discrete inputs. These connections, which suggest that designs that minimize (bounds to) the error probability also minimize (bounds to) the minimum mean-squared error and maximize (bounds to) the average mutual information, provide a unification of the behavior of key performance metrics in the asymptotic regime of high ${\sf snr}$. Overall, this high-${\sf snr}$ analysis complements the low-${\sf snr}$ insight available in~\cite{Verdu02},~\cite{Tulino03}.

\vspace{-0.35cm}

\section*{Appendix A\\Proof of Lemma \ref{mmse_deterministic}}

We determine the lower bound to the MMSE by using a genie that supplies the receiver with the true input vector and any of the other input vectors with equal probability. The genie based estimate is given by:
\begin{equation}
\hat{\x}_{genie} (\y,\{\x_i,\x_j\})=\frac{\x_i e^{-\|\y-\sqrt{\sf
snr}\HH\x_i\|^2} + \x_j e^{-\|\y-\sqrt{\sf snr}\HH\x_j\|^2}}
{e^{-\|\y-\sqrt{\sf snr}\HH\x_i\|^2}+e^{-\|\y-\sqrt{\sf
snr}\HH\x_j\|^2}}
\end{equation}
where $\x_i$ is the true input vector and $\x_j$ is the other input vector. It follows that the MMSE can be lower bounded as follows:
\begin{align}
{\sf mmse} \left({\sf snr};{\bf H}\right) &= \mathbb{E} \left\{\left\|\HH\x
- \HH\mathbb{E} \{\x|\y\}\right\|^2\big|{\bf H}\right\} \\
&= \frac{1}{{\sf M}} \sum_{i=1}^{{\sf M}} \mathbb{E} \left\{\left\|\HH\x -
\HH\mathbb{E} \{\x|\y\}\right\|^2 \big| \x = \x_i,{\bf H}\right\} \\
&\geq \frac{1}{{\sf M} ({\sf M}-1)} \sum_{i=1}^{{\sf M}} \sum_{\substack{j=1 \\ j \neq i}}^{{\sf M}} \mathbb{E} \left\{ \left\|\HH\x
- \HH \hat{\x}_{genie} (\y,\{\x_i,\x_j\})\right\|^2 \big|
\x = \x_i,{\bf H}\right\} \label{ineq_lb1} \\
&= \frac{1}{{\sf M} ({\sf M}-1)} \sum_{i=1}^{{\sf M}} \sum_{\substack{j=1 \\ j \neq i}}^{{\sf M}} \frac{\|\HH\x_i -
\HH\x_j\|^2}{4} \cdot {\sf
mmse_{BPSK}} \left(\frac{\|\HH\x_i - \HH\x_j\|^2 {\sf snr}}{4}\right) \\
&= \frac{1}{{\sf M} ({\sf M}-1)} \sum_{i=1}^{{\sf M}} \sum_{\substack{j=1 \\ j \neq i}}^{{\sf M}} \frac{{\sf d_{ij}^2} \left({\bf H}\right)}{4} \cdot {\sf
mmse_{BPSK}} \left(\frac{{\sf d_{ij}^2} \left({\bf H}\right){\sf snr}}{4}\right) \\
&\geq \frac{1}{{\sf M} ({\sf M}-1)} \sum_{i=1}^{{\sf M}} \sum_{\substack{j=1 \\ j \neq i}}^{{\sf M}} \frac{{\sf d_{ij}^2}\left({\bf H}\right)}{4} \cdot \frac{1}{2} \cdot {\sf erfc} \left(\sqrt{\frac{{\sf d_{ij}^2}\left({\bf H}\right){\sf snr}}{4}}\right) \triangleq {\sf mmse_{LB}} \left({\sf snr};{\bf H}\right) \label{ineq_lb2}
\end{align}

The first inequality in \eqref{ineq_lb1} is due to the genie based
estimator and the second inequality in \eqref{ineq_lb2} is due to the lower bound to the MMSE of BPSK:
\begin{align}
{\sf mmse_{BPSK}} (\rho) &= 1 - \int_{-\infty}^{\infty} {\sf tanh}
(2 \sqrt{\rho} \xi) \frac{e^{-(\xi - \sqrt{\rho})^2}}{\sqrt{\pi}} d
\xi \geq 1 - \int_{0}^{\infty} \frac{e^{-(\xi -
\sqrt{\rho})^2}}{\sqrt{\pi}} d \xi = 1 - \int_{-\sqrt{\rho}}^{\infty} \frac{e^{-\xi^2}}{\sqrt{\pi}} d \xi = \frac{1}{2} \cdot {\sf erfc} (\sqrt{\rho})
\end{align}

We determine the upper bound to the MMSE by using a (sub-optimal) Euclidean distance based estimator rather than the optimal conditional mean estimator. The Euclidean distance based estimate is given by:
\begin{equation}
\hat{\x}_{euc}(\y) = \arg\!\min_{\x}\left\|\y - \sqrt{{\sf snr}} \cdot \HH\x\right\|^2,
\end{equation}
It follows that the MMSE can be upper bounded as follows:
\begin{align}
{\sf mmse} \left({\sf snr};{\bf H}\right) &= \mathbb{E} \left\{\left\|\HH\x
- \HH\mathbb{E} \{\x|\y\}\right\|^2\big|{\bf H}\right\} \\
&= \frac{1}{{\sf M}} \sum_{i=1}^{{\sf M}} \mathbb{E} \left\{ \left\|\HH\x -
\HH\mathbb{E} \{\x|\y\}\right\|^2 \big| \x = \x_i,{\bf H}\right\} \\
&\leq \frac{1}{{\sf M}} \sum_{i=1}^{{\sf M}} \mathbb{E} \left\{ \left\|\HH\x
- \HH\hat{\x}_{euc}(\y)\right\|^2 \big| \x = \x_i,{\bf H}\right\} \label{ineq_ub1} \\
&= \frac{1}{{\sf M}} \sum_{i=1}^{{\sf M}} \sum_{j=1}^{{\sf M}} \mathbb{E} \left\{
\left\|\HH\x - \HH\hat{\x}_{euc}(\y)\right\|^2 \big| \x = \x_i,
\y \in \mathcal{V}_j,{\bf H}\right\} \cdot \Pr \left\{\y \in
\mathcal{V}_j \big| \x = \x_i,{\bf H}\right\} \\
&\leq \frac{1}{{\sf M}} \sum_{i=1}^{{\sf M}} \sum_{j=1}^{{\sf M}} {\sf d_{ij}^2} \left({\bf H}\right) \cdot \frac{1}{2} \cdot {\sf erfc} \left(\sqrt{\frac{{\sf d_{ij}^2}\left({\bf H}\right){\sf snr}}{4}}\right) \triangleq {\sf mmse_{UB}} \left({\sf snr};{\bf H}\right) \label{ineq_ub2}
\end{align}
where $\mathcal{V}_l$ is the Voronoi region associated with $\HH
\x_l$. The first inequality in \eqref{ineq_ub1} is due to the
(sub-optimal) Euclidean distance based estimator and the second
inequality in \eqref{ineq_ub2} is due to
\begin{align}
\Pr \left\{\y \in \mathcal{V}_j \big| \x = \x_i,{\bf H}\right\} \leq \frac{1}{2} \cdot {\sf erfc} \left(\sqrt{\frac{{\sf d_{ij}^2}\left({\bf H}\right){\sf snr}}{4}}\right)
\end{align}
and $\mathbb{E} \left\{ \left\|\HH\x - \HH\hat{\x}_{euc}(\y)\right\|^2 \big| \x = \x_i, \y \in
\mathcal{V}_j,{\bf H}\right\} = {\sf d_{ij}^2} \left({\bf H}\right)$.

\vspace{-0.35cm}

\section*{Appendix B\\Proof of Lemma \ref{mutual_information_deterministic}}

We determine the lower bound to the mutual information by using \eqref{mmse_ub} in \eqref{i-mmse-relationship} as follows:
\begin{align}
{\sf I} \left({\sf snr}; {\bf H}\right) &\geq \log {\sf M} - \int_{{\sf
snr}}^{\infty} {\sf mmse_{UB}} \left(\xi;{\bf H}\right) d \xi \\
&=\log {\sf M} - \frac{1}{{\sf M}} \sum_{i=1}^{{\sf M}} \sum_{\substack{j=1 \\ j \neq i}}^{{\sf M}} \int_{\sf snr}^{\infty} {\sf d_{ij}^2} \left({\bf H}\right) \cdot \frac{1}{2} \cdot {\sf erfc} \left(\sqrt{\frac{{\sf d_{ij}^2}\left({\bf H}\right)\xi}{4}}\right) d \xi \\
&\geq \log {\sf M} - \frac{1}{{\sf M}} \sum_{i=1}^{{\sf M}} \sum_{\substack{j=1 \\ j \neq i}}^{{\sf M}} \int_{\sf snr}^{\infty} {\sf d_{ij}^2} \left({\bf H}\right) \cdot \frac{1}{2} \cdot e^{-\frac{{\sf d_{ij}^2}\left({\bf H}\right)\xi}{4}} d \xi \label{ineq1} \\
&= \log {\sf M} - \frac{1}{{\sf M}} \sum_{i=1}^{\sf M} \sum_{\substack{j=1 \\ j \neq i}}^{\sf M} 2 \cdot
e^{-\frac{{\sf d_{ij}^2}\left({\bf H}\right) {\sf snr}}{4}} \triangleq {\sf I_{LB}} \left({\sf snr};{\bf H}\right)
\end{align}
The inequality in \eqref{ineq1} follows from the upper bound to the
complementary error function ${\sf erfc} (x) \leq e^{-x^2}$.

We determine the upper bound to the mutual information by using \eqref{mmse_lb} in \eqref{i-mmse-relationship} as follows:
\begin{align}
{\sf I} \left({\sf snr};{\bf H}\right) &\leq \log {\sf M} - \int_{{\sf
snr}}^{\infty} {\sf mmse_{LB}} \left(\xi; {\bf H}\right) d \xi \\
&=\log {\sf M} - \frac{1}{{\sf M} ({\sf M}-1)} \sum_{i=1}^{{\sf M}} \sum_{\substack{j=1 \\ j \neq i}}^{{\sf M}} \int_{\sf snr}^{\infty} \frac{{\sf d_{ij}^2} \left({\bf H}\right)}{4} \cdot \frac{1}{2} \cdot {\sf erfc} \left(\sqrt{\frac{{\sf d_{ij}^2}\left({\bf H}\right)\xi}{4}}\right) d \xi \\
&=\log {\sf M} + \frac{1}{{\sf M} ({\sf M}-1)} \sum_{i=1}^{{\sf M}} \sum_{\substack{j=1 \\ j \neq i}}^{{\sf M}} \frac{{\sf d_{ij}^2}\left({\bf H}\right) {\sf snr}}{4} \cdot \frac{1}{2} \cdot {\sf erfc} \left(\sqrt{\frac{{\sf d_{ij}^2}\left({\bf H}\right) {\sf snr}}{4}}\right) \\
&- \frac{1}{{\sf M} ({\sf M}-1)} \sum_{i=1}^{{\sf M}} \sum_{\substack{j=1 \\ j \neq i}}^{{\sf M}} \sqrt{\frac{{\sf d_{ij}^2}\left({\bf H}\right) {\sf snr}}{4}} \cdot \frac{1}{2} \cdot \frac{e^{-\frac{{\sf d_{ij}^2}\left({\bf H}\right) {\sf snr}}{4}}}{\sqrt{\pi}} \\
&- \frac{1}{{\sf M} ({\sf M}-1)} \sum_{i=1}^{{\sf M}} \sum_{\substack{j=1 \\ j \neq i}}^{{\sf M}} \frac{1}{4} \cdot {\sf erfc} \left(\sqrt{\frac{{\sf d_{ij}^2}\left({\bf H}\right) {\sf snr}}{4}}\right)\\
&\leq\log {\sf M} - \frac{1}{{\sf M} ({\sf M}-1)} \sum_{i=1}^{{\sf M}} \sum_{\substack{j=1 \\ j \neq i}}^{{\sf M}} \frac{1}{4} \cdot {\sf erfc} \left(\sqrt{\frac{{\sf d_{ij}^2}\left({\bf H}\right) {\sf snr}}{4}}\right)  \triangleq {\sf I_{UB}} \left({\sf snr}; {\bf H}\right) \label{ineq2}
\end{align}
The inequality in \eqref{ineq2} follows from the upper bound to the
complementary error function ${\sf erfc} (x) \leq e^{-x^2}/\sqrt{\pi
\cdot x^2}$.

\vspace{-0.35cm}

\section*{Appendix C\\Proof of Lemma \ref{mmse_fading}}

The lower and upper bounds to the average value of the MMSE are obtained from the lower and upper bounds to the MMSE, respectively, by averaging over the fading statistics as follows:
\begin{align}
{\sf \overline{mmse}_{LB}} \left({\sf snr}\right) &= \mathbb{E}_{{\bf H}} \left\{{\sf mmse_{LB}} \left({\sf snr};{\bf H}\right)\right\} \nonumber \\
&=  \frac{1}{{\sf snr}} \sum_{i=1}^{\sf M} \sum_{\substack{j=1 \\ j \neq i}}^{\sf M} \int_{0}^{\infty} \frac{1}{8 {\sf M} ({\sf M}-1)} \cdot {\sf d_{ij}^2} {\sf snr} \cdot {\sf erfc} \left(\sqrt{\frac{{\sf d_{ij}^2} {\sf snr}}{4}}\right) \cdot p_{{\sf d_{ij}^2}} \left({\sf d_{ij}^2}\right) d {\sf d_{ij}^2}
\label{mmse_lb1}
\end{align}
\begin{align}
{\sf \overline{mmse}_{UB}} \left({\sf snr}\right) &= \mathbb{E}_{{\bf H}} \left\{{\sf mmse_{UB}} \left({\sf snr};{\bf H}\right)\right\} \nonumber \\
&= \frac{1}{{\sf snr}} \sum_{i=1}^{\sf M} \sum_{\substack{j=1 \\ j \neq i}}^{\sf M} \int_{0}^{\infty} \frac{1}{2{\sf M}} \cdot {\sf d_{ij}^2} {\sf snr} \cdot {\sf erfc} \left(\sqrt{\frac{{\sf d_{ij}^2} {\sf snr}}{4}}\right) \cdot p_{{\sf d_{ij}^2}} \left({\sf d_{ij}^2}\right) d {\sf d_{ij}^2}
\label{mmse_ub1}
\end{align}

We establish the asymptotic expansions as ${\sf snr} \rightarrow \infty$ of the integrals that compose the lower bound and the upper bound to the average value of the MMSE in \eqref{mmse_lb1} and \eqref{mmse_ub1}, respectively, by using~\cite[Theorem 3.2]{Bleistein86}. This Theorem requires that:
\begin{equation}
\left| \int_{\lambda t}^{\infty} d t_{n-1}  \int_{t_{n-1}}^{\infty} d t_{n-2} \cdots
\int_{t_1}^{\infty} t_0 \cdot
{\sf erfc} \left(\sqrt{\frac{t_0}{4}}\right) dt_0 \right| \leq \alpha_n (t) \cdot
\phi_n (\lambda), n = 1, 2, \ldots, N+2 \label{mmse_cond1}
\end{equation}
These conditions are satisfied with $\alpha_n \left(t\right) = n \cdot 4^{n+1}$ and $\phi_n \left(\lambda\right) = 1$ because, by using the well known upper bound to the complementary error function ${\sf erfc} (x) \leq e^{-x^2}, x \geq 0$, it is possible to prove that:
\begin{align}
&\left| \int_{\lambda t}^{\infty} d t_{n-1} \cdots \int_{t_1}^{\infty} t_0 \cdot {\sf erfc} \left(\sqrt{\frac{t_0}{4}}\right) dt_0 \right| \leq \left(\lambda t + 4n\right) \cdot 4^n \cdot e^{-\lambda t/4} \leq n \cdot 4^{n+1}, n=1,2,\ldots,N+2
\end{align}
This Theorem also requires that the functions $p_{{\sf d_{ij}^2}}^{(n)} \big({\sf d_{ij}^2}\big), n = 0, 1, \ldots, N+1,$ are continuous on $[0,\infty)$, $p_{{\sf d_{ij}^2}}^{(N+2)} \big({\sf d_{ij}^2}\big)$ is piecewise continuous on $[0,\infty)$ and \footnote{The additional requirement that $\int_{0}^{\infty} \alpha_n \left({\sf d_{ij}^2}\right) p_{{\sf d_{ij}^2}}^{(n)} \left({\sf d_{ij}^2}\right) d {\sf d_{ij}^2} < \infty$ is due to the fact that the integration interval is semi-infinite.}
\begin{equation}
\int_{0}^{\infty} \alpha_n \left({\sf d_{ij}^2}\right) p_{{\sf d_{ij}^2}}^{(n)} \left({\sf d_{ij}^2}\right) d {\sf d_{ij}^2} < \infty
\end{equation}
These conditions are satisfied because, by assumption, the functions $p_{{\sf d_{ij}^2}}^{(n)} \big({\sf d_{ij}^2}\big), n = 0, 1, \ldots,$ are continuous and integrable on $[0,\infty)$.

Consequently, the asymptotic expansions as ${\sf snr} \rightarrow \infty$ of the integrals that compose the lower bound and the upper bound to the average value of the MMSE are given by:
\begin{align}
\sum_{n=0}^{N} \frac{1}{{\sf snr}^{n+1}} \cdot k_{{\sf LB}_{n+1}} \cdot p_{{\sf d_{ij}^2}}^{(n)} (0) + \mathcal{O} \left(\frac{1}{{\sf snr}^{N+2}}\right)
\label{mmse_lb_terms_expansion}
\end{align}
\begin{align}
\sum_{n=0}^{N} \frac{1}{{\sf snr}^{n+1}} \cdot k_{{\sf UB}_{n+1}} \cdot p_{{\sf d_{ij}^2}}^{(n)} (0) + \mathcal{O} \left(\frac{1}{{\sf snr}^{N+2}}\right)
\label{mmse_ub_terms_expansion}
\end{align}
where the underlying auxiliary asymptotic sequence is $\big\{(1/{\sf snr})^{n}, n = 1, 2, \ldots\big\}$ and $k_{{\sf LB}_{n}}$ and $k_{{\sf UB}_{n}}$ are given by:
\begin{align}
&k_{{\sf LB}_n} = \int_{0}^{\infty} d t_{n-1}  \int_{t_{n-1}}^{\infty} d t_{n-2} \cdots
\int_{t_1}^{\infty} \frac{1}{8{\sf M}({\sf M}-1)} \cdot t_0 \cdot
{\sf erfc} \left(\sqrt{\frac{t_0}{4}}\right) dt_0 \label{mmse_lb_iterated_integral2} \\
&k_{{\sf UB}_n} = \int_{0}^{\infty} d t_{n-1}  \int_{t_{n-1}}^{\infty} d t_{n-2} \cdots
\int_{t_1}^{\infty} \frac{1}{2{\sf M}} \cdot t_0 \cdot
{\sf erfc} \left(\sqrt{\frac{t_0}{4}}\right) dt_0 \label{mmse_ub_iterated_integral2}
\end{align}

We now establish the asymptotic expansions as ${\sf snr} \to \infty$ of the lower and the upper bounds to the average value of the MMSE in \eqref{mmse_lb1} and \eqref{mmse_ub1}, by using~\cite[Theorem 1.7.1]{Bleistein86}. The asymptotic expansion with respect to the asymptotic sequence $\big\{{\sf snr}^{-n}\big\}$ as ${\sf snr} \to \infty$ of the sum of the integrals in \eqref{mmse_lb1} and \eqref{mmse_ub1} is equal to the sum of the asymptotic expansions with respect to the asymptotic sequence $\big\{{\sf snr}^{-n}\big\}$ as ${\sf snr} \to \infty$ of the individual integrals in \eqref{mmse_lb1} and \eqref{mmse_ub1}, because \eqref{mmse_lb_terms_expansion} and \eqref{mmse_ub_terms_expansion} are asymptotic expansions of Poincar\'e type.

Consequently, the asymptotic expansions as ${\sf snr} \rightarrow \infty$ of the lower bound and the upper bound to the average value of the MMSE are given by:
\begin{align}
{\sf \overline{mmse}_{LB}} ({\sf snr}) = \sum_{n=0}^{N} \frac{1}{{\sf snr}^{n+2}} \cdot k_{{\sf LB}_{n+1}} \cdot \left(\sum_{i=1}^{{\sf M}} \sum_{\substack{j=1 \\ i \neq j}}^{{\sf M}} p_{{\sf d_{ij}^2}}^{(n)} (0)\right) + \mathcal{O} \left(\frac{1}{{\sf snr}^{N+3}}\right)
\label{mmse_lb_expansion}
\end{align}
\begin{align}
{\sf \overline{mmse}_{UB}} ({\sf snr}) = \sum_{n=0}^{N} \frac{1}{{\sf snr}^{n+2}} \cdot k_{{\sf UB}_{n+1}} \cdot \left(\sum_{i=1}^{{\sf M}} \sum_{\substack{j=1 \\ i \neq j}}^{{\sf M}} p_{{\sf d_{ij}^2}}^{(n)} (0)\right) + \mathcal{O} \left(\frac{1}{{\sf snr}^{N+3}}\right)
\label{mmse_ub_expansion}
\end{align}
where the underlying auxiliary asymptotic sequence is $\big\{(1/{\sf snr})^{n+1}, n = 1, 2, \ldots\big\}$ and $k_{{\sf LB}_{n}}$ and $k_{{\sf UB}_{n}}$ are also given by \eqref{mmse_lb_iterated_integral2} and \eqref{mmse_ub_iterated_integral2}, respectively.

The values of $k_{{\sf LB}_{n}}$ and $k_{{\sf UB}_{n}}$ can also be computed by using
\begin{align}
{\sf erfc} \left(\sqrt{\frac{t_0}{4}}\right) = \frac{2}{\pi} \int_{0}^{\infty} \frac{e^{- \frac{t_0}{4} (x^2+1) }}{x^2+1} dx
\end{align}
in \eqref{mmse_lb_iterated_integral2} and \eqref{mmse_ub_iterated_integral2}, so that:
\begin{align}
k_{{\sf LB}_n} &= \frac{2}{\pi} \int_{0}^{\infty} \frac{1}{8 {\sf M} \left({\sf M}-1\right)} \cdot \frac{n \cdot 4^{n+1}}{\left(x^2+1\right)^{n+2}} dx = \frac{1}{8 {\sf M} \left({\sf M}-1\right)} \cdot \frac{n \cdot 4^{n+1}}{\sqrt{\pi}} \cdot \frac{\Gamma \left(n + 3/2\right)}{\Gamma \left(n + 1/2\right)}
\end{align}
\begin{align}
k_{{\sf UB}_n} &= \frac{2}{\pi} \int_{0}^{\infty} \frac{1}{2 {\sf M}} \cdot \frac{n \cdot 4^{n+1}}{\left(x^2+1\right)^{n+2}} dx = \frac{1}{2 {\sf M}} \cdot \frac{n \cdot 4^{n+1}}{\sqrt{\pi}} \cdot \frac{\Gamma \left(n + 3/2\right)}{\Gamma \left(n + 1/2\right)}
\end{align}
where $\Gamma \left(\cdot\right)$ is the Gamma function.

Finally, since the functions $p_{{\sf d_{ij}^2}}^{(n)} \big({\sf d_{ij}^2}\big)$ are continuous and integrable on $[0,\infty)$ for arbitrarily large $n$ and \eqref{mmse_cond1} also holds for arbitrarily large $n$, it is possible to let $N$ go to infinity in the asymptotic expansions of the lower and upper bounds to the average value of the MMSE in \eqref{mmse_lb_expansion} and \eqref{mmse_ub_expansion}, respectively.

\vspace{-0.35cm}

\section*{Appendix D\\Proof of Lemma \ref{mutual_information_fading}}

The lower and upper bounds to the average value of the mutual information are obtained from the upper and lower bounds to the average value of the MMSE, respectively, by using \eqref{average-i-average-mmse-relationship} as follows:
\begin{align}
{\sf \bar{I}_{LB}} \left({\sf snr}\right) = \log {\sf M} - \int_{{\sf
snr}}^{\infty} {\sf \overline{mmse}_{UB}} \big(\xi\big) d\xi \\
{\sf \bar{I}_{UB}} \left({\sf snr}\right) = \log {\sf M} - \int_{{\sf
snr}}^{\infty} {\sf \overline{mmse}_{LB}} \big(\xi\big) d\xi
\end{align}

We now obtain the asymptotic expansions (as ${\sf snr} \rightarrow \infty$) of the lower bound and the upper bound to the average value of the mutual information directly from the asymptotic expansions (as ${\sf snr} \rightarrow \infty$) of the upper bound and the lower bound to the average value of the MMSE, respectively, by capitalizing on~\cite[Theorem 1.7.6]{Bleistein86}.\footnote{Note that we determine the asymptotic expansions of the lower and upper bounds to the average mutual information directly from the asymptotic expansions of the upper and lower bounds to the average MMSE in conjunction with \eqref{average-i-average-mmse-relationship}. Alternatively, we can also determine asymptotic expansions of the lower and upper bounds to the average mutual information by adopting a procedure identical to that in the average MMSE case leveraging the bounds to the mutual information in Lemma \ref{mutual_information_deterministic}.} This Theorem requires that:
\begin{align}
\int_{{\sf snr}}^{\infty} {\sf \overline{mmse}_{LB}} \big(\xi\big) d\xi < \infty \label{cond1}
\end{align}
\begin{align}
\int_{{\sf snr}}^{\infty} {\sf \overline{mmse}_{UB}} \big(\xi\big) d\xi < \infty \label{cond2}
\end{align}
and
\begin{align}
\int_{{\sf snr}}^{\infty} \frac{1}{{\sf \xi}^{n+1}} d \xi < \infty, \qquad n = 1, 2, \ldots \label{cond3}
\end{align}
It is immediate to show (e.g. by substituting \eqref{mmse_lb1} and \eqref{mmse_ub1} in \eqref{cond1} and \eqref{cond2}, respectively, and trivially bounding the value of the integrals) that the integrals \eqref{cond1} and \eqref{cond2} exist for ${\sf snr} > 0$ and, likewise, that the integral \eqref{cond3} also exists for ${\sf snr} > 0$. Consequently, the asymptotic expansion as ${\sf snr} \rightarrow \infty$ of the lower and upper bounds to the average value of the mutual information are given by:
\begin{align}
{\sf \bar{I}_{LB}} ({\sf snr}) = \log {\sf M} - \sum_{n=0}^{N} \frac{1}{{\sf snr}^{n+1}} \cdot k'_{{\sf LB}_{n+1}} \cdot \left(\sum_{i=1}^{{\sf M}} \sum_{\substack{j=1 \\ i \neq j}}^{{\sf M}} p_{{\sf d_{ij}^2}}^{(n)} (0)\right) + \mathcal{O} \left(\frac{1}{{\sf snr}^{N+2}}\right)
\label{i_lb_expansion}
\end{align}
\begin{align}
{\sf \bar{I}_{UB}} ({\sf snr}) = \log {\sf M} - \sum_{n=0}^{N} \frac{1}{{\sf snr}^{n+1}} \cdot k'_{{\sf UB}_{n+1}} \cdot \left(\sum_{i=1}^{{\sf M}} \sum_{\substack{j=1 \\ i \neq j}}^{{\sf M}} p_{{\sf d_{ij}^2}}^{(n)} (0)\right) + \mathcal{O} \left(\frac{1}{{\sf snr}^{N+2}}\right)
\label{i_ub_expansion}
\end{align}
where the underlying asymptotic sequence is $\big\{(1/{\sf snr})^{n}, n = 1, 2, \ldots\big\}$ and $k'_{{\sf LB}_{n}}$ and $k'_{{\sf UB}_{n}}$ are given by:
\begin{align}
k'_{{\sf LB}_n} = \frac{1}{n} \cdot k_{{\sf UB}_n} = \frac{1}{2 {\sf M}} \cdot \frac{4^{n+1}}{\sqrt{\pi}} \cdot \frac{\Gamma \left(n + 3/2\right)}{\Gamma \left(n + 1/2\right)} \label{i_lb_iterated_integral2}
\end{align}
\begin{align}
k'_{{\sf UB}_n} = \frac{1}{n} \cdot k_{{\sf LB}_n} = \frac{1}{8 {\sf M} \left({\sf M}-1\right)} \cdot \frac{4^{n+1}}{\sqrt{\pi}} \cdot \frac{\Gamma \left(n + 3/2\right)}{\Gamma \left(n + 1/2\right)} \label{i_ub_iterated_integral2}
\end{align}

We can also let $N$ go to infinity in the asymptotic expansions of the lower and upper bounds to the average mutual information in \eqref{i_lb_expansion} and \eqref{i_ub_expansion}, respectively.

\vspace{-0.35cm}

\section*{Appendix E\\Proofs of Theorems \ref{mmse_expansion} and \ref{mutual_information_expansion}}

The proofs capitalize on the asymptotic expansions of the upper and lower bounds to the average value of the MMSE and the average value of the mutual information in Lemmas \ref{mmse_fading} and \ref{mutual_information_fading}, respectively, given by:
\begin{align}
{\sf \overline{mmse}_{LB}} ({\sf snr}) = k_{{\sf LB}_{\sf d}} \cdot \sum_{i=1}^{{\sf M}} \sum_{\substack{j=1 \\ j \neq i}}^{{\sf M}} p_{{\sf d_{ij}^2}}^{({\sf d}-1)} (0) \cdot \frac{1}{{\sf snr}^{{\sf d}+1}} + \mathcal{O} \left(\frac{1}{{\sf snr}^{{\sf d}+2}}\right) \label{asymptotic_expansion1}
\end{align}
\begin{align}
{\sf \overline{mmse}_{UB}} ({\sf snr}) = k_{{\sf UB}_{\sf d}} \cdot \sum_{i=1}^{{\sf M}} \sum_{\substack{j=1 \\ j \neq i}}^{{\sf M}} p_{{\sf d_{ij}^2}}^{({\sf d}-1)} (0) \cdot \frac{1}{{\sf snr}^{{\sf d}+1}} + \mathcal{O} \left(\frac{1}{{\sf snr}^{{\sf d}+2}}\right) \label{asymptotic_expansion2}
\end{align}
and
\begin{align}
{\sf \bar{I}_{LB}} ({\sf snr}) = \log {\sf M} - k'_{{\sf LB}_{\sf d}} \cdot \sum_{i=1}^{{\sf M}} \sum_{\substack{j=1 \\ j \neq i}}^{{\sf M}} p_{{\sf d_{ij}^2}}^{({\sf d}-1)} (0) \cdot \frac{1}{{\sf snr}^{{\sf d}}} + \mathcal{O} \left(\frac{1}{{\sf snr}^{{\sf d}+1}}\right) \label{asymptotic_expansion3}
\end{align}
\begin{align}
{\sf \bar{I}_{UB}} ({\sf snr}) = \log {\sf M} - k'_{{\sf UB}_{\sf d}} \cdot \sum_{i=1}^{{\sf M}} \sum_{\substack{j=1 \\ j \neq i}}^{{\sf M}} p_{{\sf d_{ij}^2}}^{({\sf d}-1)} (0) \cdot \frac{1}{{\sf snr}^{{\sf d}}} + \mathcal{O} \left(\frac{1}{{\sf snr}^{{\sf d}+1}}\right) \label{asymptotic_expansion4}
\end{align}

Let us define the functions:
\begin{align}
f \left({\sf snr}\right) = {\sf snr}^{{\sf d}+1} \cdot {\sf \overline{mmse}} ({\sf snr})
\end{align}
and
\begin{align}
g \left({\sf snr}\right) = {\sf snr}^{{\sf d}} \cdot \left(\log {\sf M} - {\sf \bar{I}} ({\sf snr})\right)
\end{align}
as well as the functions:
\begin{equation}
f_1 \left({\sf snr}\right) = \left\{\begin{array}{lc}
f \left({\sf snr}\right), & k_{{\sf LB}_{\sf d}} \cdot \sum_{\substack{i,j=1 \\ i \neq j}}^{{\sf M}} p_{{\sf d_{ij}^2}}^{({\sf d}-1)} (0) \leq f \left({\sf snr}\right) \leq k_{{\sf UB}_{\sf d}} \cdot \sum_{\substack{i,j=1 \\ i \neq j}}^{{\sf M}} p_{{\sf d_{ij}^2}}^{({\sf d}-1)} (0) \\
k_{{\sf LB}_{\sf d}} \cdot \sum_{\substack{i,j=1 \\ i \neq j}}^{{\sf M}} p_{{\sf d_{ij}^2}}^{({\sf d}-1)} (0), & f \left({\sf snr}\right) \leq k_{{\sf LB}_{\sf d}} \cdot \sum_{\substack{i,j=1 \\ i \neq j}}^{{\sf M}} p_{{\sf d_{ij}^2}}^{({\sf d}-1)} (0) \\
k_{{\sf UB}_{\sf d}} \cdot \sum_{\substack{i,j=1 \\ i \neq j}}^{{\sf M}} p_{{\sf d_{ij}^2}}^{({\sf d}-1)} (0), & f \left({\sf snr}\right) \geq k_{{\sf UB}_{\sf d}} \cdot \sum_{\substack{i,j=1 \\ i \neq j}}^{{\sf M}} p_{{\sf d_{ij}^2}}^{({\sf d}-1)} (0)
\end{array}\right.
\end{equation}
\vspace{0.25cm}
\begin{equation}
f_2 \left({\sf snr}\right) = \left\{\begin{array}{lc}
f \left({\sf snr}\right) - k_{{\sf UB}_{\sf d}} \cdot \sum_{\substack{i,j=1 \\ i \neq j}}^{{\sf M}} p_{{\sf d_{ij}^2}}^{({\sf d}-1)} (0), & f \left({\sf snr}\right) \geq k_{{\sf UB}_{\sf d}} \cdot \sum_{\substack{i,j=1 \\ i \neq j}}^{{\sf M}} p_{{\sf d_{ij}^2}}^{({\sf d}-1)} (0) \\
0, & \text{otherwise}
\end{array}\right.
\end{equation}
\vspace{0.25cm}
\begin{equation}
f_3 \left({\sf snr}\right) = \left\{\begin{array}{lc}
- f \left({\sf snr}\right) + k_{{\sf LB}_{\sf d}} \cdot \sum_{\substack{i,j=1 \\ i \neq j}}^{{\sf M}} p_{{\sf d_{ij}^2}}^{({\sf d}-1)} (0), & f \left({\sf snr}\right) \leq k_{{\sf LB}_{\sf d}} \cdot \sum_{\substack{i,j=1 \\ i \neq j}}^{{\sf M}} p_{{\sf d_{ij}^2}}^{({\sf d}-1)} (0) \\
0, & \text{otherwise}
\end{array}\right.
\end{equation}
\vspace{0.25cm}
and the functions:
\vspace{0.25cm}
\begin{equation}
g_1 \left({\sf snr}\right) = \left\{\begin{array}{lc}
g \left({\sf snr}\right), & k'_{{\sf UB}_{\sf d}} \cdot \sum_{\substack{i,j=1 \\ i \neq j}}^{{\sf M}} p_{{\sf d_{ij}^2}}^{({\sf d}-1)} (0) \leq g \left({\sf snr}\right) \leq k'_{{\sf LB}_{\sf d}} \cdot \sum_{\substack{i,j=1 \\ i \neq j}}^{{\sf M}} p_{{\sf d_{ij}^2}}^{({\sf d}-1)} (0) \\
k'_{{\sf UB}_{\sf d}} \cdot \sum_{\substack{i,j=1 \\ i \neq j}}^{{\sf M}} p_{{\sf d_{ij}^2}}^{({\sf d}-1)} (0), & g \left({\sf snr}\right) \leq k'_{{\sf UB}_{\sf d}} \cdot \sum_{\substack{i,j=1 \\ i \neq j}}^{{\sf M}} p_{{\sf d_{ij}^2}}^{({\sf d}-1)} (0) \\
k'_{{\sf LB}_{\sf d}} \cdot \sum_{\substack{i,j=1 \\ i \neq j}}^{{\sf M}} p_{{\sf d_{ij}^2}}^{({\sf d}-1)} (0), & g \left({\sf snr}\right) \geq k'_{{\sf LB}_{\sf d}} \cdot \sum_{\substack{i,j=1 \\ i \neq j}}^{{\sf M}} p_{{\sf d_{ij}^2}}^{({\sf d}-1)} (0)
\end{array}\right.
\end{equation}
\vspace{0.25cm}
\begin{equation}
g_2 \left({\sf snr}\right) = \left\{\begin{array}{lc}
g \left({\sf snr}\right) - k'_{{\sf LB}_{\sf d}} \cdot \sum_{\substack{i,j=1 \\ i \neq j}}^{{\sf M}} p_{{\sf d_{ij}^2}}^{({\sf d}-1)} (0), & g \left({\sf snr}\right) \geq k'_{{\sf LB}_{\sf d}} \cdot \sum_{\substack{i,j=1 \\ i \neq j}}^{{\sf M}} p_{{\sf d_{ij}^2}}^{({\sf d}-1)} (0) \\
0, & \text{otherwise}
\end{array}\right.
\end{equation}
\vspace{0.125cm}
and
\vspace{0.125cm}
\begin{equation}
g_3 \left({\sf snr}\right) = \left\{\begin{array}{lc}
- g \left({\sf snr}\right) + k'_{{\sf UB}_{\sf d}} \cdot \sum_{\substack{i,j=1 \\ i \neq j}}^{{\sf M}} p_{{\sf d_{ij}^2}}^{({\sf d}-1)} (0), & g \left({\sf snr}\right) \leq k'_{{\sf UB}_{\sf d}} \cdot \sum_{\substack{i,j=1 \\ i \neq j}}^{{\sf M}} p_{{\sf d_{ij}^2}}^{({\sf d}-1)} (0) \\
0, & \text{otherwise}
\end{array}\right.
\end{equation}

In view of the asymptotic expansions in \eqref{asymptotic_expansion1}, \eqref{asymptotic_expansion2}, \eqref{asymptotic_expansion3} and \eqref{asymptotic_expansion4}, it is possible to bound $f \left({\sf snr}\right)$ and $g \left({\sf snr}\right)$ for ${\sf snr} > {\sf snr_0}$, where ${\sf snr_0}$ is a sufficiently high value of ${\sf snr}$, as follows:
\begin{align}
k_{{\sf LB}_{\sf d}} \cdot \sum_{i=1}^{{\sf M}} \sum_{\substack{j=1 \\ j \neq i}}^{{\sf M}} p_{{\sf d_{ij}^2}}^{({\sf d}-1)} (0) - {\sf c_{LB}} \cdot \frac{1}{{\sf snr}}  \leq f \left({\sf snr}\right) \leq k_{{\sf UB}_{\sf d}} \cdot \sum_{i=1}^{{\sf M}} \sum_{\substack{j=1 \\ j \neq i}}^{{\sf M}} p_{{\sf d_{ij}^2}}^{({\sf d}-1)} (0) + {\sf c_{UB}} \cdot \frac{1}{{\sf snr}} \label{avg_mmse_dimension_bounds}
\end{align}
and
\begin{align}
k'_{{\sf UB}_{\sf d}} \cdot \sum_{i=1}^{{\sf M}} \sum_{\substack{j=1 \\ j \neq i}}^{{\sf M}} p_{{\sf d_{ij}^2}}^{({\sf d}-1)} (0) - {\sf c'_{LB}} \cdot \frac{1}{{\sf snr}}  \leq g \left({\sf snr}\right) \leq k'_{{\sf LB}_{\sf d}} \cdot \sum_{i=1}^{{\sf M}} \sum_{\substack{j=1 \\ j \neq i}}^{{\sf M}} p_{{\sf d_{ij}^2}}^{({\sf d}-1)} (0) + {\sf c'_{UB}} \cdot \frac{1}{{\sf snr}} \label{avg_mutual_information_dimension_bounds}
\end{align}
where ${\sf c_{LB}}$, ${\sf c_{UB}}$, ${\sf c'_{LB}}$ and ${\sf c'_{UB}}$ are positive constants. Note thus that $f_2 \left({\sf snr}\right)$, $f_3 \left({\sf snr}\right)$, $g_2 \left({\sf snr}\right)$ and $g_3 \left({\sf snr}\right)$ can be bounded as follows:
\begin{align}
\left|f_2 \left({\sf snr}\right)\right| = f_2 \left({\sf snr}\right) \leq {\sf c_{UB}} \cdot \frac{1}{{\sf snr}} = \mathcal{O} \left(\frac{1}{{\sf snr}}\right)
\end{align}
\begin{align}
\left|f_3 \left({\sf snr}\right)\right| = f_3 \left({\sf snr}\right) \leq {\sf c_{LB}} \cdot \frac{1}{{\sf snr}} = \mathcal{O} \left(\frac{1}{{\sf snr}}\right)
\end{align}
\begin{align}
\left|g_2 \left({\sf snr}\right)\right| = g_2 \left({\sf snr}\right) \leq {\sf c'_{UB}} \cdot \frac{1}{{\sf snr}} = \mathcal{O} \left(\frac{1}{{\sf snr}}\right)
\end{align}
\begin{align}
\left|g_3 \left({\sf snr}\right)\right| = g_3 \left({\sf snr}\right) \leq {\sf c'_{LB}} \cdot \frac{1}{{\sf snr}} = \mathcal{O} \left(\frac{1}{{\sf snr}}\right)
\end{align}
Note also that $f_1 \left({\sf snr}\right)$ and $g_1 \left({\sf snr}\right)$ are piecewise infinitely differentiable because $f \left({\sf snr}\right)$ and $g \left({\sf snr}\right)$ are infinitely differentiable and
\begin{align}
k_{{\sf LB}_{\sf d}} \cdot \sum_{\substack{i,j=1 \\ i \neq j}}^{{\sf M}} p_{{\sf d_{ij}^2}}^{({\sf d}-1)} (0) \leq f_1 \left({\sf snr}\right) \leq k_{{\sf UB}_{\sf d}} \cdot \sum_{\substack{i,j=1 \\ i \neq j}}^{{\sf M}} p_{{\sf d_{ij}^2}}^{({\sf d}-1)} (0)
\end{align}
and
\begin{align}
k'_{{\sf UB}_{\sf d}} \cdot \sum_{\substack{i,j=1 \\ i \neq j}}^{{\sf M}} p_{{\sf d_{ij}^2}}^{({\sf d}-1)} (0) \leq g_1 \left({\sf snr}\right) \leq k'_{{\sf LB}_{\sf d}} \cdot \sum_{\substack{i,j=1 \\ i \neq j}}^{{\sf M}} p_{{\sf d_{ij}^2}}^{({\sf d}-1)} (0)
\end{align}
Therefore, Theorems \ref{mmse_expansion} and \ref{mutual_information_expansion} follow immediately by noting that:
\begin{align}
f \left({\sf snr}\right) = f_1 \left({\sf snr}\right) + f_2 \left({\sf snr}\right) - f_3 \left({\sf snr}\right) = f_1 \left({\sf snr}\right) + \mathcal{O} \left(\frac{1}{{\sf snr}}\right) \label{expansion1}
\end{align}
and
\begin{align}
g \left({\sf snr}\right) = g_1 \left({\sf snr}\right) + g_2 \left({\sf snr}\right) - g_3 \left({\sf snr}\right) = g_1 \left({\sf snr}\right) + \mathcal{O} \left(\frac{1}{{\sf snr}}\right) \label{expansion2}
\end{align}
and, in view of \eqref{expansion1} and \eqref{expansion2} together with \eqref{limsup_mmse}, \eqref{liminf_mmse}, \eqref{limsup_i} and \eqref{liminf_i},
\begin{align}
\limsup_{{\sf snr} \to \infty} f \left({\sf snr}\right) = \limsup_{{\sf snr} \to \infty} f_1 \left({\sf snr}\right) = \overline{\epsilon}_{\sf d} ~~~~~~~~~ \text{and} ~~~~~~~~~ \liminf_{{\sf snr} \to \infty} f \left({\sf snr}\right) = \liminf_{{\sf snr} \to \infty} f_1 \left({\sf snr}\right) = \underline{\epsilon}_{\sf d}
\end{align}
and
\begin{align}
\limsup_{{\sf snr} \to \infty} g \left({\sf snr}\right) = \limsup_{{\sf snr} \to \infty} g_1 \left({\sf snr}\right) = \overline{\epsilon}'_{\sf d} ~~~~~~~~~ \text{and} ~~~~~~~~~ \liminf_{{\sf snr} \to \infty} g \left({\sf snr}\right) = \liminf_{{\sf snr} \to \infty} g_1 \left({\sf snr}\right) = \underline{\epsilon}'_{\sf d}
\end{align}

\vspace{-0.35cm}

\section*{Appendix F\\Proof of Lemma \ref{pe_deterministic}}

We obtain the lower bound to the error probability by using a genie that supplies the receiver with the true input vector and any of the other input vectors with equal probability. Then,
\begin{align}
{\sf P_{e}} \left({\sf snr};{\bf H}\right) &= \Pr \left(e|{\bf H}\right) = \frac{1}{{\sf M}} \sum_{i=1}^{{\sf M}} \Pr \left(e|{\bf x} = {\bf x}_i,{\bf H}\right) \nonumber \\
&\geq \frac{1}{{\sf M} ({\sf M}-1)} \sum_{i=1}^{{\sf M}} \sum_{\substack{j=1 \\ j \neq i}}^{{\sf M}} \Pr \left({\bf x}_i \rightarrow {\bf x}_j | {\bf x} = {\bf x}_i,{\bf H}\right) \nonumber \\
&= \frac{1}{{\sf M} ({\sf M}-1)} \sum_{i=1}^{{\sf M}} \sum_{\substack{j=1 \\ j \neq i}}^{{\sf M}} \frac{1}{2} \cdot {\sf erfc} \left(\sqrt{\frac{{\sf d_{ij}^2} \left({\bf H}\right){\sf snr}}{4}}\right) \triangleq {\sf P_{e_{LB}}} \left({\sf snr};\left({\bf H}\right)\right)
\end{align}
where $\Pr \left(e|{\bf x} = {\bf x}_i,{\bf H}\right)$ is the probability of error when ${\bf x}_i$ is transmitted given the channel matrix and $\Pr \left({\bf x}_i \rightarrow {\bf x}_j | {\bf x} = {\bf x}_i,{\bf H}\right)$ is the probability of choosing ${\bf x}_j$ over ${\bf x}_i$ when ${\bf x}_i$ is transmitted given the channel matrix. The inequality is due to the use of a genie.

We obtain the upper bound to the error probability by using the well-known union bound~\cite{Proakis08}. Then,
\begin{align}
{\sf P_{e}} \left({\sf snr};{\bf H}\right) &= \Pr \left(e|{\bf H}\right) = \frac{1}{{\sf M}} \sum_{i=1}^{{\sf M}} \Pr \left(e|{\bf x} = {\bf x}_i,{\bf H}\right) \nonumber \\
&\leq \frac{1}{{\sf M}} \sum_{i=1}^{{\sf M}} \sum_{\substack{j=1 \\ j \neq i}}^{{\sf M}} \Pr \left({\bf x}_i \rightarrow {\bf x}_j | {\bf x} = {\bf x}_i,{\bf H}\right) \nonumber \\
&= \frac{1}{{\sf M}} \sum_{i=1}^{{\sf M}} \sum_{\substack{j=1 \\ j \neq i}}^{{\sf M}} \frac{1}{2} \cdot {\sf erfc} \left(\sqrt{\frac{{\sf d_{ij}^2}\left({\bf H}\right){\sf snr}}{4}}\right) \triangleq {\sf P_{e_{UB}}} \left({\sf snr};{\bf H}\right)
\end{align}
where $\Pr \left(e|{\bf x} = {\bf x}_i,{\bf H}\right)$ is the probability of error when ${\bf x}_i$ is transmitted given the channel matrix and $\Pr \left({\bf x}_i \rightarrow {\bf x}_j | {\bf x} = {\bf x}_i,{\bf H}\right)$ is the probability of choosing ${\bf x}_j$ over ${\bf x}_i$ when ${\bf x}_i$ is transmitted given the channel matrix. The inequality is due to the union bound.

\vspace{-0.65cm}

\section*{Appendix G\\Proof of Lemma \ref{pe_fading}}

We follow the previous procedure to determine the asymptotic expansions of the lower and upper bounds to the average error probability. We obtain the lower and upper bounds to the average value of the error probability from the lower and upper bounds to the error probability by averaging over the fading statistics. The bounds are given by:
\begin{align}
{\sf \bar{P}_{e_{LB}}} \big({\sf snr}\big) &= \mathbb{E}_{{\bf H}} \left\{{\sf P_{e_{LB}}} \left({\sf snr};{\bf H}\right)\right\} =  \sum_{i=1}^{{\sf M}} \sum_{\substack{j=1\\j\neq i}}^{{\sf M}} \int_{0}^{\infty} \frac{1}{2 {\sf M} ({\sf M}-1)} \cdot {\sf erfc} \left(\sqrt{\frac{{\sf d_{ij}^2}{\sf snr}}{4}}\right) \cdot p_{{\sf d_{ij}^2}} \left({\sf d_{ij}^2}\right) d {\sf d_{ij}^2} \label{pe_lb} \\
{\sf \bar{P}_{e_{UB}}} \big({\sf snr}\big) &= \mathbb{E}_{{\bf H}} \left\{{\sf P_{e_{LB}}} \left({\sf snr};{\bf H}\right)\right\} = \sum_{i=1}^{{\sf M}} \sum_{\substack{j=1\\j\neq i}}^{{\sf M}} \int_{0}^{\infty} \frac{1}{2{\sf M}} \cdot {\sf erfc} \left(\sqrt{\frac{{\sf d_{ij}^2}{\sf snr}}{4}}\right) \cdot p_{{\sf d_{ij}^2}} \left({\sf d_{ij}^2}\right) d {\sf d_{ij}^2} \label{pe_ub}
\end{align}

We obtain the asymptotic expansions as ${\sf snr} \rightarrow \infty$ of the lower and upper bounds to the average value of the error probability by capitalizing on~\cite[Theorem 3.2]{Bleistein86} and \cite[Theorem 1.7.1]{Bleistein86}. The asymptotic expansions are given by:
\begin{align}
{\sf \bar{P}_{e_{LB}}} \big({\sf snr}\big) & = \sum_{n=0}^{N} \frac{1}{{\sf snr}^{n+1}} \cdot k''_{{\sf LB}_{n+1}} \cdot \left(\sum_{i=1}^{{\sf M}} \sum_{\substack{j=1 \\ j \neq i}}^{{\sf M}} p_{{\sf d_{ij}^2}}^{(n)} (0)\right) + \mathcal{O} \left(\frac{1}{{\sf snr}^{N+2}}\right)\\
{\sf \bar{P}_{e_{UB}}} \big({\sf snr}\big) & = \sum_{n=0}^{N} \frac{1}{{\sf snr}^{n+1}} \cdot k''_{{\sf UB}_{n+1}} \cdot \left(\sum_{i=1}^{{\sf M}} \sum_{\substack{j=1 \\ j \neq i}}^{{\sf M}} p_{{\sf d_{ij}^2}}^{(n)} (0)\right) + \mathcal{O} \left(\frac{1}{{\sf snr}^{N+2}}\right)
\end{align}
where the underlying auxiliary asymptotic sequence is $\big\{(1/{\sf snr})^{n}, n = 1, 2, \ldots\big\}$ and $k''_{{\sf LB}_n}$ and $k''_{{\sf UB}_n}$ are given by:
\begin{align}
k''_{{\sf LB}_n} = \int_{0}^{\infty} d t_{n-1}  \int_{t_{n-1}}^{\infty} d t_{n-2} \cdots
\int_{t_1}^{\infty} \frac{1}{2 {\sf M} ({\sf M}-1)} \cdot {\sf erfc} \left(\sqrt{\frac{t_0}{4}}\right) dt_0 \label{pe_lb_iterated_integral}
\end{align}
\begin{align}
k''_{{\sf UB}_n} = \int_{0}^{\infty} d t_{n-1}  \int_{t_{n-1}}^{\infty} d t_{n-2} \cdots
\int_{t_1}^{\infty} \frac{1}{2{\sf M}} \cdot {\sf erfc} \left(\sqrt{\frac{t_0}{4}}\right) dt_0 \label{pe_ub_iterated_integral}
\end{align}
The values of $k''_{{\sf LB}_{n}}$ and $k''_{{\sf UB}_{n}}$ can also be computed by using
\begin{align}
{\sf erfc} \left(\sqrt{\frac{t_0}{4}}\right) = \frac{2}{\pi} \int_{0}^{\infty} \frac{e^{- \frac{t_0}{4} (x^2+1) }}{x^2+1} dx
\end{align}
in \eqref{pe_lb_iterated_integral} and \eqref{pe_ub_iterated_integral}, so that:
\begin{align}
k''_{{\sf LB}_n} &= \frac{1}{\pi} \int_{0}^{\infty} \frac{1}{{\sf M} \left({\sf M}-1\right)} \cdot \frac{4^{n}}{\left(x^2+1\right)^{n+1}} dx = \frac{1}{2 {\sf M} ({\sf M}-1)} \cdot \frac{4^n}{\sqrt{\pi}} \cdot \frac{\Gamma \left(n+1/2\right)}{\Gamma \left(n+1\right)}
\end{align}
\begin{align}
k''_{{\sf UB}_n} &= \frac{1}{\pi} \int_{0}^{\infty} \frac{1}{{\sf M}} \cdot \frac{4^{n}}{\left(x^2+1\right)^{n+1}} dx = \frac{1}{2 {\sf M}} \cdot \frac{4^n}{\sqrt{\pi}} \cdot \frac{\Gamma \left(n+1/2\right)}{\Gamma \left(n+1\right)}
\end{align}
where $\Gamma \left(\cdot\right)$ is the Gamma function. These expansions hold, once again, for arbitrarily large $N$.

It is also simple to verify the conditions that justify the application of~\cite[Theorem 3.2]{Bleistein86} and \cite[Theorem 1.7.1]{Bleistein86}. For the asymptotic expansion of the individual integrals in \eqref{pe_lb} and \eqref{pe_ub}, it can be shown that
\begin{equation}
\Bigg| \int_{\lambda t}^{\infty} d t_{n-1}  \int_{t_{n-1}}^{\infty} d t_{n-2} \cdots
\int_{t_1}^{\infty} {\sf erfc} \left(\sqrt{\frac{t_0}{4}}\right) dt_0 \Bigg| \leq \alpha_n (t) \cdot \phi_n (\lambda), \qquad n = 1, 2, \ldots \label{pe_cond}
\end{equation}
with $\alpha_n (t) = 4^{n}$ and $\phi_n (\lambda) = 1$. Furthermore, by assumption, the functions $p_{{\sf d_{ij}^2}}^{(n)} \big({\sf d_{ij}^2}\big), n = 0, 1, \ldots,$ are continuous and integrable on $[0,\infty)$ and
\begin{equation}
\int_{0}^{\infty} \alpha_n \left({\sf d_{ij}^2}\right) p_{{\sf d_{ij}^2}}^{(n)} \left({\sf d_{ij}^2}\right) d {\sf d_{ij}^2} < \infty
\end{equation}
For the asymptotic expansion of the lower and upper bounds to the average value of the error probability, we use the fact that the asymptotic expansions are of Poincar\'e type to write the asymptotic expansion of the sum as the sum of the asymptotic expansions.

\vspace{-0.5cm}

\section*{Acknowledgements}

\vspace{-0.2cm}

The author acknowledges fruitful discussions with Sergio Verd\'u.

\vspace{-0.5cm}

\bibliographystyle{IEEEtran}

\bibliography{references}

\end{document}